\RequirePackage{silence}
\documentclass[10pt,aps,pra,twocolumn,showpacs,superscriptaddress,nofootinbib]{revtex4-1}
\usepackage[english]{babel} 
\usepackage[usenames, dvipsnames]{color} 
\usepackage{graphicx}
\usepackage{bm}
\usepackage{amsmath,amssymb}
\usepackage{times,verbatim}
\usepackage[pdftex,colorlinks=true,urlcolor=blue,linkcolor=Blue,
            citecolor=RedViolet]{hyperref}
\newcommand{\id}{{\sf 1 \hspace{-0.3ex} \rule{0.1ex}{1.52ex}\rule[-.01ex]{0.3ex}{0.1ex}}}

\begin{document}

\title{Exact Work Characteristic Functions for Time-Dependent Bosonic Quadratic Hamiltonians}
\author{F. Nicacio}
\email{nicacio@if.ufrj.br  }
\affiliation{
Instituto de F\'isica,
Universidade Federal do Rio de Janeiro, Rio de Janeiro, RJ 21941-972, Brazil
}                          
\author{R. N. P. Maia}
\email{pupio@macae.ufrj.br}
\affiliation{
Instituto Polit\'ecnico,
Universidade Federal do Rio de Janeiro, Maca\'e, RJ 27930-560, Brazil
}
              
\date{\today}


\begin{abstract}
\noindent 
We derive an exact analytical expression for the characteristic function of work in
closed many-body bosonic systems governed by general time-dependent quadratic
Hamiltonians.
This is made possible by adopting a unified prescription in which
all operators appearing in the characteristic function
are treated as elements of the inhomogeneous Metaplectic group.
The resulting expression involves a single symplectic
matrix, a single phase-space displacement, and a single accumulated phase.
It holds for any number of degrees of freedom, arbitrary time dependence of the
quadratic Hamiltonian, and any initial thermal equilibrium state of the system.
The formalism trades the explicit time-ordering problem associated
with the unitary evolution for the determination of a symplectic path in phase space,
from which the corresponding driving Hamiltonian can be reconstructed.
We also discuss the generalization to arbitrary nonequilibrium initial states
within the two-projective-measurement scheme for quantum work.
The same construction yields Loschmidt echoes and mixed-state total phases.
Using the derived characteristic function, we obtain formulas for
the Helmholtz free energy, the Jarzynski equality, the mean work,
the work variance, and the nonequilibrium lag.
Undisplaced Hamiltonians, sudden quenches, constant-Hessian protocols,
and the adiabatic limit follow as special cases
of the general result.
As an illustration, we analyze a one-dimensional system
constructed from an unconventional symplectic path
and exhibiting physically relevant limiting cases.
We then show how affine and purely symplectic contributions
affect work statistics and irreversibility.
\end{abstract}

\maketitle

\section{Introduction}   
Fluctuation theorems~\cite{campisi2011, esposito2009, goold2016}
occupy a central position in classical and quantum nonequilibrium thermodynamics
because they provide exact relations for the statistics of thermodynamic quantities
beyond the average constraints of the second law.
In classical systems, these relations apply to stochastic trajectories governed
by Langevin or master equations,
while in quantum systems they extend to closed and open dynamics
and to measurement processes.
Their validity has been experimentally investigated in platforms
including trapped ions~\cite{huber2008,an2015},
optical tweezers~\cite{martins2025},
superconducting circuits and quantum computers~\cite{masuyama2018,hahn2023},
and electronic conductors~\cite{utsumi2010,nakamura2010}.

A central result in this context is the Jarzynski equality,
originally derived for classical systems~\cite{jarzynski1997}
and later extended to quantum processes
through the two-projective-measurement scheme~\cite{talkner2007,talkner2008}.
It relates the exponential average of the work performed during an arbitrary
nonequilibrium protocol to the equilibrium free-energy difference,
\begin{equation}\label{jarzynski:eq} \left\langle e^{-\beta
W}\right\rangle={\rm e}^{-\beta\Delta F},
\end{equation}
where $\beta=(k_{\mathrm B}T)^{-1}$ and $\Delta F$
is the difference between the equilibrium free
energies associated with the final and initial Hamiltonians.
This relation enables free-energy recon-
struction in nanoscale, single-molecule, and biophysical sys-
tems~\cite{jarzynski2011,hoang2018,hummer2001,liphardt2002,collin2005}.

The characteristic function (CF) ---
the Fourier transformed probability density of work ---
encodes the complete work statistics, and previous work has computed
it for a number of driven quantum system including
parametric and forced harmonic
oscillators~\cite{engel2007,deffner2008,martins2025,talkner2008b,deffner2010},
coupled oscillators driven by a quantum agent~\cite{silva2023}, and harmonic
chains~\cite{paternostro2019}.
These models share an important structural property:
their Hamiltonians are quadratic in the canonical position and momentum operators.
Quadratic Hamiltonians (QH) generate affine symplectic transformations in phase space,
represented quantum mechanically by Weyl and Metaplectic
operators~\cite{littlejohn1986,almeida1998,gossonbook2006}.
This correspondence suggests that work statistics for the entire class
can be treated within a unified phase-space framework.

The main obstacle concerns computability: determining the CF for a specific
system requires an explicit expression for the evolution operator,
which in turn demands a temporal ordering \cite{sakuray},
together with its interplay with other operators in the definition of that function.
Existing analytical treatments therefore tend to focus
on specific models~\cite{talkner2008b,galve2009}
or on sudden or adiabatic regimes~\cite{paternostro2019,Ma}.
A general closed-form expression valid for arbitrarily time-dependent
many-body quadratic Hamiltonians is therefore desirable.

In this work, we derive such an expression for closed bo\-so\-nic systems with
any number of degrees of freedom. We represent the evolution operator,
the energy exponentials, and the thermal operator within
the inhomogeneous Metaplectic group \cite{littlejohn1986,gossonbook2006,nicacio2021}.
Their composition reduces the operator entering the CF to a single Metaplectic operator,
a single Weyl displacement operator, and an accumulated phase.
The resulting CF is then evaluated through the Weyl
representation \cite{littlejohn1986,almeida1998,gossonbook2006}.
We thus exchange the explicit quantum time-ordering problem
for the problem of finding a symplectic path in phase space \cite{gosson2015}.
In this inverse construction adopted here, one specifies this path and reconstructs
the corresponding time-dependent QH.
However,
for a prescribed Hamiltonian, the same formulation remains
applicable once one solves the associated symplectic equations
analytically or numerically.

The general treatment yields the Helmholtz free energy,
the Jarzynski equality, the mean work, the work variance,
and the nonequilibrium lag in terms of symplectic matrices,
phase-space displacements, and covariance matrices.
We also extend the construction to arbitrary nonequilibrium initial
states within the two-projective-measurement scheme,
in which the initial-state dependence is encoded by a single Weyl symbol.
Undisplaced Hamiltonians, sudden quenches, constant-Hessian protocols,
and the adiabatic limit follow as special cases.
The same algebraic structure also provides Loschmidt echoes and mixed-state
total phases.

As an illustration,
we consider a one-dimensional time-de\-pen\-dent dynamics constructed from an
unconventional symplectic path with physically relevant limiting cases.
This example allows us to distinguish the effects of the affine and purely symplectic
components on the work statistics and the nonequilibrium lag,
and to compare the exact dynamics with sudden-quench and adiabatic descriptions.

This paper is organized as follows.
Section \ref{sec:stow} reviews the driven thermodynamic process and the work
fluctuation properties available in the literature.
Section \ref{sec:MB} describes the mathematical tools that our calculations
require: the symplectic group, the inhomogeneous Metaplectic representation, the
Weyl symbol, the Wick rotation that produces thermal operators, and the
Williamson theorem.
Section \ref{sec:cfq} presents our general result for the work CF of quadratic
Hamiltonians.
Section \ref{sec:ssp} derives the resulting thermostatistical properties,
while Sec.~\ref{sec:DA} collects the limiting cases and dynamical approximations,
and Sec.~\ref{sec:rel} connects the formalism to Loschmidt echoes and geometric
phases.
Section \ref{application} works out the work statistics of a time-dependent
one-dimensional harmonic oscillator described by a non-trivial symplectic matrix.
Section \ref{conc} closes the paper with our conclusions and a discussion of the
present approach.

\section{Work fluctuation properties} \label{sec:stow} 
Consider a system initially in equilibrium with a thermal reservoir
at inverse temperature $\beta = 1/(k_\text{B} T)$, where $k_\text{B}$
is the Boltzmann constant.
A Hamiltonian operator $\hat H(\lambda_t)$ governs the dynamics of the
system, and depends on time through a control parameter
$\lambda : t \in \mathbb{R} \mapsto \lambda_t \in \mathbb{R}$.
%
%
At time $t=0$ the system is separated from the thermal reservoir and
undergoes a thermodynamic process during which some external agent 
changes the control parameter from $\lambda_0$ at time $t=0$ to $\lambda_\tau$
at time $t=\tau$.
During this interval, the system and the external agent exchange
an amount of energy identified as the work $W$.
The quantum nature of the system and the initial equilibrium
state are two independent sources of randomness in this process,
therefore, the work $W$ is a stochastic variable associated with a probability
distribution $P(W)$.

The CF associated with $P(W)$ is a complex-valued function,
$\chi : u \in \mathbb{R} \mapsto \chi (u) \in \mathbb{C}$,
defined as the expectation value 
\begin{gather}\label{wcf}
\chi (u) = \int dW \, {\rm e}^{i u W / \hbar} \, P(W)
\, ,
\end{gather}
which uniquely determines the distribution via a Fourier transform \cite{vanKampen}.
Furthermore, it allows for efficient computation of statistical moments through
differentiation \cite{vanKampen}:
\begin{gather}\label{eq:Wmoments}
\overline{ W^k }=
(-i\hbar)^k \frac{\partial^k}{\partial u^k} \chi (u) \Big|_{u=0}\, ,
\end{gather}
where the overline indicates a mean with respect to the
probability $P(W)$.

The time-ordered unitary operator
describes the evolution of the system state from $t = 0$ to $t = \tau$,
\begin{gather}\label{eq:Utdep}
\hat U_\tau = \overset{\rightharpoonup}{T}
{\rm e}^{ -\frac{i}{\hbar}\, \int_0^\tau \! \hat{\mathcal H}_t \, dt }
\Longleftrightarrow
i\hbar(\tfrac{\partial}{\partial t}\hat U_t ) \,
\hat U_t^{-1} = \hat{{\mathcal H}}_t
\, ,
\end{gather}
{\it i.e.}, the solution of the Schrödinger equation
%
%
for the system Hamiltonian $\hat{\mathcal H}_t \equiv \hat{\mathcal H}(\lambda_t)$.
Under the protocol described above, and assuming that the system is
initially prepared in an instantaneous thermal equilibrium state,
\begin{equation}\label{eq:ThermalState}
\hat \rho_t = \frac{{\rm e}^{-\beta \hat {\mathcal H}(\lambda_t)}}{Z(\lambda_t)} \, ,
\,\,\,
Z(\lambda_t) :=
\text{Tr} [ {\rm e}^{-\beta \hat {\mathcal H}(\lambda_t)} ] \, ,
\end{equation}
the CF in \eqref{wcf} takes the form \cite{talkner2007}
%
\begin{gather}\label{eq:chi}
\chi_\tau (u) = \text{Tr} \left[
\, \hat U_\tau^\dagger \, {\rm e}^{iu \hat {\mathcal H} (\lambda_\tau) /\hbar} \,
\hat U_\tau \, {\rm e}^{-i u \hat {\mathcal H} (\lambda_0) /\hbar } \, \hat \rho_0
\right] \, .
\end{gather}
The Jarzynski equality follows directly from this expression.
Indeed, evaluating the CF at
the purely imaginary argument $u=i\hbar\beta$,
one obtains \cite{talkner2007}
\begin{gather}\label{eq:ebetaW}
\begin{aligned}
\overline{{\rm e}^{-\beta W}}  =
\chi_\tau (u = i\hbar \beta) =
\frac{\mathcal Z(\lambda_\tau)}{\mathcal Z(\lambda_0)} \, .
\end{aligned}
\end{gather}
In Eq.~(\ref{jarzynski:eq}),
the quantity $\Delta F = F[\lambda_\tau] - F[\lambda_0]$ denotes the equilibrium
free-energy difference between the initial and final configurations of the control
parameter $\lambda_t$, where the Helmholtz free energy is defined as
$F[\lambda_t] = -\beta^{-1} \ln [ Z(\lambda_t) ]$.
Note that, while the initial state is an equilibrium thermal state,
the final state $\hat \rho_\tau = \hat U_\tau \hat \rho_0^\text{th}\hat U_\tau^\dag$
is generally not the thermal equilibrium state associated with the final Hamiltonian
$\hat {\mathcal H}(\lambda_\tau)$.

As a consequence of the convexity of the exponential function,
the Jarzynski equality implies the inequality
$\overline{W} \geq \Delta F$,
which states the second law of thermodynamics in statistical terms \cite{jarzynski1997}.
Since $\Delta F$ is an equilibrium quantity,
it is naturally associated with isothermal transformations.
The dissipated work \cite{goold2016},
$ W_\text{diss} =  \overline{W}  - \Delta F \geq 0$,
or, equivalently, the nonequilibrium lag \cite{paternostro2019},
\begin{gather}\label{eq:noneqlag}
L = \beta ( \overline{W} - \Delta F ) \geq 0
\, ,
\end{gather}
quantifies the degree of irreversibility of the nonequilibrium process.

All statistical properties of nonequilibrium work processes are therefore
encoded in the work distribution $P(W)$, or, equivalently,
in the CF defined in \eqref{eq:chi}.
Among these properties,
one may also derive relations connecting the statistics of forward and backward
processes, such as the Tasaki–Crooks fluctuation theorem \cite{tasaki-crooks}.

\section{Mathematical Background: 
Symplectic Geometry and Quantum Mechanics}\label{sec:MB} 
The association between phase-space symplectic geometry and unitary operators
in quantum mechanics \cite{almeida1998, gosson2015, combescure2005, nicacio2021}
provides the necessary framework to determine the CF
for the broad class of many-body systems governed by general
time-dependent QH.
To this end, in what follows we will present the necessary mathematical tools.

As a matter of notation, the dot product between
$u,v \in \mathbb R^n$ is defined by
$u \cdot v := u^\top v = \sum_{i = 1 }^n u_i v_i \in \mathbb R$
and the set of all $n \times n$ matrices is denoted by
$\text{Mat}(n, {\mathbb K})$,
where ${\mathbb K} = {\mathbb R, \mathbb C}$.
Consider the matrix
\begin{gather}\label{matJ}
{\sf J} =
\begin{pmatrix} \mathbf{0}_n & \mathbf{I}_n \\
               -\mathbf{I}_n & \mathbf{0}_n
\end{pmatrix} \in \text{Mat}(2n,\mathbb{R}) \, ,
\,\,\,
{\sf J}^\top = {\sf J}^{-1} = - {\sf J} \, ,
\end{gather}
where ${\bf 0}_n$ and ${\bf I}_n$ are, respectively,
the null and identity matrices in $\text{Mat}(n, \mathbb K)$.
The identity matrix in $\text{Mat}(2n, \mathbb K)$
will be denoted by ${\sf I}_{2n}$, or simply by $I$,
and we note that ${\sf I}_{2n} = {\bf I}_n \oplus {\bf I}_n$.
The real symplectic group is defined as
$\text{Sp}(2n,\mathbb{R}) =
\{ \mathsf{S} \in \text{Mat}(2n, \mathbb R) |
   \mathsf{S}^\top \! {\sf J} \mathsf{S} = {\sf J} \}$.
The matrix $\mathsf J$ belongs to $\text{Sp}(2n,\mathbb{R})$
and defines the antisymmetric (symplectic) bilinear form
$\mathsf J u \cdot v = - \mathsf J v \cdot u$,
which is invariant under symplectic transformations:
$\mathsf J u \cdot v = \mathsf J {\mathsf S} u \cdot {\mathsf S} v$,
$\forall {\sf S} \in \text{Sp}(2n,\mathbb{R})$.

The $2n$-component column-vector operator
\begin{gather} \label{eq:vecx}
\hat x =
( \hat q_1 , \ldots , \hat q_n , \hat p_1 , \ldots , \hat p_n )^\dagger
\end{gather}
describes a system with $n$ degrees of freedom, where
$\hat q_i$ and $\hat p_i$
are position and canonically conjugate momentum operators, respectively.
These operators satisfy the canonical commutation relations
$[\hat x_j, \hat x_k] = i\hbar {\sf J}_{jk}$,
where $\hat x_j$ denotes $j^\text{th}$ component
of the vector in (\ref{eq:vecx}).

\subsection{Time-dependent Quadratic Hamiltonians}\label{sec:TDQH}
Consider a generic system with $n$ degrees of freedom described by a quadratic
Hamiltonian%
\footnote{Notation remark:
Generic Hamiltonians are denoted in this text by $\hat {\mathcal H}$,
whereas quadratic Hamiltonians case denoted as $\hat H$.}
\begin{equation}\label{eq:QuadHam}
\hat H_t = \tfrac{1}{2}\, (\hat x - \xi_t) \cdot {\bf H}_{t} \,
                          (\hat x - \xi_t) + h_t \, .
\end{equation}
In this equation, the subscript $t$ indicates that the Hamiltonian is time‑dependent
through the functions
$\xi : t \in {\mathbb R} \mapsto \xi_t \in {\mathbb R}^{2n} $
(a $2n$-dimensional real column vector),
through the Hessian matrix
${\bf H}: t \in {\mathbb R} \mapsto {\bf H}_t \in \text{Mat}(2n, \mathbb R)$,
a $2n \times 2n$ real symmetric matrix, and also through
$h: t \in {\mathbb R} \mapsto h_t \in {\mathbb R}$,
a real scalar function representing a time-dependent offset of
the system energy.
Our main objective here is to compute the solution of
the Schrödinger equation (\ref{eq:Utdep})
for the Hamiltonian $\hat H_t$ in (\ref{eq:QuadHam}).

Let us define a family of matrices
${\sf S}_t \in {\rm Sp}(2n,\mathbb R)$
and a column vector
$z:t \in {\mathbb R} \mapsto z_t \in {\mathbb R}^{2n} $
as the solutions of the differential equations
\begin{equation}\label{eq:dif_eq}
\dot{\sf S}_t  = {\sf J}{\bf H}_t {\sf S}_t \, , \,\,\,
\dot{z}_t = - {\sf S}_t^{-1}{\sf J}{\bf H}_{t}\xi_t \, , 
\end{equation}
subjected to the initial conditions 
${\sf S}_{0} = {\sf I}_{2n}$ and
$z_0 = 0$.
The first equation generally requires a time-ordering due to the explicit
time dependence of ${\bf H}_t$,
whereas the second can be solved by direct integration,
\begin{equation}\label{eq:dif_eq_sol}
z_t = - {\int_0^t \! {\sf S}_{t'}^{-1}{\sf J}{\bf H}_{t'}\xi_{t'} \, dt'} \, ,
\end{equation}
Independently of analytical solvability,
both equations represent canonical transformations
in phase space:
$z_t$ describes a translation (affine transformation),
while ${\sf S}_t$ corresponds to a linear canonical transformation,
{\it i.e.}, a symplectic matrix \cite{arnold1989}.
Using these parameters,
the Hamiltonian in Eq.(\ref{eq:QuadHam}) can be rewritten as
\begin{equation}\label{eq:ImpQuadHam}
\hat H_t = -
\tfrac{1}{2} \hat x \cdot {\sf J} \dot{\sf S}_t {\sf S}_t^{-1} \hat x
- (\hat x - \tfrac{1}{2} \xi_t) \cdot {\sf J}{\sf S}_t\dot{z}_t + h_t \, .
\end{equation}

According to \cite{littlejohn1986,gosson2015},
this Hamiltonian generates a unitary evolution operator of the form
\begin{gather}\label{eq:UtdepQuad}
\hat U_t =
\overset{\rightharpoonup}{T}
{\rm e}^{ -\frac{i}{\hbar} \int_0^t \! \hat{H}_t dt} =
{\rm e}^{- \frac{i}{\hbar} \varphi_t }
\hat M_{{\sf S}_t}
\hat T_{z_t} \, ,
\end{gather}
where $\varphi: t \in \mathbb R \mapsto \varphi_t \in \mathbb R $
is a real phase given by
\begin{equation}\label{eq:phase}
\varphi_t =
\int_0^t \left[ h_{t'} + \tfrac{1}{2} \xi_{t'} \cdot
                         {\sf J} \tfrac{d}{ds}({\sf S}_{t'} z_{t'}) \right] ds \, ,
\end{equation}
or, equivalently, it is the solution of
$i\hbar (\tfrac{d}{dt} {\rm e}^{- \frac{i}{\hbar} \varphi_t }) \,
{\rm e}^{\frac{i}{\hbar} \varphi_t } = \dot{\varphi}_t$;
the operator $ \hat T_{z_t} $ is a Weyl (displacement)
operator \cite{littlejohn1986}, defined by
\begin{equation} \label{eq:weyl}
i\hbar (\tfrac{d}{dt} \hat T_{z_t} ){ \hat T_{z_t} }^{-1} =
{\sf J} (\hat x - \tfrac{1}{2} z_t ) \cdot \dot{z}_t
\Longleftrightarrow
\hat T_{z_t} =
{\rm e}^ {\tfrac{i}{\hbar} \hat x \cdot {\sf J} z_t}  \, ;
\end{equation}
and $\hat M_{{\sf S}_t}$ is a Metaplectic operator
\cite{littlejohn1986}, a solution of
\begin{equation}\label{eq:met_tdep}
i\hbar (\tfrac{d}{dt}\hat M_{{\sf S}_t}){\hat M_{{\sf S}_t}}^{-1} =
\tfrac{1}{2} \hat x \cdot (-{\sf J}\dot{\sf S}_t {\sf S}_t^{-1}) \hat x \, ,
\end{equation}
or $\hat M_{{\sf S}_t} =
\overset{\rightharpoonup}{T}
{\rm e}^{ -\frac{i}{2\hbar} \int_0^t \! \hat x \cdot
(-{\sf J}\dot{\sf S}_{t'} {\sf S}_{t'}^{-1}) \hat x \, ds} $, equivalently.

By taking the temporal derivative of the product in (\ref{eq:UtdepQuad})
and using the explicit expressions in Eqs.~(\ref{eq:phase})-(\ref{eq:met_tdep}),
one verifies that the Hamiltonian in (\ref{eq:ImpQuadHam}) generates the
Schrödinger equation (\ref{eq:Utdep}),
as shown in \cite{littlejohn1986}.

Arguably, at this stage, the unitary operator in (\ref{eq:UtdepQuad})
does not yet provide an explicit solution of the Schrödinger equation for
a generic QH of the form in (\ref{eq:QuadHam}).
Indeed, the unitary evolution operator in Eq.(\ref{eq:UtdepQuad})
is written in terms of the solutions of the differential equations
in Eqs.(\ref{eq:dif_eq}), where the determination of the symplectic matrix
generally also requires a time ordering due to the explicit
time dependence of the Hessian ${\bf H}_{t}$.

In practical situations, numerical methods furnish approximate solutions for ${\mathsf S}_t$
\cite{blanes}.
Nevertheless, we will circumvent the matrix time-ordering problem
by reversing the logic.
Instead of starting with a given time‑dependent Hessian ${\bf H}_t$
and solving the differential equation in Eq.(\ref{eq:dif_eq}) for ${\sf S}_t$,
one may first design a path ${\sf S}_t$ in the group $\text{Sp}(2n, \mathbb R)$
and then determine the corresponding ${\bf H}_t$ through Eq.(\ref{eq:dif_eq}).
For every differentiable symplectic path ${\sf S}_t$,
the matrix ${\bf H}_t = -{\sf J} \dot{\sf S}_t {\sf S}_t^{-1}$ is symmetric.
Conversely, every sufficiently regular quadratic Hamiltonian defines such
a path through Eq.(\ref{eq:dif_eq}).
Note that the dimension of $\text{Sp}(2n,{\mathbb R})$
is equal to the dimension of the subset of real symmetric matrices
in $\text{Mat}(2n,{\mathbb R})$, both being equal to $n(2n+1)$.
Once ${\sf S}_t$ is fixed,
one solves the second equation in (\ref{eq:dif_eq}) explicitly for $z_t$
and rewrites the original Hamiltonian in Eq.(\ref{eq:QuadHam})
in the equivalent form of Eq.(\ref{eq:ImpQuadHam}).
In this way, the evolution operator in Eq.(\ref{eq:UtdepQuad})
provides an exact representation of the quantum dynamics generated
by a generic time‑dependent QH,
entirely in terms of symplectic matrices and phase‑space translations \cite{gosson2015}.

\subsection{Subgroups of Unitary Operators}\label{sec:SUO}  
The Weyl operators defined in (\ref{eq:weyl}) belong to the Heisenberg group,
defined through the composition rule
\begin{equation}\label{eq:WeylComp}
\hat T_{z'}\hat T_{z''} = {\rm e}^{\frac{i}{2\hbar} {\sf J} z' \cdot z''} \,
                           \hat T_{z'+z''}
\end{equation}
together with the identities $\hat T_{0} = \hat {\id}$ and
$\hat T_{z}^\dagger = \hat T_{z}^{-1} = \hat T_{-z}$,
where $z, z', z'' \in \mathbb R^{2n}$ are column vectors.
These operators, also known as displacement operators
in the quantum-optics literature \cite{QuantOptics},
represent a simultaneous translation in position and momentum
\cite{littlejohn1986,almeida1998}:
\begin{equation}\label{eq:TZxTZ}
\hat T_{z}^\dagger \, \hat x \, \hat T_{z} = \hat x + z \, .
\end{equation}

Given ${\sf S} \in {\rm Sp}(2n,\mathbb R)$
and the vector $\hat x$ in (\ref{eq:vecx}),
a Metaplectic operator associated with ${\sf S}_t$ is defined by
\cite{littlejohn1986,gossonbook2006}%
\begin{gather}\label{eq:MSxMS}
\hat M_\mathsf{S}^\dagger \, \hat x \, \hat M_\mathsf{S} =
\mathsf{S}\, \hat x \, .
\end{gather}
For two matrices ${\sf S'}, {\sf S''} \in {\rm Sp}(2n,\mathbb R)$,
the Metaplectic group is defined by the composition rule
\begin{gather}\label{eq:met_composition}
\hat{M}_{\mathsf{S}'}\hat{M}_{\mathsf{S}''}
= \hat{M}_{\mathsf{S}' \mathsf{S}''} \, ,
\end{gather}
by the inversion formula
$\hat{M}_{\mathsf{S}}^\dagger = \hat{M}_{\mathsf{S}}^{-1} =
\hat{M}_{\mathsf{S}^{-1}}$,
and by the identity $\hat{M}_{{\sf I}_{2n}} = \hat {\id}$.
These rules define the Metaplectic group as a unitary
representation of ${\rm Sp}(2n,\mathbb R)$.
Furthermore, using Eqs.(\ref{eq:weyl}) and (\ref{eq:MSxMS}),
one can show that Metaplectic and Weyl operators satisfy the covariance relation
\cite{littlejohn1986,gossonbook2006}
\begin{equation}\label{MSTMS}
\hat M_\mathsf{S} \, \hat T_z \, \hat M_\mathsf{S}^\dagger =
\hat T_{{\sf S}z} \, .
\end{equation}

The composition laws described above define the
inhomogeneous Metaplectic group \cite{littlejohn1986,gossonbook2006},
whose elements are described by products of the form
$ {\rm e}^{i \phi } \hat M_{{\sf S}} \hat T_{z}$,
with $\phi \in \mathbb R$, ${\sf S} \in {\rm Sp}(2n, \mathbb R)$
and $z \in \mathbb R^{2n}$, as already anticipated in Eq.(\ref{eq:UtdepQuad}).

\subsection{Weyl Representation}\label{sec:WR} 
The set of operators acting on the Hilbert space of
a continuous variable quantum system with $n$ degrees of freedom,
described by the canonical operator $\hat x$, is a vector space \cite{nicacio2021}.
The set of all Weyl operators provides a complete basis in this operator space,
allowing any operator $\hat O$ to be uniquely expanded as \cite{almeida1998}
\begin{equation}\label{eq:WeylExp}
\hat O = \int_{\mathbb R^{2n}} \frac{d^{2n}\eta}{(2\pi\hbar)^n}
\, \hat T_\eta \, O(\eta) \, ,
\,\,\,
O(\eta) = {\rm Tr}(\hat O \hat T_\eta^\dagger) \, ,
\end{equation}
where $O(\eta)$ is the Weyl symbol of the operator $\hat O$,
defined through the Hilbert–Schmidt inner product with the Weyl operator
$\hat T_\eta$.
For the Metaplectic operator $\hat M_{{\sf S}}$,
the corresponding Weyl symbol reads \cite{mehlig,almeida1998,gossonbook2006}
\begin{equation} \label{eq:Hsmet}
{M}_\mathsf{S}(\eta) =
\frac{
\exp\left[
-\frac{i}{4\hbar} \eta \cdot
{\sf J}\mathbf{C}_{{\sf S}}^{-1}
{\sf J}\eta\right]}
{\sqrt{\det\left({\sf S} - {\sf I}\right)}},
\end{equation}
where ${\bf C}_\mathsf{S} \in {\rm Mat}(2n,\mathbb R)$
denotes the {\it Cayley parametrization} of ${\sf S} \in {\rm Sp}(2n,\mathbb R)$,
defined as
\begin{equation}\label{eq:Cayley}
{\bf C}_\mathsf{S} :=
-{\sf J} \frac{(\mathsf{S} - {\sf I})}
{(\mathsf{S} + {\sf I})}
= {\bf C}_\mathsf{S}^\top = - {\bf C}_\mathsf{S^{-1}}.
\end{equation}
Note that, since $\hat T_0 = \hat \id$, one has $O(0) = {\rm Tr}(\hat O)$.
In particular, one obtains the trace of a Metaplectic operator by evaluating
its Weyl symbol at the origin
\begin{equation}\label{eq:TrM}
{\rm Tr}(\hat M_{{\sf S}}) = {M}_\mathsf{S}(0) =
\frac{1}{\sqrt{\det({\sf S} - {\sf I})}} \, .
\end{equation}

Ultimately, in the present work,
we express the work CF in Eq.(\ref{eq:chi}) in
terms of Weyl symbols of Metaplectic operators of the form in Eq.(\ref{eq:Hsmet}).

\subsection{Time-Independent Quadratic Hamiltonians}\label{sec:TIQH} 
The QH (\ref{eq:QuadHam}) is time-independent
if ${\bf H}_t = {\bf H}$, $\xi_t = \xi$, $h_t = h$, $\forall t$
and we write $\hat H$ instead of $\hat H_t$.
In this case, the differential equations (\ref{eq:dif_eq})
are analytically solvable without the need
for time ordering, yielding
\begin{equation}\label{eq:Sz_tindep}
{\sf S}_t = \text{e}^{t {\sf J} {\bf H}} \in {\rm Sp}(2n,{\mathbb R}) \, ,
\,\,\,
z_t = ({\sf S}_{-t} - {\sf I}) \xi \in \mathbb R^{2n}\, .
\end{equation}
The symplectic matrix ${\sf S}_t$ thus defines a one-parameter subgroup of
$\text{Sp}(2n,{\mathbb R})$ satisfying
${\sf S}_t^{-1} = {\sf J}{\sf S}_t^{\top}{\sf J}^\top = {\sf S}_{-t}$.
Using the explicit expressions for ${\sf S}_t$ and $z_t$,
one can explicitly integrate, for time-independent parameters,
the equations that define the phase $\varphi_t$,
the Weyl operator $\hat T_{z_t}$, and the Metaplectic operator $\hat{M}_{{\sf S}_t}$
in Eqs.(\ref{eq:phase}-\ref{eq:met_tdep}).
Consequently, the unitary evolution operator in Eq.(\ref{eq:UtdepQuad}) reduces to
\begin{equation} \label{eq:Utindep}
\hat U_t = {\rm e}^{ -\frac{i}{\hbar} \hat{H} t } =
{\rm e}^{- \frac{i}{\hbar} \varphi_t }
\hat M_{{\sf S}_t}
\hat T_{z_t} \, ,
\end{equation}
where
\begin{equation} \label{eq:UtindOp}
\begin{aligned}
&\varphi_t = h t +  \tfrac{1}{2} {\sf J}\xi \cdot {\sf S}_t \xi \, , \\
&\hat T_{z_t} = {\rm e}^ {\tfrac{i}{\hbar} \hat x \cdot {\sf J} z_t}  \, , \\
&\hat M_{{\sf S}_t} =
{\rm e}^{ -\frac{i}{2\hbar} \hat x \cdot {\bf H}\hat x \, t } \, .
\end{aligned}
\end{equation}
Alternatively, using (\ref{eq:TZxTZ}), one can write that
\begin{equation} \label{eq:Utindep2}
\hat H =
\tfrac{1}{2} \hat{T}_{\xi} (\hat x \cdot
{\bf H} \hat x)\hat{T}_{\xi}^\dag + h
\Longleftrightarrow
\hat U_t =
{\rm e}^{- \frac{i}{\hbar} h t }
\hat T_{\xi} \hat M_{{\sf S}_t}
\hat T_{\xi}^\dagger
\end{equation}
and applying the covariance relation (\ref{MSTMS}) in the form
$\hat T_{\xi} \hat M_{{\sf S}_t}$ $= \hat M_{{\sf S}_t} \hat T_{{\sf S}_t\xi}$,
the composition in (\ref{eq:WeylComp}), and noting that ${\sf J}\xi \cdot \xi = 0$ for any $\xi$,
recovers the unitary operator in (\ref{eq:Utindep}).

In the subsequent analysis of work statistics,
we further need to consider solutions of a ``$u$-independent''
Schrödinger equation for a time-dependent Hamiltonian:
\[
i\hbar (\tfrac{d}{du} {\rm e}^{-iu \hat H_t /\hbar} )
{\rm e}^{i u \hat H_t /\hbar} = \hat H_t \, ,
\]
which, for the QH (\ref{eq:QuadHam})
and by analogy with (\ref{eq:Utindep}), may be written as
\begin{equation} \label{eq:Unit_u}
{\rm e}^{-\frac{i}{\hbar} u \hat H_t} =
{\rm e}^{- \frac{i}{\hbar} \varphi_u^{(t)} }
\hat M_{{\sf S}_u^{(t)}}
\hat T_{z_u^{(t)}} =
{\rm e}^{- \frac{i}{\hbar} h_t u }
\hat T_{\xi_t} \hat M_{{\sf S}_u^{(t)}}
\hat T_{\xi_{t}}^\dagger \, .
\end{equation}
The second equality above is obtained in analogy to Eq.(\ref{eq:Utindep2})
and, analogously to Eq.(\ref{eq:UtindOp}), one has
\begin{equation} \label{eq:Utindep3}
\begin{aligned}
&\varphi_u^{(t)} =
h_t u  +  \tfrac{1}{2}{\sf J} \xi_t \cdot {\sf S}_u^{(t)} \xi_t \, , \\
&\hat T_{z_u^{(t)}} =
{\rm e}^ {\tfrac{i}{\hbar} \hat x \cdot {\sf J} z_u^{(t)}}  \, , \\
&\hat M_{{\sf S}_u^{(t)}} =
{\rm e}^{ -\frac{i}{2\hbar} \hat x \cdot {\bf H}_t\hat x \, u } \, .
\end{aligned}
\end{equation}
The parameters contained in the above operators,
according to Eqs.(\ref{eq:dif_eq}) and (\ref{eq:Sz_tindep}),
are solutions of the differential equations
\begin{equation}\label{eq:difSz_uindep}
\tfrac{d}{du}{\sf S}_u^{(t)} =
{\sf J}{\bf H}_t {\sf S}_u^{(t)} \, , \,\,\,
%
\tfrac{d}{du}{z}_u^{(t)} =
- {\sf J}{\bf H}_{t} {\sf S}_{-u}^{(t)} \xi_t \, ,
\end{equation}
that is,
\begin{equation}\label{eq:Sz_uindep}
\begin{aligned}
{\sf S}_u^{(t)}  =
\text{e}^{u {\sf J} {\bf H}_t } \in \text{Sp}(2n,\mathbb R) , \,
z_u^{(t)} =   ({\sf S}_{-u}^{(t)} - {\sf I}) \xi_t \in {\mathbb R}^{2n} .
\end{aligned}
\end{equation}

This representation will play a crucial role in the construction of
the work CF,
as it allows all exponentials of QHs to be expressed
in a unified Metaplectic‑Weyl form.

\subsection{Wick Rotation and Thermal Operators}\label{sec:WRTO} 
Expanding the Metaplectic operator in (\ref{eq:UtindOp}),
one can directly verify the relation in (\ref{eq:MSxMS}) \cite{littlejohn1986}:
\begin{equation*} 
\tfrac{d}{dt} ( \hat M_{\mathsf{S}_t}^\dagger \, \hat x \,
                \hat M_{\mathsf{S}_t}) = {\sf J}{\bf H} \, \hat x
\Longrightarrow \hat M_{\mathsf{S}_t}^\dagger \, \hat x \,
                \hat M_{\mathsf{S}_t} = \mathsf{S}_t \hat x \, ,
\end{equation*}
and, thanks to this relation, we can also write
\[
\tfrac{d}{d\beta} ( \hat M_{{\rm S}_\beta}^\dagger \, \hat x \,
                \hat M_{{\rm S}_\beta})
= -i\hbar{\sf J}{\bf H} \, \hat x
\Longrightarrow
\hat M_{{\rm S}_\beta}^\dagger \, \hat x \,
\hat M_{{\rm S}_\beta} = {\rm S}_\beta \hat x \, .
\]
Consequently, for time‑independent QHs, performing a Wick rotation
$t \mapsto - i\hbar \beta$ in the Metaplectic operator $\hat M_{{\sf S}_t}$,
one obtains the thermal operator
\begin{gather} \label{eq:met_thl}
{\rm e}^{ -\frac{\beta}{2} \hat x \cdot {\bf H} \hat x } =
\hat M_{{\rm S}_\beta} =
{\rm e}^{ -\frac{i}{2\hbar} \hat x \cdot {\bf H} \hat x \, t } |_{t = - i\hbar \beta} \, ,
\end{gather}
which corresponds to the Wick-rotated symplectic matrix \cite{nicacio2021}
\begin{equation} \label{eq:wsymp}
{\rm S}_\beta := {\sf S}_{t}\big|_{t = - i\hbar \beta}
= \text{e}^{- i\hbar \beta {\sf J} {\bf H} } \, .
\end{equation}
Note that ${\rm S}_\beta^{-1} = {\rm S}_{-\beta} = {\rm S}_\beta^{\ast}$,
while ${\rm S}_\beta^{\dagger} = {\sf J} {\rm S}_\beta{\sf J}^\top$ and, therefore,
${\rm S}_\beta \in {\rm Sp}(2n,\mathbb C)$.

In the CF (\ref{eq:chi}),
as well as in the calculations involving thermodynamic quantities,
we must consider thermal operators associated with instantaneous
Gibbs states of time-dependent Hamiltonians of the form (\ref{eq:QuadHam}).
Following the same reasoning above, performing the Wick rotation
$u \mapsto - i\hbar \beta$ in (\ref{eq:Unit_u})
yields the instantaneous thermal operator
\begin{equation}\label{eq:th_oper}
{\rm e}^{ -\beta \hat{H}_t } =
{\rm e}^{- \beta h_t }
\hat T_{\xi_t} \hat M_{{\rm S}_\beta^{ (t) }} \hat T_{\xi_t}^\dagger \, ,
\,\,\,
{\rm S}_\beta^{(t)} := \text{e}^{- i\hbar \beta {\sf J} {\bf H}_t } \, .
\end{equation}
From this operator, one can determine the partition function
$\mathcal{Z}_t := {\rm Tr} \, {\rm e}^{ -\beta \hat{H}_t }$,
which,
employing the cyclicity of the trace,
the unitarity of the Weyl operators
$\hat T_{\xi_t}^\dagger \hat T_{\xi_t} = \hat \id $,
and the formula for the trace of a Metaplectic operator (\ref{eq:TrM}),
becomes
\begin{align} \label{eq:pf}
\mathcal{Z}_t =
{\rm e}^{- \beta h_t } \, {\rm Tr} \hat M_{{\rm S}_\beta^{ (t) }}
= {\rm e}^{- \beta h_t }
     | \det ( {\rm S}_\beta^{(t)} - {\sf I} ) |^{-\frac{1}{2}}   \, .
\end{align}
The appearance of the modulus in this formula is related to the choice of
the square root branch in (\ref{eq:Hsmet}).
This will be discussed in Sec.\ref{sec:watcf}.
Consequently, the instantaneous equilibrium state
for the Hamiltonian in (\ref{eq:QuadHam}) is
\begin{equation} \label{eq:thermHt}
\hat \rho_t^\text{th} = \frac{{\rm e}^{-\beta \hat H_t}}{\mathcal Z_t} =
\frac{ \hat T_{\xi_t} \hat M_{{\rm S}_\beta^{ (t) }} \hat T_{\xi_t}^\dagger }
     { | \det ( {\rm S}_\beta^{(t)} - {\sf I}
                   ) |^{-\frac{1}{2}} } \, .
\end{equation}
%

The Wick rotation in (\ref{eq:met_thl}) can formally be performed for any Hamiltonian.
However, the convergence of the partition function defined in
Eq.(\ref{eq:ThermalState}) requires a Hamiltonian bounded from below.
For QHs, this condition is equivalent to requiring
that the Hessian be positive definite, ${\bf H}_t > 0,\forall t$.
We refer to such Hamiltonians as {\it elliptic} \cite{nicacio2021}.

\subsection{Williamson Theorem}                                  

For any real positive-definite matrix, and hence for the positive-definite
Hessians considered here,
one can perform a symplectic diagonalization by
invoking the Williamson theorem \cite{williamson1936,arnold1989,nicacio17}:
If ${\bf H}_t > 0,\forall t$,
there exists ${\sf S}_{{\bf H}_t} \in {\rm Sp}(2n,\mathbb R)$
such that
\begin{equation} \label{eq:sympdiag}
{\sf S}_{{\bf H}_t}^\top {\bf H}_t {\sf S}_{{\bf H}_t} =
{{\bf \Lambda}_{{\bf H}_t}}  \, ,
\end{equation}
where
${{\bf \Lambda}_{{\bf H}_t}} =
{\rm Diag}(\mu_1,...,\mu_n,\mu_1,...,\mu_n) \in {\rm Mat}(2n,\mathbb R)$
is the diagonal matrix containing the symplectic spectrum of ${\bf H}_t$,
composed by the positive symplectic eigenvalues $\mu_i > 0, \forall i $.
These eigenvalues are the roots of the secular equation
\begin{equation}\label{eq:secular}
\det[{\sf J}{\bf H}_t \pm i\mu{\sf I}] = 0 \, ,
\end{equation}
for $\sf J$ in (\ref{matJ}).
One obtains the diagonalizing symplectic matrix  ${\sf S}_{{\bf H}_t}$
following the procedure described in \cite{nicacio17}.

From this theorem,
we can obtain the eigenvalues of the complex symplectic matrix
${\rm S}_\beta^{(t)}$ defined in Eq.(\ref{eq:th_oper}),
\begin{equation}\label{eq:secular2}
\det[{\rm S}_\beta^{(t)} - {\rm e}^{\pm \hbar \beta \mu_i} {\sf I}] = 0 \, ,
\end{equation}
which shows that the spectrum of ${\rm S}_\beta^{(t)}$
consists exclusively of positive eigenvalues \cite{nicacio2021}.
These eigenvalues guarantee the convergence
of the partition function in (\ref{eq:ThermalState})
for elliptic Hamiltonians --- by virtue of the Williamson theorem,
this partition function coincides with that of a collection of
$n$ independent harmonic oscillators with frequencies $\mu_1,...,\mu_n$ \cite{nicacio2021}.
From Eq.(\ref{eq:sympdiag}),
we can obtain the explicit diagonalization of ${\rm S}_\beta^{(t)}$ \cite{nicacio2021},
\begin{equation*} 
\begin{aligned}
{\sf S}_{{\bf H}_t}^{-1} {\rm S}_\beta^{(t)} {\sf S}_{{\bf H}_t} & =
  \text{e}^{- i\hbar \beta {\sf J} {\sf S}_{{\bf H}_t}^{\top} {\bf H}_t {\sf S}_{{\bf H}_t} }
= \text{e}^{- i\hbar \beta {\sf J} {{\bf \Lambda}_{{\bf H}_t}} } \\
& = \cosh(\beta\hbar{{\bf \Lambda}_{{\bf H}_t}})
- i {\sf J} \sinh(\beta\hbar{{\bf \Lambda}_{{\bf H}_t}}) \, ,
\end{aligned}
\end{equation*}
where in the first equality we used the definition
in Eq.(\ref{eq:th_oper})
alongside the symplecticity of ${\sf S}_{{\bf H}_t}$;
in the second equality we used Eq.(\ref{eq:sympdiag}),
while in the third, we Taylor expanded the exponential,
noting that
$[{\sf J},{{\bf \Lambda}_{{\bf H}_t}}] = 0$.

The quantum version of the Williamson decomposition
is obtained considering the Metaplectic representation
$\hat M_{{\sf S}_{{\bf H}_t}}$ of the symplectic matrix ${{\sf S}_{{\bf H}_t}}$
in (\ref{eq:sympdiag}).
Using the Weyl and
Metaplectic covariance respectively expressed as
(\ref{eq:TZxTZ}) and (\ref{eq:MSxMS}),
we may rewrite the Hamiltonian in (\ref{eq:QuadHam}) as
\begin{equation}\label{eq:H_decomp}
\begin{aligned}
\hat H_t & = \tfrac{1}{2}
\hat T_{\xi_t} (\hat x \cdot
{\sf S}_{{\bf H}_t}^{-\top} {\bf \Lambda}_{{\bf H}_t}{\sf S}_{{\bf H}_t}^{-1} \hat x )
\hat T_{\xi_t}^\dag  + h_t \\
& = \tfrac{1}{2}\hat T_{\xi_t} ({\sf S}_{{\bf H}_t}^{-1} \hat x \cdot
 {\bf \Lambda}_{{\bf H}_t}{\sf S}_{{\bf H}_t}^{-1} \hat x )
\hat T_{\xi_t}^\dag  + h_t \\
& = \hat T_{\xi_t} \hat M_{{\sf S}_{{\bf H}_t}}
\hat H^\text{oh} \,
\hat M_{{\sf S}_{{\bf H}_t}}^\dag \hat T_{\xi_t}^\dag  + h_t \, ,
\end{aligned}
\end{equation}
where we defined
\begin{equation} \label{eq:H_OH}
\hat H^\text{oh}(t) :=
\tfrac{1}{2}\, \hat x  \cdot {\bf \Lambda}_{{\bf H}_{t}}\hat x  =
\sum_{j=1}^n \tfrac{1}{2}\mu_j(t) (\hat q_j^2 + \hat p_j^2)
\end{equation}
as the sum of $n$ independent harmonic oscillators,
since ${\bf \Lambda}_{{\bf H}_{t}}$ is diagonal,
see Eq.(\ref{eq:sympdiag}) and Eq.(\ref{eq:vecx}).
The time-dependent symplectic eigenvalues $\mu_j$
are normal-mode frequencies of the system,
associated with the time-dependent eigenvalues
$E^{(j)}_{m_j} (t) = {\hbar \mu_j(t) (m_j + \tfrac{1}{2})}$
of the $j^\text{th}$ Hamiltonian operator
$\tfrac{1}{2}\mu_j(t) (\hat q_j^2 + \hat p_j^2)$.
These eigenvalues corresponds to
time-{\it independent} eigenvectors $|{m_j}\rangle$,
which belong to a Fock space for each $j = 1,..., n$.

The eigenvectors of $\hat H^\text{oh}$ are
tensor product of the $n$ Fock states $|m_j\rangle$.
We write them and their corresponding eigenvalues as
\begin{equation}\label{eq:Spec_H_OH}
|{\mathfrak m}\rangle :=
\bigotimes_{j=1}^{n}|m_j\rangle
\, , \,\,\,
E_{\mathfrak m}(t) := \sum_{j=1}^n E^{(j)}_{m_j}(t) \, .
\end{equation}
According to (\ref{eq:H_decomp}),
$\hat H_t$ and $\hat H^\text{oh}(t)$ are similar operators,
so the eigenvectors and eigenvalues of $\hat H_t$ read,
respectively,
\begin{equation} \label{eq:specQuad}
|\tilde{\mathfrak m}(t)\rangle :=
\hat T_{\xi_t} \hat M_{{\sf S}_{{\bf H}_t}} |{\mathfrak m}\rangle
\, , \,\,\,
E_{\mathfrak m}(t) +  h_t \, .
\end{equation}
Further, when we write this Hamiltonian as in (\ref{eq:ImpQuadHam}),
the unitary operator in (\ref{eq:UtdepQuad}) governs the evolution.

\subsection{What about the Characteristic Function?} \label{sec:watcf}
We now employ all the mathematical results presented so far in
the calculation of the CF in (\ref{eq:chi}).
The central idea is to express all operators appearing in that equation
as products of phase factors, Weyl operators, and Metaplectic operators.
Subsections \ref{sec:TDQH}, \ref{sec:TIQH}, and \ref{sec:WRTO}
develop this procedure systematically.
We compose these operators using the group properties
discussed in Subsec.\ref{sec:SUO}.
As a result, the total operator inside the trace in Eq.(\ref{eq:chi})
reduces to a product of a single phase factor,
a single Weyl operator, and a single Metaplectic operator.
Finally, we evaluate the trace in Eq.(\ref{eq:chi}) with the Weyl‑representation
techniques of Subsec.\ref{sec:WR}.
In this way, the CF takes the form of the Weyl symbol of an effective
Metaplectic operator, from which one derives several nonequilibrium
statistical properties of the system.

At this point,
it is appropriate to highlight some caveats regarding
the analytical derivation of the CF.
One may formally construct the instantaneous thermal operator in
(\ref{eq:th_oper}) through a Wick rotation for any quadratic Hamiltonian of the
form in Eq.(\ref{eq:QuadHam}).
However, the associated partition function converges only for elliptic
Hamiltonians, as discussed in Subsec.\ref{sec:WRTO}.
Furthermore, due to Eq.(\ref{eq:secular2}),
both the symbol in Eq.(\ref{eq:Hsmet}) and the trace formula in Eq.(\ref{eq:TrM})
converge for the symplectic matrix appearing in (\ref{eq:th_oper}) when the Hamiltonian
is elliptic.
For nonelliptic Hamiltonians, the partition function exhibits isolated
singularities at specific temperatures, as shown in Ref.~\cite{nicacio2021}.
Although the formal developments presented here are not affected by these singularities,
the physical interpretation of the corresponding thermal operators
as representing equilibrium states becomes delicate in such cases.

If the symplectic matrix $\mathsf S$ entering in Eqs.(\ref{eq:Hsmet}) and (\ref{eq:TrM})
possesses an eigenvalue equal to unity,
these formulas are no longer directly applicable in their present form.
Nevertheless, this is only a manageable artifact of the Weyl‑symbol representation.
Since the treatment of these singular cases is technically involved,
we do not pursue it here and instead refer the reader to Ref.\cite{nicacio2021}
for a detailed discussion.
Throughout this work, we therefore assume
that none of the effective symplectic matrices entering
the Weyl-symbol and trace formulas has an eigenvalue equal to unity.

Finally, it is necessary to clarify a subtle point concerning
the Metaplectic representation:
it is not a faithful representation of the symplectic group,
in the sense that each symplectic matrix $\sf S$ corresponds to two operators
$\pm \hat{M}_{\mathsf{S}}$.
As a consequence, the group composition law should, in general, be written as
$\hat{M}_{\mathsf{S}'} \hat{M}_{\mathsf{S}''} = \pm \hat{M}_{\mathsf{S}' \mathsf{S}''}$,
where the sign is determined by a
Maslov‑type index \cite{littlejohn1986,gossonbook2006}.
In this work, all formulas should therefore be understood modulo a global Maslov phase,
later fixed by continuity and normalization.
At the appropriate stage, physical constraints fix the overall sign
of operator compositions. For a comprehensive discussion of the representation theory of
Metaplectic operators and the associated Maslov indices, we refer the reader to
Ref.\cite{gossonbook2006}.

\section{The Characteristic Function for General Quadratic Hamiltonians} \label{sec:cfq}
As in Sec.~\ref{sec:stow}, a generic quench process
--- understood as continuous temporal change
in the Hamiltonian of the system or a driving protocol
--- is described by the dependence of the Hamiltonian on
a time‑varying control parameter $\lambda_t$.
For a quadratic time-dependent Hamiltonian (\ref{eq:QuadHam}),
one parametrizes the change protocol by
$\lambda_t = (\xi_t, {\bf H}_t, h_t)$ for $ 0 \le t \le \tau$.
From this point onward,
we will use  $\hat{H}(\lambda_t)$ or $\hat{H}_t$ interchangeably
for the same quadratic Hamiltonian in (\ref{eq:QuadHam}),
{\it i.e.}, $\hat{H}(\lambda_t) = \hat{H}_t$, $\forall t$.

\subsection{Characteristic Function for an Initial Thermal State}\label{sec:CFITS} 
We now compute explicitly the CF in Eq.(\ref{eq:chi})
for a generic quadratic Hamiltonian in Eq.(\ref{eq:QuadHam}).

Using the expression of the
instantaneous thermal operator derived in Eq.(\ref{eq:thermHt}),
the initial thermal equilibrium state entering the work protocol is
\begin{equation}\label{eq:thermH0}
\hat \rho_0^\text{th} = \frac{{\rm e}^{-\beta \hat H_0}}{\mathcal Z_0} =
\frac{ \hat T_{\xi_0} \hat M_{{\rm S}_\beta^{ (0) }} \hat T_{\xi_0}^\dagger }
     { | \det ( {\rm S}_\beta^{(0)} - {\sf I}
                   ) |^{-\frac{1}{2}} } \, ,
\end{equation}
where ${\rm S}_\beta^{(0)} := \text{e}^{- i\hbar \beta {\sf J} {\bf H}_0 }$.
We express each of the remaining operators in (\ref{eq:chi})
through the Metaplectic--Weyl decomposition of Sec.\ref{sec:MB}.
Specifically, we employ Eq.(\ref{eq:UtdepQuad}) for the evolution operator $\hat U_\tau$
and Eq.(\ref{eq:Unit_u}) for both operators  ${\rm e}^{i u \hat H_t /\hbar}$ and
${\rm e}^{-i u \hat H_0 /\hbar}$.

In this way, the operator product inside the trace in Eq.(\ref{eq:chi})
reads explicitly
\begin{equation}\label{eq:prod1}
\begin{aligned}
\hat U_\tau^\dagger \, {\rm e}^{iu \hat H_\tau /\hbar} \,
\hat U_\tau \, {\rm e}^{-i u \hat H_0 /\hbar } \, \hat \rho_0^\text{th} =
\frac{{\rm e}^{\frac{i}{\hbar}(\varphi_u^{(\tau)}-\varphi_u^{(0)})}}
     {  | \det ( {\rm S}_\beta^{(0)} - {\sf I}
                   )|^{-\frac{1}{2}}} \, \hat O \,
\end{aligned}
\end{equation}
where $\hat O$ denotes a product of
Weyl and Metaplectic operators
\[
\hat O =
\hat T_{z_\tau}^\dagger \hat M_{{\sf S}_\tau}^\dagger
\hat T_{z_u^{(\tau)}}^\dagger \hat M_{{\sf S}_u^{(\tau)}}^\dagger
\hat M_{{\sf S}_\tau} \hat T_{z_\tau}
\hat M_{{\sf S}_u^{(0)}}\hat T_{z_u^{(0)}}
\hat T_{\xi_0} \hat M_{{\rm S}_\beta^{ (0) }} \hat T_{\xi_0}^\dagger \, .
\]

Using repeatedly the covariance relation in Eq.(\ref{MSTMS}),
the group composition laws in
Eqs.(\ref{eq:WeylComp}) and (\ref{eq:met_composition}),
and rearranging the resulting expression,
this operator product reduces to the form
(see Appendix \ref{ap:QuadChiFunc})
\[
\hat O = \hat M_{{\bf S}_\beta^{(u,\tau)}} \, \hat T_{\zeta_\beta^{(u,\tau)}}
\]
up to a global phase.
Taking the trace and collecting all phase contributions,
the CF assumes the form
\begin{gather}\label{eq:chiQuadFunc}
\chi_\tau (u) = \frac{ {\rm e}^{\frac{i}{\hbar} \Phi_\beta^{(u,\tau)} }}
                     {  | \det ( {\rm S}_\beta^{(0)} - {\sf I}
                   ) |^{-\frac{1}{2}}} \,
\text{Tr} \left[ \hat M_{{\bf S}_\beta^{(u,\tau)}} \,
                 \hat T_{\zeta_\beta^{(u,\tau)}} \right] \, .
\end{gather}
Here, the symplectic matrix ${\bf S}_\beta^{(u,\tau)}$,
the phase-space vector $\zeta_\beta^{(u,\tau)}$,
and the accumulated phase $\Phi_\beta^{(u,\tau)}$ are given by
\begin{widetext}
\begin{equation}\label{eq:chiparam}
\begin{aligned}
&{\bf S}_\beta^{(u,\tau)} = {\sf S}_\tau^{-1}
{\sf S}_{-u}^{(\tau)} {\sf S}_\tau {\sf S}_u^{(0)}
{\rm S}_\beta^{(0)} \, , \\
&\zeta_\beta^{(u,\tau)}  =
[-( {\bf S}_\beta^{(u,\tau)} )^{-1} +
   {\rm S}_{-\beta}^{(0)}{\sf S}_{-u}^{(0)} ] \, z_\tau
- ( {\bf S}_\beta^{(u,\tau)} )^{-1}{\sf S}_\tau^{-1}
( {\sf S}_{-u}^{(\tau)} - {\sf I} ) \xi_\tau
+ ( {\rm S}_{-\beta}^{(0)} {\sf S}_{-u}^{(0)} - {\sf I} ) \xi_0 \, , \\
& \Phi_\beta^{(u,\tau)} = (h_\tau - h_0) u
+  \tfrac{1}{2}{\sf J} \xi_\tau \cdot {\sf S}_u^{(\tau)} \xi_\tau
- \tfrac{1}{2}{\sf J}\xi_0 \cdot {\sf S}_{u}^{(0)}{\rm S}_\beta^{(0)}\xi_0 \\
& \hspace{1.0cm} + \tfrac{1}{2}{\sf J} {\sf S}_{\tau} z_\tau \cdot
[ {\sf S}_{u}^{(\tau)} {\sf S}_\tau z_\tau
- ({\sf S}_{u}^{(\tau)}  - {\sf S}_{-u}^{(\tau)}  )\xi_\tau]
+ \tfrac{1}{2}{\sf J}\xi_0 \cdot
( {\sf I} - {\rm S}_u^{(0)}{\rm S}_\beta^{(0)})\zeta^{(u,\tau)}_\beta \, ,
\end{aligned}
\end{equation}
\end{widetext}
where $z_t$ and ${\sf S}_t$ are the solutions of
the differential equations in (\ref{eq:dif_eq}),
${\rm S}_\beta^{(t)} = {\exp[-i\hbar \beta{\sf J} {\bf H}_t]}$
given by (\ref{eq:wsymp}), and
${\sf S}_{u}^{(t)}$ given in (\ref{eq:Sz_uindep}).
The functions $h_t$, $\xi_t$, and ${\bf H}_t$
are the Hamiltonian parameters, see Eq.(\ref{eq:QuadHam}).

At this stage,
the CF can be evaluated explicitly using the Weyl representation
of the Metaplectic operator. Employing the Weyl expansion in Eq.(\ref{eq:WeylExp})
alongside the symbol for a Metaplectic operator (\ref{eq:Hsmet}),
one finally obtains
\begin{gather}\label{eq:finalQuadChi}
\chi_\tau (u) =
\frac{ {\rm e}^{ \frac{i}{\hbar} \Phi_\beta^{(u,\tau)}
                 -\frac{i}{4\hbar} \zeta_\beta^{(u,\tau)} \cdot
                  {\sf J}\mathbf{C}_{{\bf S}_\beta^{(u,\tau)}}^{-1}
                  {\sf J}\zeta_\beta^{(u,\tau)}  }
                                                                         }
     { \sqrt{\det[ {\bf S}_\beta^{(u,\tau)} - {\sf I}] /
             \det[ {\rm S}_\beta^{(0)} - {\sf I}]}           }
\, ,
\end{gather}
where $\mathbf{C}_{{\bf S}_\beta^{(u,\tau)}}$
is the Cayley parametrization defined by formula (\ref{eq:Cayley}).
We choose the square-root branch in Eq.(\ref{eq:finalQuadChi})
by continuation from $u=0$ or $\tau = 0$,
where the normalization condition
fixes the physical sign.

It is worth emphasizing that we have deliberately
avoided tracking the Maslov‑type sign ambiguities associated
with the Metaplectic representation.
One fixes such signs {\it a posteriori}
by imposing the normalization condition
$\chi_\tau(u = 0) = \chi_0(u) = {\rm Tr}\hat \rho_0^\text{th} = 1$,
from the definition in (\ref{eq:chi}),
which uniquely determines the physical branch
of the square roots appearing in Eq.(\ref{eq:finalQuadChi}).
Indeed, setting $u = 0$ one finds ${\sf S}_{0}^{(t)} =  {\sf I}, \forall t$,
which implies ${\bf S}_\beta^{(0,\tau)} = {\rm S}_\beta^{(0)}$ and
\[
\zeta_\beta^{(0,\tau)}  =
( {\rm S}_{-\beta}^{(0)} - {\sf I} ) \xi_0 \, , \,\,\,
\Phi_\beta^{(0,\tau)}  = -\tfrac{1}{2}
{\sf J} \xi_0 \cdot {\rm S}_\beta^{(0)}\xi_0 \, ,
\]
which in turn, when inserted in (\ref{eq:finalQuadChi}),
gives $\chi_\tau(0) = 1$.
Further, for the CF regarded as a function of $\tau$,
we have
\[
{\bf S}_\beta^{(u,0)} = {\bf S}_\beta^{(0,\tau)}, \,\,\,
\zeta_\beta^{(u,0)} = \zeta_\beta^{(0,\tau)} , \,\,\,
\Phi_\beta^{(u,0)} = \Phi_\beta^{(0,\tau)} \, ,
\]
thus, by the same arguments as for $u=0$,
$\chi_0(u) = 1$.

\subsection{General Initial State} 
When the state at the beginning of the work protocol is not in thermal equilibrium
with a heat bath, the CF is no longer given by Eq.(\ref{eq:chi}).
Consider instead a generic initial state $\hat \rho$ and the spectral decomposition
of the initial Hamiltonian $\hat {\mathcal H}(\lambda_0) = \hat {\mathcal H}_0$,
\[
\hat {\mathcal H}_0 = \sum_{\mathfrak{n}_0}
E_{\mathfrak{n}_0} |\mathfrak{n}_0\rangle\langle \mathfrak{n}_0| \, ,
\]
where $E_{\mathfrak{n}_0}$ and $|\mathfrak{n}_0\rangle$ denote, respectively,
the eigenvalues and the eigenvectors of $\hat {\mathcal H}_0$,
assumed to be nondegenerate.
Following Ref.~\cite{talkner2008},
the CF corresponding to a work protocol
$\lambda_t = (\xi_t, {\bf {\mathcal H}}_t, h_t)$ for $ 0 \le t \le \tau$
reads
\begin{gather}\label{chi:c2}
\chi_\tau (u) = \text{Tr} \left[
\, \hat U_\tau^\dagger \, {\rm e}^{iu \hat {\mathcal H} (\lambda_\tau) /\hbar} \,
\hat U_\tau \, {\rm e}^{-i u \hat {\mathcal H} (\lambda_0) /\hbar } \, \bar \rho_0
\right] \, ,
\end{gather}
where $\bar\rho_0 :=
\sum_{\mathfrak{n}_0}
\langle \mathfrak{n}_0| \hat \rho | \mathfrak{n}_0 \rangle
|\mathfrak{n}_0\rangle \langle \mathfrak{n}_0|$
is the initial state $\hat \rho$ projected onto the
eigenbasis of the initial Hamiltonian $\hat {\mathcal H}_0$.

It is still possible to derive an analytic expression for
the CF when the work protocol is governed by a
time‑dependent quadratic Hamiltonian of the form in (\ref{eq:QuadHam}).
To this end, we consider the Weyl expansion (\ref{eq:WeylExp})
for the projected density operator:
\begin{equation}\label{eq:WeylExp2}
\bar \rho_0 = \int_{\mathbb R^{2n}} \frac{d^{2n}\eta}{(2\pi\hbar)^n}
\, \hat T_\eta \, \Psi_0(\eta) \, ,
\,\,\,
\Psi_0(\eta) = {\rm Tr}(\bar \rho_0 \hat T_\eta^\dagger) \, .
\end{equation}
The function $\Psi_0(\eta)$ is the Weyl symbol of the state $\bar \rho_0$,
also commonly referred to as the CF of the quantum state,
since it corresponds to the symplectic Fourier transform of
its Wigner function \cite{almeida1998}.

The replacement of the initial state $\hat\rho$ by its diagonal projection
$\bar\rho_0$ reflects the two-projective-measurement prescription for quantum work.
The first energy measurement dephases the state in the eigenbasis of $\hat H_0$,
so that coherences between distinct initial-energy eigenspaces are erased before
the unitary evolution takes place. Consequently, the CF depends
only on the initial energy populations and not on the off-diagonal elements of
$\hat\rho$ in this basis. In the phase-space formulation adopted below, the Weyl symbol
$\Psi_0(\eta)$ of $\bar\rho_0$ encodes these populations.

Inserting (\ref{eq:WeylExp2}) into (\ref{chi:c2}),
one obtains
\begin{eqnarray*} 
\chi_\tau (u)  &=& \int_{\mathbb R^{2n}} \frac{d^{2n}\eta}{(2\pi\hbar)^n}
\Psi_0(\eta) \times  \\
&\times& \text{Tr} \left[
\, \hat U_\tau^\dagger \, {\rm e}^{iu \hat H (\lambda_\tau) /\hbar} \,
\hat U_\tau \, {\rm e}^{-i u \hat H (\lambda_0) /\hbar } \, \hat T_\eta
\right] \, .
\end{eqnarray*}
One then evaluates the trace inside the integral
following the same reasoning as in Sec.\ref{sec:cfq},
and the CF becomes
\[
\chi_\tau (u)  =
\int_{\mathbb R^{2n}} \frac{d^{2n}\eta}{(2\pi\hbar)^n} \Psi_0(\eta)
{\rm e}^{\frac{i}{\hbar} \tilde{\Phi}}
\text{Tr} \left[ \hat M_{{\bf S}_0^{(u,\tau)}} \,
                 \hat T_{\tilde{\zeta}_0^{(u,\tau)}} \right] \, ,
\]
which,
upon using the Weyl representation of
the Metaplectic operator (\ref{eq:Hsmet}), yields
\begin{equation}\label{eq:chi_GenState}
\chi_\tau (u) = \int_{\mathbb R^{2n}}
\frac{ d^{2n}\!\eta \,
       {\rm e}^{ \frac{i}{\hbar} \tilde{\Phi} -\frac{i}{4\hbar} \tilde{\zeta} \cdot
       {\sf J}\mathbf{C}_{\tilde{\bf S}}^{-1} {\sf J} \tilde{\zeta}
                 }}
{(2\pi\hbar)^n \sqrt{\det(\tilde{\bf S} - {\sf I})}} \Psi_0(\eta) \, ,
\end{equation}
where we have defined
\begin{equation} \label{eq:chiparam3}
\begin{aligned}
\tilde{\bf S} & := {\sf S}_\tau^{-1} {\sf S}_{-u}^{(\tau)}
{\sf S}_\tau {\sf S}_u^{(0)} \, , \,\,\,
\tilde{\zeta} := {\zeta}_0^{(u,\tau)} + \eta \, , \\
\tilde{\Phi} & := \Phi_0^{(u,\tau)} + \tfrac{1}{2} {\sf J}\zeta_0^{(u,\tau)} \cdot \eta
\end{aligned}
\end{equation}
for ${\zeta}_0^{(u,\tau)}$ and $\Phi_0^{(u,\tau)}$
obtained by setting $\beta = 0$ (or ${\rm S}_\beta^{0} = {\sf I}$)
in the corresponding expressions in Eqs.(\ref{eq:chiparam}).
All temperature dependence in this expression enters
exclusively through the function $\Psi_0(\eta)$,
which fully characterizes the initial quantum state.
When the initial state is a thermal equilibrium state,
$\bar \rho_0 = \hat \rho_0^\text{th}$,
the integral expression in (\ref{eq:chi_GenState}) reduces exactly
to the formula in  (\ref{eq:finalQuadChi}).

The linear dependence of the parameters in Eq.(\ref{eq:chiparam3}) on
the integration variable $\eta$ implies that the kernel of the
integral in (\ref{eq:chi_GenState}) is Gaussian. Consequently,
once the function $\Psi_0(\eta)$ is known, the integral may be evaluated analytically.
One determines the function $\Psi_0(\eta)$ either directly from its definition
in (\ref{eq:WeylExp}) for $\hat O = \bar\rho_0$ or, equivalently,
as the symplectic Fourier transform of the Wigner function associated with the
state $\bar\rho_0$ \cite{almeida1998}.

\section{System Statistical Properties} \label{sec:ssp}
In this section, we analyze some thermostatistical properties of
systems described by general quadratic Hamiltonians.
Some of these properties arise as direct mathematical consequences
of the quadratic structure of the Hamiltonian,
while others follow from the general expression for the work characteristic
function derived in the previous section.

\subsection{Covariance Properties and Equilibrium States}
All thermodynamic quantities can be derived from the partition function.
For a quadratic Hamiltonian, Eq.(\ref{eq:pf}) shows that one can express these
quantities entirely in terms of the
symplectic matrices ${\rm S}_\beta^{(t)}$ \cite{nicacio2021}.
In what follows, we will explore some invariant properties of the state arising
from the symplectic invariance of the partition function.

Using Eqs.(\ref{eq:TZxTZ}) and (\ref{eq:MSxMS}), we have
\[
\hat M_{{\sf S}}\hat T_{\eta} \hat x \hat T_{\eta}^\dagger\hat M_{{\sf S}}^\dagger  =
{\sf S}^{-1}\hat x - \eta
\]
for any matrix ${\sf S} \in {\rm Sp}(2n,\mathbb R)$
and any vector $\eta \in {\mathbb R}^{2n}$.
As a consequence, the transformed Hamiltonian
\begin{equation} \label{eq:covHam}
\hat H'  :=
\hat M_{{\sf S}}\hat T_{\eta} \hat H_t \hat T_{\eta}^\dagger\hat M_{{\sf S}}^\dagger
= \tfrac{1}{2} \, (\hat x - \xi'_t) \cdot {\bf H}'_{t} (\hat x - \xi'_t) + h_t \, ,
\end{equation}
is of the same nature, a quadratic Hamiltonian in (\ref{eq:QuadHam})
but with
$\xi'_t = {\sf S}(\xi_t + \eta)$ and
${\bf H}'_{t} = {\sf S}^{-\top}{\bf H}_{t}{\sf S}^{-1}$.
While the Hamiltonian is covariant under
Weyl and Metaplectic conjugations,
the partition function remains invariant, since
${\rm Tr} \, {\rm e}^{ -\beta \hat{H}_t } =
 {\rm Tr} \, {\rm e}^{ -\beta \hat{H}'_t }$.
Consequently, these transformations preserve
all thermodynamic and statistical properties of the system \cite{nicacio2021}.
Note that this invariance is simply a manifestation
of the unitary invariance of the partition function due to trace cyclicity.
However, within the symplectic formalism,
one can thus characterize entire families of quadratic Hamiltonians
related by Weyl and Metaplectic transformations that share
the same thermodynamic behavior.
In Eq.(\ref{eq:pf}), this invariance is reflected in the identity
\begin{equation} \label{eq:invpf}
\det ( {\sf S}{\rm S}_\beta^{(t)}{\sf S}^{-1} - {\sf I} ) =
\det ( {\rm S}_\beta^{(t)} - {\sf I} ) \, ,
\end{equation}
since $\hat M_{\sf S} \hat M_{{\rm S}_\beta^{ (t) }} \hat M_{\sf S}^\dag =
\hat M_{{\sf S}{\rm S}_\beta^{(t)}{\sf S}^{-1}}$.

Note that, from Eq.(\ref{eq:secular2}),
the partition function in (\ref{eq:pf}) can be expressed in terms of the
symplectic eigenvalues of ${\bf H}_{t}$.
Equation (\ref{eq:secular}) also shows that these eigenvalues remain the same
for the whole class of quadratic Hamiltonians in (\ref{eq:covHam}),
so they constitute the truly relevant quantities that fix the thermodynamic
behavior of the system \cite{nicacio2021}.

For a generic quantum state $\hat \rho$,
we define the mean-value vector
$\langle \hat x \rangle = {\rm Tr}(\hat x \hat\rho) \in {\mathbb R}^{2n}$
and the real $2n \times 2n$ (symmetric) covariance matrix ${\bf V}$ with elements
\begin{equation}\label{eq:cm}
\mathbf V_{\! jk}  =
\tfrac{1}{2} {\rm Tr}
\left[
      \left\{ \hat x_j - \langle \hat x_j \rangle ,
      \hat x_k - \langle \hat x_k \rangle \right\}
      \hat \rho\right]
\end{equation}
for $j,k = 1,...,2n$.
For a system governed by the quadratic Hamiltonian (\ref{eq:QuadHam}),
one writes the internal energy $\langle\hat H \rangle = {\rm Tr}(\hat H_t \hat\rho)$
entirely in terms of these mean quantities
\begin{equation}\label{eq:meanenerg}
\begin{aligned}
\langle\hat H_t \rangle &= \tfrac{1}{2} \sum_{j,k = 1}^{2n} ({\bf H}_t)_{jk}
                                       {\rm Tr}[(\hat x-\xi_t)_j
                                                (\hat x-\xi_t)_k \hat \rho] + h_t \\
&= \tfrac{1}{2} {\rm Tr}({\bf H}_t {\bf V}) +
   \tfrac{1}{2} (\langle \hat x  \rangle- \xi_t) \cdot{\bf H}_t
                (\langle \hat x \rangle - \xi_t) + h_t \, ,
\end{aligned}
\end{equation}
where we used that ${\bf H}_t = {\bf H}_t^\top$
and replaced $\hat x_j \hat x_k$ from (\ref{eq:cm}).
This expression is manifestly covariant under the transformation in (\ref{eq:covHam}).

The instantaneous thermal equilibrium state in (\ref{eq:thermHt}),
associated with the Hamiltonian in (\ref{eq:QuadHam}),
is a Gaussian state with mean value and covariance matrix \cite{banchi}
\begin{equation} \label{eq:mvcmDefs}
\langle \hat x \rangle_t^\text{th} :=
{\rm Tr} \left(\hat x \hat \rho^\text{th}_t \right) = \xi_t \, , \,\,
{\mathbf V}_t^\text{th}  =
\tfrac{i}{2} \hbar \, {\rm coth}(\tfrac{i}{2}\hbar\beta {\sf J} {\bf H}_t) {\sf J} \, .
\end{equation}
Using the Williamson diagonalization in Eq.(\ref{eq:sympdiag}),
expanding the function ${\rm coth}(x)$ in Taylor series,
taking into account that $[{\sf J},{\bf \Lambda}_t] = 0$,
and using the symplecticity of ${\sf S}_{{\bf H}_t}$,
one rewrites the covariance matrix as
\begin{equation} \label{eq:cm_t}
{\mathbf V}_t^\text{th}  =
\tfrac{\hbar}{2} {\sf S}_{{\bf H}_t}
{\rm coth}(\tfrac{1}{2}\hbar\beta
{\bf \Lambda}_{{\bf H}_t}) {\sf S}_{{\bf H}_t}^\top
\end{equation}
and the corresponding mean energy (\ref{eq:meanenerg})
follows entirely in terms of the symplectic eigenvalues
\[
\langle\hat H_t \rangle^\text{th}
= \tfrac{1}{2} {\rm Tr}({\bf H}_t {\bf V}_t^\text{th}) =
\tfrac{\hbar}{2}\sum_{i = 1}^n \mu_i{\rm coth}(\tfrac{1}{2}\hbar\beta\mu_i)
+ h_t \, .
\]

Let us now consider the state obtained by evolving the initial thermal equilibrium
state $\hat \rho_0^\text{th}$ under the unitary dynamics generated by
the Hamiltonian (\ref{eq:QuadHam}):
\begin{equation}\label{eq:evolrho0}
\hat \rho_\tau = \hat U_\tau \hat \rho_0^\text{th} \hat U_\tau^\dag =
{\mathcal Z_0}^{-1}{\rm e}^{-\beta \, \hat U_\tau \hat H_0 \hat U_\tau^\dag} \, ,
\end{equation}
where we used the expression in (\ref{eq:thermHt}) for $t=0$.
Expressing $\hat U_\tau$ as in (\ref{eq:UtdepQuad}),
using the covariance of the Hamiltonian (\ref{eq:covHam})
[or directly the rules in (\ref{eq:TZxTZ}) and (\ref{eq:MSxMS})],
and the definitions in Eq.(\ref{eq:mvcmDefs}),
$\hat \rho_\tau$ is again a Gaussian state with mean-value vector and covariance matrix
\begin{equation} \label{eq:mvcmEvol0}
\begin{aligned}
\langle \hat x \rangle^\tau & =
{\sf S}_\tau(z_\tau + \langle \hat x \rangle_0^\text{th}) =
{\sf S}_\tau(z_\tau + \xi_0) \, ,  \\
\mathbf V^\tau  & =
{\sf S}_\tau {\bf V}_0^\text{th} {\sf S}_\tau^{\top} =
\tfrac{i}{2} \hbar \, {\rm coth}( \tfrac{i}{2}\hbar\beta
                                  {\sf S}_\tau{\sf J} {\bf H}_0 {\sf S}_\tau^{-1})
                                  {\sf J} \, ,
\end{aligned}
\end{equation}
where ${\sf S}_\tau$ and $z_\tau$ are the solutions of Eqs.(\ref{eq:dif_eq}).

Although $\hat\rho_\tau$ in (\ref{eq:evolrho0})
is unitarily equivalent to the initial thermal state
and is also a Gibbs state of the transformed Hamiltonian
$\hat U_\tau\hat H_0\hat U_\tau^\dagger$, it does not generally coincide
with the instantaneous equilibrium state (\ref{eq:thermHt}) for $t = \tau$.
This distinction is central to the emergence of
irreversibility in nonequilibrium work protocols.
The state in (\ref{eq:thermHt}) for $t = \tau$ is an
instantaneous equilibrium state associated with
the Hamiltonian $\hat H_\tau$,
while $\hat \rho_\tau$ is the result of the work protocol, dictated by $\hat U_t$,
applied to the initial thermal equilibrium state $\hat \rho_0^\text{th}$.


Thermal equilibrium states for QH have been described by
complex symplectic matrices ${\rm S}_{\beta}^{(t)}$,
which is essential for deriving the general results in previous sections
through the Weyl-Metaplectic setup.
A generic Gaussian state is uniquely specified by a covariance matrix and
a mean value vector which, for thermal equilibrium states, are given
in (\ref{eq:mvcmDefs}). The algebraic identity
\[
(1-{\rm e}^{2x})^{-1} = \tfrac{1}{2}[1 - {\rm coth}(x)]
\]
shows the equivalence of the two descriptions.
From the definition of ${\rm S}_{\beta}^{(t)}$ in (\ref{eq:th_oper}) and
noting that ${\bf H}_t$ is non-singular (positive-definite), then
\begin{equation} \label{eq:ident_S-V}
\begin{aligned}
( {\sf I}_{2n} - {\rm S}_{-\beta}^{(t)})^{-1} & =
\tfrac{1}{2}[{\sf I}_{2n} - {\rm coth}(\tfrac{i}{2}\hbar\beta {\sf J} {\bf H}_t) ] \\
&= \tfrac{1}{2}[{\sf I}_{2n} - 2\tfrac{i}{\hbar}{\mathbf V}_t^\text{th}{\sf J}] \\
&= -\tfrac{i}{\hbar}[{\mathbf V}_t^\text{th} - \tfrac{i\hbar}{2} {\sf J}]{\sf J} \, ,
\end{aligned}
\end{equation}
where in the second equality, we used the expression for
${\mathbf V}_t^\text{th}$ in Eq.(\ref{eq:mvcmDefs}).
Remarkably,
the matrix in brackets in the last equality in the above equation
encodes the Robertson-Schrödinger (Heisenberg)
uncertainty principle \cite{nicacio17},
derived from the non-commutative relation
between position and momentum operators through
${\mathbf V} \pm \tfrac{i\hbar}{2} {\sf J} \ge 0$,
valid for any covariance matrix.

\subsection{Free energy and Jarzynski Equality} 
The Helmholtz free energy associated with the instantaneous Hamiltonian $\hat H_t$
is defined in terms of the partition function as
\begin{gather}\label{eq:def_fe}
F(\lambda_t) = - \frac{1}{\beta} \ln ( \mathcal{Z}_t )  =
\frac{1}{2\beta} \, \ln | \text{det}( {\rm S}_\beta^{(t)} - {\sf I})|
+ h_t \, ,
\end{gather}
where we used the expression in (\ref{eq:pf}) and
${\rm S}_\beta^{(t)}$ is the symplectic matrix defined in (\ref{eq:th_oper}).
Consequently,
a driven process induces a net change in free energy given by
$\Delta F = F (\lambda_\tau ) - F (\lambda_0 )$,
which, when inserted in the Jarzynski equality \eqref{jarzynski:eq},
reads
\begin{gather}\label{expW:cl}
\overline{{\rm e}^{-\beta W}} =
\sqrt{
\frac{ |\det ( {\rm S}_\beta^{(0)} - {\sf I})|}
     { |\det ( {\rm S}_\beta^{(\tau)} - {\sf I} )| } } \,
{\rm e}^{-\beta (h_\tau - h_0)} \, .
\end{gather}
This result casts the Jarzynski equality entirely in terms of the
initial and final symplectic matrices defined in Eq.(\ref{eq:th_oper}) and
associated with the QHs governing the work protocol.
This relation can be rewritten for the covariance matrices of the equilibrium
states using the identity (\ref{eq:ident_S-V}):
\[
\overline{{\rm e}^{-\beta W}} =
\sqrt{ \frac{ \det ({\mathbf V}_\tau^\text{th} - \tfrac{i\hbar}{2} {\sf J} )}
     {        \det ({\mathbf V}_0^\text{th} - \tfrac{i\hbar}{2} {\sf J})}}  \,
{\rm e}^{-\beta (h_\tau - h_0)} \, .
\]
Needless to say,
both the Jarzynski equality and the Helmholtz free energy in (\ref{eq:def_fe})
are invariant under Weyl and Metaplectic conjugation due to (\ref{eq:invpf}).
The same Jarzynski equality can also be recovered directly
from the CF derived in Eq.(\ref{eq:finalQuadChi})
by evaluating it at the imaginary argument $u = i\hbar\beta$,
as shown explicitly in Appendix \ref{ap:Jar}
--- this provides a nontrivial consistency check of
the general formalism developed in this work.

It is worth emphasizing that,
in writing Eq.(\ref{eq:def_fe}), we have implicitly chosen the appropriate branch of the
logarithm. This choice is necessary in order to ensure that the free energy remains a
real‑valued quantity, consistent with its thermodynamic interpretation.
In the language of the Metaplectic representation,
this corresponds to fixing the Maslov‑type index associated with the symplectic
matrix ${\rm S}_\beta^{(t)}$.
As discussed in \cite{nicacio2021}, this sign ambiguity is intimately
related to the Conley–Zehnder index appearing in
the trace formula for Metaplectic operators \cite{gossonbook2006}.
The requirement that the partition function $\mathcal Z_t$ be positive
uniquely fixes the physical branch and therefore removes any residual
ambiguity in the definition of the free energy.

Finally, we note that since $\overline{{\rm e}^{-\beta W}} > 0$, by definition,
the right‑hand side of Eq.(\ref{jarzynski:eq}) is also strictly positive,
as required for thermodynamic consistency.

\subsection{Mean Value of Work and Irreversibility }  
Differentiating the CF according to (\ref{eq:Wmoments}) yields directly the
mean work performed during the protocol:
\begin{equation}\label{eq:GenW}
\begin{aligned}
\overline{ W } &=
{\rm Tr}[(\hat U_\tau^\dag \hat H_\tau\hat U_\tau-\hat H_0) \hat\rho_0^\text{th}] \\
& = {\rm Tr}(\hat H_\tau \hat\rho_\tau)
- {\rm Tr}(\hat H_0 \hat\rho_0^\text{th}) \, ,
\end{aligned}
\end{equation}
which shows that the average work corresponds to
the energy variation induced by the work protocol,
since $\hat\rho_\tau$ is the evolution of the initial state,
see Eq.(\ref{eq:evolrho0}).
We remark that this mean value is a continuous function of the final time $\tau$.

For QHs, using Eqs.(\ref{eq:meanenerg}) and (\ref{eq:mvcmEvol0}),
the above mean value becomes
\begin{equation}\label{eq:Wmv}
\begin{aligned}
\overline{ W }  = (h_\tau-h_0) & +
\tfrac{1}{2} (\langle \hat x \rangle^\tau - \xi_\tau) \cdot
{\bf H}_\tau (\langle \hat x \rangle^\tau - \xi_\tau) \\
& + \tfrac{1}{2} \text{Tr}({\bf H}_\tau {\bf V}^\tau - {\bf H}_0{\bf V}_0^\text{th} )  \, .
\end{aligned}
\end{equation}
One can also obtain this result by explicitly differentiating the
CF in Eq.(\ref{eq:finalQuadChi}),
but the derivation becomes more transparent directly in terms of state averages.

Appendix \ref{ap:meanvalue} nevertheless presents that differentiation
as another nontrivial consistency check of the formalism developed here.

Let us analyze separately the physical origin of
each contribution to the mean value (\ref{eq:Wmv}):
\begin{itemize}
\item[$(i)$]
The term $(h_\tau-h_0)$ corresponds to shifts in the reference energy
of the Hamiltonians and reflects a change in the potential energy ground.
\item[$(ii)$]
The second term is the energy contribution due to
the evolution of the state mean value. Here,
$\langle \hat x \rangle^\tau$ is the evolved mean vector
of the initial thermal state $\hat\rho_0^\text{th}$
given in (\ref{eq:mvcmEvol0}),
while $\xi_\tau$ is the mean value of the instantaneous
thermal equilibrium state $\hat\rho_\tau^\text{th}$ in (\ref{eq:thermHt}), that is,
$\langle \hat x \rangle_\tau^\text{th} =
{\rm Tr}(\hat x \hat\rho_\tau^\text{th}) =
\xi_\tau \ne \langle \hat x \rangle^\tau$.
This term is a scalar product and
its sign will depend on the nature of the Hessian matrix ${\bf H}_t$;
in particular, for positive-definite Hessians,
it yields a positive contribution.
Note that the contribution of the mean value of the initial state is
solely encapsulated in $\langle \hat x \rangle^\tau$,
since $\langle \hat x \rangle_0^\text{th} = \xi_0$ and the mean-value contribution
vanishes in (\ref{eq:meanenerg}).
\item[$(iii)$]
The third term accounts for changes in correlations encoded in the covariance matrix.
The transformation property (\ref{eq:mvcmEvol0}) recasts this term as
\[
\qquad \text{Tr}({\bf H}_\tau {\bf V}^\tau - {\bf H}_0{\bf V}_0^\text{th} ) =
\text{Tr}[( {\sf S}_{\tau}^\top  {\bf H}_\tau {\sf S}_{\tau}
                        - {\bf H}_0 ){\bf V}_0^\text{th} ] \, .
\]
Since ${\bf V}_0$ is always positive definite,
the inertia of the matrix
$( {\sf S}_{\tau}^\top  {\bf H}_\tau {\sf S}_{\tau} - {\bf H}_0 )$
determines by itself the sign of this term \cite{horn2013}.
\end{itemize}

The nonequilibrium lag defined in (\ref{eq:noneqlag}) quantifies the degree
of irreversibility of the work protocol
for the mean value of work in (\ref{eq:Wmv}) and Eq.(\ref{expW:cl}).
If the evolved state (\ref{eq:evolrho0}) coincides with the instantaneous
equilibrium state (\ref{eq:thermHt}) at $t = \tau$,
due to the arbitrariness of $\tau$,
the system remains in equilibrium throughout the whole evolution
and no work is dissipated.
Indeed, $\hat \rho_\tau = \hat \rho_\tau^\text{th}$
if and only if $\hat U_\tau \hat H_0 \hat U_\tau^\dag = \hat H_\tau - c \hat\id$,
where $c$ is a constant,
or equivalently, due to (\ref{eq:covHam}),
\begin{equation*} 
{\bf H}_{0}  = {\sf S}_\tau^{\top} {\bf H}_{\tau}{\sf S}_\tau \,,  \,\,\,
{\sf S}_\tau(z_\tau + \xi_0) = \xi_\tau \, , \,\,\, h_0 = h_\tau + c \, .
\end{equation*}
Substituting these relations into the expressions in (\ref{eq:mvcmDefs})
and in (\ref{eq:mvcmEvol0}), from Eq.(\ref{eq:Wmv}),
one finds $\overline{ W } = c$.
The change in the free energy is also trivial:
inserting (\ref{eq:th_oper}) in the free energy definition Eq.(\ref{eq:def_fe}) and
using that ${\rm S}_\beta^{(\tau)} =
{\sf S}_\tau {\rm S}_\beta^{(0)}{\sf S}_\tau^{-1}$,
we find $\Delta F = F(\lambda_\tau) - F(\lambda_0) = c$.
Consequently, the nonequilibrium lag vanishes identically.
Nonetheless, this is not the only situation with a
vanishing nonequilibrium lag:
other quasi-static parameter changes also imply the nontrivial
$\Delta F = \overline{ W } $.

\section{Dynamical Approximations for the Characteristic Function}  \label{sec:DA}
In this section, we present several relevant limiting cases and dynamical
approximations of the general expressions derived in the previous sections.
These approximations allow for a clearer physical interpretation of
the CF and provide direct connections
with results available in the literature.

\subsection{Undisplaced Hamiltonian}
Let us consider a QH (\ref{eq:QuadHam}) with vanishing displacement
$\xi_t = 0, \forall t$.
As a consequence, from (\ref{eq:Sz_uindep}),
one has $z_t = z_u^{(t)} =0, \forall t$, which significantly simplifies
the general expressions for the CF.
In particular, the parameters appearing in Eq.(\ref{eq:chiparam}) reduce to
$\zeta_\beta^{(u,\tau)} = 0$ and $\Phi_\beta^{(u,\tau)} = ( h_\tau - h_0)u$.

For this set of parameters,
Eq.(\ref{eq:chiQuadFunc}) in the light of Eq.(\ref{eq:TrM}) shows that
the CF (\ref{eq:finalQuadChi}) becomes
\begin{gather}\label{eq:chinullxi}
\begin{aligned}
\chi_\tau (u) &=
\frac{ M_{{\bf S}_\beta^{(u,\tau)}}(0) }
     { M_{{\rm S}_\beta^{(0)}} (0)} \,
    {\rm e}^{ \frac{i}{\hbar} (h_\tau - h_0) u}  \\
& = \sqrt{ \frac{ \det ( {\rm S}_\beta^{(0)} - {\sf I})}
               { \det ({\bf S}_\beta^{(u,\tau)} - {\sf I})}  }
{\rm e}^{ \frac{i}{\hbar} (h_\tau - h_0) u} \, .
\end{aligned}
\end{gather}
This expression is basically the ratio between the traces of two Metaplectic operators,
which highlights the purely symplectic origin of the work statistics in this case.
Substituting $u = i\hbar \beta$ into the above function,
we recover straightforwardly the Jarzynski equality in (\ref{expW:cl}).
From the first equation in (\ref{eq:mvcmEvol0}),
the mean value of work (\ref{eq:Wmv}) receives no contributions
from the state mean-value evolution:
\begin{equation}\label{eq:Wmv_nullxi}
\begin{aligned}
& \overline{ W }  = (h_\tau-h_0) +
\tfrac{1}{2} \text{Tr}[ {\bf \Delta}{\bf H}_\tau{\bf V}_0^\text{th} ]  \,, \\
&{\Delta}{\bf H}_\tau :=
{\sf S}_{\tau}^\top  {\bf H}_\tau {\sf S}_{\tau} - {\bf H}_0 \,.
\end{aligned}
\end{equation}

We evaluate the variance
$\sigma_W^2 := \overline{ W^2 } - \overline{W}^2$
of the work probability distribution by squaring (\ref{eq:Wmv_nullxi})
and determining the second moment $k=2$ in (\ref{eq:Wmoments}) for
CF in (\ref{eq:chinullxi}).
Following similar steps as the ones in Appendix \ref{ap:meanvalue},
and using explicitly Eqs.(\ref{eq_ap:derinv}) and (\ref{eq_ap:jacobi}),
we find
\begin{equation}\label{eq:varW_nullxi}
\begin{aligned}
\sigma_W^2
&= \tfrac{1}{2}
\text{Tr}[({\Delta}{\bf H}_\tau {\bf V}_0^\text{th})^2 ] +
\tfrac{\hbar^2}{8} \text{Tr}[({\sf J} \, {\Delta}{\bf H}_\tau )^2 ] \, ,
\end{aligned}
\end{equation}
where we used the definition of ${\Delta}{\bf H}_\tau$ in (\ref{eq:Wmv_nullxi}).
As expected, this variance increases with the difference
between the initial and final Hamiltonians.
Notably,
the term in $\hbar^2$ only depends on the dynamics (the work protocol)
since there is no information about the system state in this part.

\subsection{Sudden Evolution} \label{sec:SE}
A sudden evolution (or a sudden quench) is a dynamical approximation for
the unitary evolution in which the system Hamiltonian $\hat {\mathcal H}(\lambda_0)$
changes abruptly to another configuration $\hat H(\lambda_\tau)$.
In this idealized limit, we assume the protocol
duration to be negligibly short compared with all intrinsic dynamical timescales of the system.
As a consequence, the evolution operator in (\ref{eq:Utdep}) is the identity operator,
$\hat U_t = \hat \id$,
and the CF in (\ref{eq:chi}) simplifies to
\begin{gather} \label{eq:suddenchi}
\chi_\tau (u) = \text{Tr} \left[
\, {\rm e}^{iu \hat {\mathcal H} (\lambda_\tau) /\hbar}
\, {\rm e}^{-i u \hat {\mathcal H} (\lambda_0) /\hbar } \, \hat \rho_0^\text{th} \right] \, .
\end{gather}
Here the indices $0$ and $\tau$
only discriminate the system before and after the sudden quench
and the Hamiltonians are time-independent.

For QHs, the sudden change of the parameters
$\lambda_0 = (\xi_0,{\bf H}_0, h_0) \to \lambda_\tau = (\xi_\tau,{\bf H}_\tau,h_\tau)$
represents the change between (time-independent) QHs
in (\ref{eq:QuadHam})
for $t=0$ and $t=\tau$, while $\hat U_t = \hat \id$ corresponds to
$\varphi_t = 0, z_t = 0, {\mathsf S}_t = {\sf I}, \forall t$,
according to Eq.(\ref{eq:UtdepQuad}).
These, when replaced in (\ref{eq:chiparam}), give
a CF in (\ref{eq:finalQuadChi}) with
\begin{equation}\label{eq:chiparam2}
\begin{aligned}
{\bf S}_\beta^{(u,\tau)} & =
{\sf S}_{-u}^{(\tau)} {\sf S}_u^{(0)} {\rm S}_\beta^{(0)} \, , \\
\zeta_\beta^{(u,\tau)} & =
{\rm S}_{-\beta}^{(0)}{\sf S}_{-u}^{(0)}
( {\sf S}_{u}^{(\tau)} - {\sf I}) \xi_\tau
 + ( {\rm S}_{-\beta}^{(0)}{\sf S}_{-u}^{(0)}- {\sf I}) \xi_0\, , \\
\Phi_\beta^{(u,\tau)} &= (h_\tau - h_0) u
+ \tfrac{1}{2}{\sf J} \xi_\tau \cdot {\sf S}_u^{(\tau)} \xi_\tau
- \tfrac{1}{2}{\sf J} \xi_0 \cdot {\sf S}_{u}^{(0)}{\rm S}_{\beta}^{(0)} \xi_0 \\
& + \tfrac{1}{2}{\sf J}\xi_0 \cdot
( {\rm S}_{-\beta}^{(0)} {\sf S}_{-u}^{(0)} - {\sf I} )
( {\sf S}_{u}^{(\tau)} - {\sf I}) \xi_\tau  \, ,
\end{aligned}
\end{equation}
where some simplifications were performed.

The mean work performed during
the protocol follows directly from Eq.(\ref{eq:GenW}) with $\hat U_t = \hat \id$,
\begin{equation}\label{eq:sudden_meanW}
\begin{aligned}
\overline{W} &=
{\rm Tr}[(\hat H_\tau - \hat H_0) \hat\rho_0^\text{th}] \\
&= (h_\tau-h_0) +
\tfrac{1}{2} (\xi_\tau -\xi_0)\cdot {\bf H}_\tau (\xi_\tau -\xi_0) \\
& \hspace{1.91cm} + \tfrac{1}{2} \text{Tr}[({\bf H}_\tau - {\bf H}_0){\bf V}_0^\text{th} ] \, ,
\end{aligned}
\end{equation}
where the last equality follows by setting
$z_t = 0$ and ${\sf S}_t = {\sf I}$ in (\ref{eq:Wmv}).
One can also obtain this mean value from (\ref{eq:Wmoments})
for the CF in (\ref{eq:finalQuadChi}),
but for the parameters of Eq.(\ref{eq:chiparam2}).

When additionally $\xi_0 = 0$, we recover the sudden-change
second moment $k = 2$ in (3) from the second derivative of
scenario for an undisplaced Hamiltonian, so the sudden-change CF described by
the parameters in (\ref{eq:chiparam2}) becomes the one in (\ref{eq:chinullxi})
with ${\sf S}_\tau = {\sf I}$.
Likewise, Eq.(\ref{eq:sudden_meanW}) becomes the one in (\ref{eq:Wmv_nullxi})
for ${\sf S}_\tau = {\sf I}$, as a consequence of the relation between
the respective CFs.
This limiting case includes the scenario studied in \cite{paternostro2019},
where a set of initially decoupled oscillators is
suddenly coupled to form an interacting harmonic chain.

\subsection{Constant Hessian} 
Another relevant situation corresponds to a work protocol in which
the Hessian remains constant throughout the evolution, namely
$\lambda_t = (\xi_t,{\bf H}_0, h_t)$ for $ 0 \le t \le \tau$.
This class of protocols includes, for instance, forced harmonic oscillators,
such as those discussed in \cite{talkner2008b}.

The unitary evolution operator in the present case is (\ref{eq:UtdepQuad}),
however the strict quadratic part in (\ref{eq:met_tdep}) is time independent,
that is, the first differential equation in Eq.(\ref{eq:dif_eq})
has the explicit solution ${\sf S}_t = {\rm e}^{{\sf J}{\bf H}_0 t}$ and
the solution in (\ref{eq:dif_eq_sol}) yields
\begin{equation}\label{eq:mvH0}
\textstyle{
z_\tau = {\sf J}^\top{\bf H}_{0} \int_0^\tau \!dt \, {\sf S}_{t}^{-1}\xi_t
=  {\sf S}_{\tau}^{-1} \xi_\tau - \xi_0 -
\int_0^\tau \! dt \, {\sf S}_{t}^{-1} \dot\xi_t}   \, ,
\end{equation}
where we performed an integration by parts.
Moreover, since the Hessian is time independent, one has
${\sf S}_u^{(\tau)} = {\sf S}_u^{(0)} = {\rm e}^{{\sf J}{\bf H}_0 u}$
and $[{\sf S}_u^{(\tau)},{\sf S}_t ]$ = 0,
which leads to simplifications of the parameters in (\ref{eq:chiparam}):
\[
\begin{aligned}
{\bf S}_\beta^{(u,\tau)} & = {\rm S}_\beta^{(0)} \, , \\
\zeta_\beta^{(u,\tau)} & =
{\rm S}_{-\beta}^{(0)} ( {\sf S}_{-u}^{(0)} - {\sf I} )z_\tau
- {\rm S}_{-\beta}^{(0)}{\sf S}_\tau^{-1}
( {\sf S}_{-u}^{(0)} - {\sf I} ) \xi_\tau \\
&+( {\rm S}_{-\beta}^{(0)} {\sf S}_{-u}^{(0)} - {\sf I} ) \xi_0 \\
\Phi_\beta^{(u,\tau)} & = (h_\tau - h_0) u
+  \tfrac{1}{2}{\sf J} \xi_\tau \cdot {\sf S}_u^{(0)} \xi_\tau
-  \tfrac{1}{2}{\sf J} \xi_0 \cdot {\sf S}_{u}^{(0)} {\rm S}_\beta^{(0)}\xi_0 \\
& + \tfrac{1}{2} {\sf J} z_\tau \cdot
[ {\sf S}_{u}^{(0)} z_\tau
- {\sf S}_{-\tau}({\sf S}_{u}^{(0)}  - {\sf S}_{-u}^{(0)}  )\xi_\tau] \\
& + \tfrac{1}{2}{\sf J}\xi_0 \cdot
( {\sf I} - {\rm S}_u^{(0)}{\rm S}_\beta^{(0)})\zeta^{(u,\tau)}_\beta \, .
\end{aligned}
\]

Since the Hessian does not change during the protocol,
it does not contribute to the average work.
Indeed, using that
${\sf S}_{\tau}^\top  {\bf H}_0 {\sf S}_{\tau} = {\bf H}_0$,
Eq.(\ref{eq:Wmv}) becomes
\[
\overline{W}  = (h_\tau-h_0) +
\tfrac{1}{2} (\langle \hat x \rangle^\tau - \xi_\tau) \cdot
{\bf H}_0 (\langle \hat x \rangle^\tau - \xi_\tau) \, .
\]
From Eq.(\ref{eq:mvcmEvol0}),
the difference between the evolved mean value and the instantaneous
equilibrium displacement becomes
\[
\langle \hat x \rangle^\tau - \xi_\tau =
- \int_0^\tau \!\!\! dt \, {\sf S}_{\tau-t} \dot\xi_t \, ,
\]
where we used Eq.(\ref{eq:mvH0}).

If one further considers a sudden quench
($z_\tau = 0$ and ${\sf S}_\tau = {\sf I}$),
the CF parameters reduce to
\[
\begin{aligned}
\zeta_\beta^{(u,\tau)} & =
- {\rm S}_{-\beta}^{(0)} ( {\sf S}_{-u}^{(0)} - {\sf I} ) \xi_\tau
+( {\rm S}_{-\beta}^{(0)} {\sf S}_{-u}^{(0)} - {\sf I} ) \xi_0 \\
\Phi_\beta^{(u,\tau)} & = (h_\tau - h_0) u
+  \tfrac{1}{2}{\sf J} \xi_\tau \cdot {\sf S}_u^{(0)} \xi_\tau
-  \tfrac{1}{2}{\sf J} \xi_0 \cdot {\sf S}_{u}^{(0)} {\rm S}_\beta^{(0)}\xi_0 \\
& + \tfrac{1}{2}{\sf J}\xi_0 \cdot
( {\sf I} - {\rm S}_u^{(0)}{\rm S}_\beta^{(0)})\zeta^{(u,\tau)}_\beta \, ,
\end{aligned}
\]
and the mean value of work follows simply from
$(\langle \hat x \rangle^\tau - \xi_\tau) = \xi_0 - \xi_\tau$.

\subsection{Adiabatic Approximation}\label{sec:AA}
In contrast to a sudden quench, the adiabatic theorem
describes a sufficiently slow variation of the control parameter $\lambda_t$
of a nondegenerate Hamiltonian \cite{sakuray}:
If $| {\mathfrak n}(\lambda_t) \rangle$ denotes an eigenstate of
$\hat {\mathcal H}(\lambda_t)$,
then,
\begin{equation} \label{eq:adiab_ev}
\hat U_t | \mathfrak{n}(\lambda_0) \rangle =
\exp[i \theta_\mathfrak{n}(t)] | \mathfrak{n}(\lambda_t) \rangle
\, , \,\,\, \theta_\mathfrak{n}(t) \in {\mathbb R},
\end{equation}
where $\hat U_t$ is the unitary evolution generated by $\hat {\mathcal H}(\lambda_t)$,
the one in (\ref{eq:Utdep}). 

Consider the instantaneous spectral decomposition of the Ha\-mil\-tonian
\[
\hat {\mathcal H}(\lambda_t) = \sum_{\mathfrak{n}} E_{\mathfrak{n}}(\lambda_t)
|\mathfrak{n}(\lambda_t)\rangle\langle \mathfrak{n}(\lambda_t)| \, ,
\]
where we assume the eigenvalues $E_\mathfrak{n}(\lambda_t)$ to be non-de\-ge\-ne\-ra\-te.
Under an adiabatic evolution, see Eq.(\ref{eq:adiab_ev}),
over the interval $[0,\tau]$, we define
\begin{equation}\label{eq:H_adiab}
\hat {\mathcal H}^0_t := \hat U_t^\dag \hat {\mathcal H}(\lambda_t) \hat U_t =
\sum_\mathfrak{n} E_\mathfrak{n}(\lambda_t) | \mathfrak{n}(\lambda_0) \rangle \langle
\mathfrak{n}(\lambda_0)|
\end{equation}
as a Hamiltonian with the same eigenvalues as
the final Hamiltonian $\hat {\mathcal H}(\lambda_t)$
but corresponding to the eigenvectors of the initial one $\hat {\mathcal H}(\lambda_0)$.
Note that $[\hat {\mathcal H}^0_t,\hat {\mathcal H}(\lambda_0)]=0$.
Using this result,
the CF in Eq.(\ref{eq:chi})
for an initial thermal state reduces to
\begin{equation} \label{eq:adiabchi}
\begin{aligned}
\chi_\tau (u) &= \frac{1}{\mathcal Z(\lambda_0)}
\sum_\mathfrak{n}
{\rm e}^{ i u [E_\mathfrak{n}(\lambda_\tau) - E_\mathfrak{n}(\lambda_0) ]/\hbar -
\beta E_\mathfrak{n}(\lambda_0) } \\
&= \text{Tr}\!\left[ {\rm e}^{i u \hat {\mathcal H}^0_\tau /\hbar }
 {\rm e}^{- i u \hat {\mathcal H}(\lambda_0) /\hbar } \, \hat \rho_0^\text{th} \right] \, .
\end{aligned}
\end{equation}
Note that this CF is the one of a sudden quench,
see Eq.(\ref{eq:suddenchi}), between
$\hat {\mathcal H}(\lambda_0)$ and $\hat{\mathcal H}^0_\tau$.

We now specialize to QHs of the form (\ref{eq:QuadHam}).
To this end,
we need to determine the operator $\hat H^0_\tau$ for
such Hamiltonians considering the established relation in (\ref{eq:specQuad})
among the eigenvectors of the general QH $\hat H_t$
(denoted by $|\tilde{\mathfrak m}(\tau) \rangle$)
and those of a harmonic oscillator Hamiltonian in Eq.(\ref{eq:H_OH}),
which are denoted by $|\mathfrak{m}\rangle$, see Eq.(\ref{eq:Spec_H_OH}).

According to the adiabatic evolution in (\ref{eq:adiab_ev})
over the interval $[0,\tau]$,

\begin{equation}\label{eq:H0_eigvec}
\begin{aligned}
|\tilde{\mathfrak m}(0) \rangle &=
{\rm e}^{-i \theta_{\mathfrak{m}}(\tau)}
\hat U_\tau^\dagger |\tilde{\mathfrak m}(\tau) \rangle \\
&= {\rm e}^{-i \theta_{\mathfrak{m}}(\tau)}
\hat U_\tau^\dagger \,
\hat T_{\xi_\tau} \hat M_{{\sf S}_{{\bf H}_\tau}}|{\mathfrak m}\rangle \\
& = {\rm e}^{-i \theta_{\mathfrak{m}}(\tau)}
\hat T_{\xi_0} \hat M_{{\sf S}_{{\bf H}_0}}|{\mathfrak m}\rangle \, ,
\end{aligned}
\end{equation}
which are thus eigenvectors of $\hat H_0$.
Therefore, from (\ref{eq:H_adiab}),
\[
\hat H^0_\tau = \left(\sum_{j=1}^n \sum_{m_j=0}^{\infty} E_{m_j}^{(j)}(\tau)\right)
\hat T_{\xi_0} \hat M_{{\sf S}_{{\bf H}_0}}|{\mathfrak m}\rangle
\! \langle {\mathfrak m}|\hat M_{{\sf
S}_{{\bf H}_0}}^\dag\hat T_{\xi_0}^\dag
\]
is a Hamiltonian with the same eigenvalues of (\ref{eq:H_decomp}),
which are also equal to the eigenvalues of (\ref{eq:H_OH}) apart from the offset
$h_\tau$,
but associated with the eigenvectors of $\hat H_0$ in (\ref{eq:H0_eigvec}).
Therefore, it can be written as
\begin{equation}\label{eq:H_decomp2}
\begin{aligned}
\hat {H}_\tau^0 & =
\hat T_{\xi_0} \hat M_{{\sf S}_{{\bf H}_0}}
\hat H^\text{oh}(\tau) \,
\hat M_{{\sf S}_{{\bf H}_0}}^\dag \hat T_{\xi_0}^\dag  + h_\tau \\
& = \hat T_{\xi_0} [\tfrac{1}{2} {\sf S}_{{\bf H}_0}^{-1} \hat x
\cdot {\bf \Lambda}_{{\bf H}_{\tau}} {\sf S}_{{\bf H}_0}^{-1}\hat x]
\hat T_{\xi_0}^\dag  + h_\tau \\
& = \hat T_{\xi_0} [\tfrac{1}{2} \hat x
\cdot {\bf H}_\tau^0 \, \hat x] \hat T_{\xi_0}^\dag  + h_\tau  \\
& = \tfrac{1}{2} (\hat x - \xi_0) \cdot {\bf H}_\tau^0 (\hat x-\xi_0) + h_\tau \, ,
\end{aligned}
\end{equation}
where we defined the Hessian matrix
\begin{equation} \label{eq:Hess_adiab}
{\bf H}_\tau^0  :=
{\sf S}_{{\bf H}_0}^{-\top} {\bf \Lambda}_{{\bf H}_{\tau}} {\sf S}_{{\bf H}_0}^{-1} \,.
\end{equation}
We remark that the relation between $\hat {\bf H}_\tau^0$ and $\hat H^\text{oh}(\tau)$
established above is only possible due to the time-independence
of the eigenvectors $|\mathfrak m \rangle$ of $\hat H^\text{oh}(\tau)$.

The CF (\ref{eq:adiabchi}) for quadratic
Hamiltonians can now be evaluated
using the Metaplectic–Weyl representation described in Sec.\ref{sec:MB}.
The exponential of $\hat H^0_\tau$ in Eq.(\ref{eq:adiabchi}),
for the QH in (\ref{eq:H_decomp2}), can be obtained by
formula in (\ref{eq:Unit_u}):
\[
{\rm e}^{\frac{i}{\hbar} u \hat {\bf H}_\tau^0} =
{\rm e}^{\frac{i}{\hbar} h_\tau u }
\hat T_{\xi_0} \hat M_{\tilde{\sf S}_{-u}^{(\tau)}}
\hat T_{\xi_0}^\dagger \, ,
\]
for
\begin{equation}\label{eq:tildeS0tau}
\tilde {\sf S}_{u}^{(t)} := {\rm e}^{u {\sf J}{\bf H}_t^0} =
{\sf S}_{{\bf H}_0} {\rm e}^{u {\sf J}{\bf \Lambda}_{{\bf H}_{t}}}
{\sf S}_{{\bf H}_0}^{-1} \, ,
\end{equation}
where we used (\ref{eq:Hess_adiab}) and the symplecticity of ${\sf S}_{{\bf H}_0}$;
the same formula (\ref{eq:Unit_u}) for $t=0$ determines
the second exponential in (\ref{eq:adiabchi})
and $\hat\rho_0^\text{th}$ is in (\ref{eq:thermH0});
inserting all of these into (\ref{eq:adiabchi}) and using the cyclicity of the trace,
we attain
\begin{equation} \label{eq:adiabchiquad}
\begin{aligned}
\chi_\tau (u) & =
{\rm Tr}\left[ \hat M_{\tilde{\sf S}_{-u}^{(\tau)}}
               \hat M_{{\sf S}_u^{(0)}} \hat M_{ {\rm S}_\beta^{(0)}}\right]
               {\rm e}^{ \frac{i}{\hbar} (h_\tau - h_0) u} \\
&= \sqrt{ \frac{ \det ( {\rm S}_\beta^{(0)} - {\sf I})}
{ \det ( \tilde{\sf S}_{-u}^{(\tau)} {\sf S}_u^{(0)} {\rm S}_\beta^{(0)} - {\sf I})} } \,
{\rm e}^{ \frac{i}{\hbar} (h_\tau - h_0) u} \, ,
\end{aligned}
\end{equation}
where we obtained the final form from (\ref{eq:met_composition}) and (\ref{eq:TrM}).
Following the guidelines of Appendix \ref{ap:QuadChiFunc},
one also obtains Eq.(\ref{eq:adiabchiquad}) by simplifying
and adjusting the general case of Sec.\ref{sec:CFITS}.

Interestingly enough,
the adiabatic CF (\ref{eq:adiabchiquad}) and
the derived work statistics do not depend on displacements
$\xi_t$ of the QH.
Noting that the CF in (\ref{eq:adiabchi})
is written solely in terms of energy eigenvalues,
this independence is not a particularity of QHs,
since displacements are due to similarity transformations (through Weyl operators),
which leaves the eigenvalues of any Hamiltonian invariant.
For the quadratic case, this is exemplified in Eqs.(\ref{eq:H_OH}-\ref{eq:Spec_H_OH}).

As another remarkable property,
the CF in (\ref{eq:adiabchiquad}) depends
solely on the symplectic eigenvalues of the Hamiltonians,
which are symplectically invariant \cite{nicacio17}.
From Eq.(\ref{eq:sympdiag}), one can write
${\sf S}_{u}^{(0)} = {\rm e}^{u {\sf J}{\bf H}_0} =
{\sf S}_{{\bf H}_0} {\rm e}^{u {\sf J}{\bf \Lambda}_{{\bf H}_{0}}}
{\sf S}_{{\bf H}_0}^{-1}$, which also holds similarly for ${\rm S}_\beta^{(0)}$
[see Eq.(\ref{eq:wsymp})],
and using (\ref{eq:tildeS0tau}),
the second expression in (\ref{eq:adiabchiquad}) becomes

\[
\chi_\tau (u)  =
\sqrt{ \frac{ \det ( {\rm e}^{-i\hbar\beta {\sf J}{\bf \Lambda}_{{\bf H}_{0}}} - {\sf I})}
{ \det [ {\rm e}^{u {\sf J}({\bf \Lambda}_{{\bf H}_{0}} - {\bf \Lambda}_{{\bf H}_{\tau}})
-i\hbar \beta {\sf J} {\bf \Lambda}_{{\bf H}_{0}} } - {\sf I} ]} } \,
{\rm e}^{ \frac{i}{\hbar} (h_\tau - h_0) u} \, .
\]
Just as ordinary eigenvalues are invariant under similarity transformations,
symplectic eigenvalues are invariant under symplectic congruences,
as can be seen from Eq.(\ref{eq:secular}); see Ref.\cite{nicacio17} for details.
Again, this is not an exclusive property of QHs; rather,
it is actually a manifestation of the similarity invariance of (\ref{eq:adiabchi})
for the quadratic case.

The independence on displacements in the adiabatic case
brings the corresponding CF
to the form of the one for non-displaced Hamiltonians in (\ref{eq:chinullxi}).
For this reason,
the adiabatic $\chi_\tau (u)$ in (\ref{eq:adiabchiquad})
follows directly from the substitution
${\bf S}_\beta^{(u,\tau)} \rightarrow
\tilde{\sf S}_{-u}^{(\tau)} {\sf S}_u^{(0)} {\rm S}_\beta^{(0)}$
into (\ref{eq:chinullxi}).
Additionally, Eq.(\ref{eq:adiabchiquad})
is equivalent to the CF of a sudden change,
between the Hamiltonians $\tfrac{1}{2}\hat x \cdot {\bf H}_0 \hat x$ and
$\tfrac{1}{2}\hat x \cdot {\bf H}_0^\tau \hat x$, see Eq.(\ref{eq:suddenchi}).
One thus obtains the mean value and the variance of work in the adiabatic limit
by substituting
${\sf S}_{\tau}^\top  {\bf H}_\tau {\sf S}_{\tau} \rightarrow {\bf H}_\tau^0$
respectively in Eqs.(\ref{eq:Wmv_nullxi}) and (\ref{eq:varW_nullxi}):
\begin{equation}\label{eq:adiabW}
\begin{aligned}
\overline{W} & = (h_\tau-h_0) +
\tfrac{1}{2} \text{Tr}[\Delta{\bf H}^0_\tau{\bf V}_0^\text{th} ] \, , \\
\,\,\,
\sigma_W^2 &= \tfrac{1}{2} \text{Tr}[(\Delta{\bf H}^0_\tau {\bf V}_0^\text{th})^2 ]
+ \tfrac{\hbar^2}{8} \text{Tr}[({\sf J} \Delta {\bf H}^0_\tau)^2 ] \, ,
\end{aligned}
\end{equation}
where we defined $\Delta {\bf H}^0_\tau := {\bf H}_\tau^0 - {\bf H}_0$.
Both the mean value and the variance are also symplectically
invariant and depend only on symplectic eigenvalues, since
\[
\begin{aligned}
\text{Tr}[\Delta {\bf H}^0_\tau {\bf V}_0^\text{th} ] & =
\tfrac{\hbar}{2} \text{Tr}[({\bf \Lambda}_{{\bf H}_\tau} - {\bf \Lambda}_{{\bf H}_0} )
{\rm coth}(\tfrac{1}{2}\hbar\beta {\bf \Lambda}_{{\bf H}_0})] \\
& = \hbar \sum_{j=1}^n [\mu_j(\tau)-\mu_j(0)]{\rm coth}[\tfrac{1}{2}\mu_j(0)\hbar\beta] \, ,
\end{aligned}
\]
where we used (\ref{eq:Hess_adiab}) and (\ref{eq:cm_t}).

The mean energy of the system in the adiabatic limit becomes
\begin{equation}\label{eq:admeanenerg}
\begin{aligned}
\langle\hat H_\tau \rangle & =
{\rm Tr} [\hat H_\tau \hat U_\tau \hat\rho_0^\text{th} \hat U_\tau^\dag]
= {\rm Tr} [ \hat U_\tau^\dag \hat H_\tau \hat U_\tau \hat\rho_0^\text{th}] \\
& = {\rm Tr} [ \hat H^0_\tau \hat\rho_0^\text{th}]
= \tfrac{1}{2} {\rm Tr}({\bf H}_\tau^0 {\bf V}_0^\text{th}) + h_\tau \, ,
\end{aligned}
\end{equation}
where we used the cyclicity of the trace,
the adiabatic result in (\ref{eq:H_adiab}),
and the mean-energy formula in Eq.(\ref{eq:meanenerg}).
The covariance matrix ${\bf V}_0^\text{th}$ is defined in (\ref{eq:mvcmEvol0}).
The mean-value of the initial thermal state ($t=0$) is $\xi_0$, see Eq.(\ref{eq:mvcmEvol0}),
however it cancels with the displacement of the Hamiltonian in (\ref{eq:admeanenerg}).
Remarkably, the work in (\ref{eq:adiabW}) does not depend on $\xi_t$,
because the characteristic function itself does not.

\section{Relation to Other Physical Quantities}\label{sec:rel}
Several physical quantities of interest can be expressed as special cases or direct
generalizations of the operator structure appearing in the CF
defined in Eq.(\ref{eq:chi}).
In this subsection, we illustrate this connection by discussing two
representative examples that naturally fit within the symplectic–Metaplectic framework
developed in this work.

Let us consider a system first evolving forward in time under a
Hamiltonian $\hat {\mathcal H}(\lambda_0)$ during a time interval $t \in [0,\tau]$.
At $t = \tau$, a sudden quench is applied,
$\hat {\mathcal H}(\lambda_0) \rightarrow \hat {\mathcal H}(\lambda_\tau)$,
after which the system evolves backward under the Hamiltonian
$\hat {\mathcal H}(\lambda_\tau)$.
A quantity that measures the sensitivity of the dynamics to this quench
is the {\it Loschmidt echo}, $L(t) = |G(t)|^2$,
for the probability amplitude
\begin{gather*}
G(t) = \left\langle {\rm e}^{i \hat {\mathcal H}(\lambda_\tau) t/\hbar}
                    {\rm e}^{-i \hat {\mathcal H}(\lambda_0) t/\hbar}
\right\rangle \, .
\end{gather*}
If the expectation value is taken with respect to a thermal equilibrium state
corresponding to the Hamiltonian $\hat{{\mathcal H}}(\lambda_0)$,
then this probability amplitude is readily obtained from (\ref{eq:suddenchi}):
$G(t) = [\chi_\tau(u=t)]$.
For QHs, this correspondence becomes explicit by inserting
$u = t$ into the parameters defined in (\ref{eq:chiparam2})
and evaluating the CF using (\ref{eq:finalQuadChi}).
For a non-thermal initial state and a QH,
the expression for the echo is given by $\chi_{\tau}(u=t)$
in (\ref{eq:chi_GenState}), but with $\Psi_0(\eta)$ replaced by the Weyl representation of
the considered initial state and ${\sf S}_\tau$, $z_\tau$, and $\phi_\tau$ all in
Eqs.(\ref{eq:chiparam3}) replaced, respectively,
by the identity matrix, the zero vector, and zero.
This structural equivalence underlies several recent studies linking
nonequilibrium work statistics,
dynamical quantum phase transitions, and Loschmidt echoes \cite{silva2008}.
For a generic initial state that is not thermal, one obtains the Loschmidt amplitude
by combining the general expression for the CF in Eq. (\ref{chi:c2})
with the sudden‑quench approximation. This amounts to setting
$\phi_t = 0, z_t = 0, {\sf S}_t = {\sf I}, \forall t$
(sudden-quench approximation) in (\ref{eq:chiparam3}),
replacing the Weyl symbol $\Psi_0(\eta)$ in (\ref{eq:chi_GenState})
by that of the desired initial state, and finally identifying
$G(t) = \chi_\tau(u=-t)$.

Another quantity that fits naturally within the present fra\-me\-work
is the {\it total phase} accumulated by a quantum state under unitary evolution.
For pure states, this phase decomposes into
a dynamical and geometric (Berry) contributions \cite{sakuray}.
For mixed states, Sjöqvist {\it et al.} proposed a consistent generalization \cite{sjoqvist}.
Given an initial density operator $\hat \rho$ undergoing a unitary evolution generated by
$\hat {\mathcal H}_t$, the total phase is defined as
$\phi(t) = \text{Arg}[{\rm Tr}(\hat U_t \hat\rho )]$.
The present formalism computes this quantity immediately.
Indeed, it corresponds to a special case of the CF in
(\ref{eq:chi_GenState}) upon replacing the Weyl symbol $\Psi_0(\eta)$ by that of the state
$\hat \rho$, setting $\hat {\mathcal H}_\tau = 0$, and identifying $u = -t$.
This construction recovers, as a particular instance, the results
of Ref.\cite{nicacio13} for time-independent quadratic Hamiltonians.

In this sense,
the characteristic‑function formalism provides a unified description of work
statistics, Loschmidt echoes, and geometric‑phase‑like
quantities within a single algebraic framework.

\section{Example: Time-Dependent Symplectic Dynamics} \label{application}
As an illustration of the general formalism,
we consider a one-dimensional system ($n=1$) with dynamics described by
the symplectic matrix
\begin{equation}\label{ex:sympmat}
{\sf S}_t
=
\begin{pmatrix}
\cos\theta_t
&
{\alpha_t}^{-1}{\sin\theta_t}
\\
-\alpha_t\sin\theta_t
&
\cos\theta_t
\end{pmatrix} \, , \,\,\, \alpha_t : = m_t\theta_t \, ,
\end{equation}
where $\theta_t$ and $m_t$ are, in principle, generic temporal functions.
This symplectic matrix generates the time-dependent Hessian through
the first equation in (\ref{eq:dif_eq}):
\begin{equation}\label{ex:Ht}
\begin{aligned}
{\bf H}_t & = {\sf J}^\top \dot{\sf S}_t {\sf S}_t^{-1} \\
& =
\begin{pmatrix}
\alpha_t\dot\theta_t
+
\tfrac{1}{2}\sin(2\theta_t)\dot\alpha_t
&
-
\dfrac{\sin^2\theta_t}{\alpha_t}\dot\alpha_t
\\
-
\dfrac{\sin^2\theta_t}{\alpha_t}\dot\alpha_t
&
\dfrac{\dot\theta_t}{\alpha_t}
-
\dfrac{\sin(2\theta_t)}{2\alpha_t^2}\dot\alpha_t
\end{pmatrix}.
\end{aligned}
\end{equation}

Consider also the time-dependent vector
\begin{equation}\label{ex:xit}
\xi_t
=
-a_0\,({\sf J}{\bf H}_t)^{-1}{\sf S}_t
\begin{pmatrix}
1\\
1
\end{pmatrix}, \,\,\, a_0 \in \mathbb R \, ,
\end{equation}
explicitly chosen to give a simple form for
the solution of the second equation in (\ref{eq:dif_eq}), that is,
\begin{equation}\label{ex:vecz}
\dot z_t
=
a_0
\begin{pmatrix}
1\\
1
\end{pmatrix} \Longleftrightarrow
z_t
=
a_0 t
\begin{pmatrix}
1\\
1
\end{pmatrix}.
\end{equation}
Note that $z_0 = 0$,
as required for the solution of the differential equation in (\ref{eq:dif_eq}).
With all the above elements,
we construct the Hamiltonian of the system in (\ref{eq:ImpQuadHam}) with $h_t = 0$.

If we consider two functions $\theta_t$ and $m_t$
such that $\alpha_t = 1 \forall t$ (in some particular system of units),
the symplectic matrix in Eq.(\ref{ex:sympmat}) is a genuine rotation and the Hessian in
(\ref{ex:Ht}) becomes that of an isotropic harmonic oscillator
${\bf H}_t = \dot{\theta}_t {\sf I}_2$.
That is, the system evolution is governed by the Hamiltonian
$\hat H_t = \tfrac{1}{2}\dot{\theta}_t( \hat p^2 + \hat q^2)$.
The unitary evolution operator for this Hamiltonian is particularly
simple to obtain, since $[\hat H_{t'}, \hat H_t] = 0$, $\forall t', t$ \cite{sakuray}.
This means that we obtain the solution ${\sf S}_t$ of Eq.(\ref{eq:dif_eq}),
which coincides with Eq.(\ref{ex:sympmat}),
analytically without needing a symplectic path.
However, once we obtain this solution, the formalism developed here applies
immediately and determines all the work statistics and thermodynamic quantities.

From the irreversibility point of view,
an interesting situation emerges when the protocol is cyclic
and the associated symplectic path satisfies ${\sf S}_\tau = {\sf I}_2$.
If we chose the final protocol time $\tau$ to be equal to this period,
then $z_\tau = z_0$, $h_\tau = h_0$, ${\bf H}_\tau = {\bf H}_0$,
and ${\sf S}_\tau = {\sf S}_0 = {\sf I}_2$,
and the physical system recovers its initial state.
All statistical and thermodynamical properties will be periodic,
including the nonequilibrium Lag in (\ref{eq:noneqlag}) that will vanish at this instant,
since $\Delta F = 0$ and $\overline{W} = 0$ from (\ref{eq:Wmv}).
At any other value of $\tau$,
the degree of irreversibility will be non-null.
For the symplectic matrix in (\ref{ex:sympmat}), we obtain a periodic dynamics
by setting $\alpha_t = \alpha_0 {\rm e}^{\sin(\gamma t)}$ and
$\theta_t = \gamma t$. Despite the non-trivial path in phase-space,
the system returns to its initial state after $\tau = 2\pi/\gamma$.

Now we move to our main example, whose purpose is threefold.
First, we illustrate the direct computation of the characteristic function
from Eq.(\ref{eq:finalQuadChi}).
Second, we examine the role of phase-space displacements on work fluctuations.
Third, we compare the exact dynamics with the sudden-quench
and adiabatic approximations developed in Sec.\ref{sec:DA}.

Now let us consider the particular functions
\begin{equation}\label{ex:momega}
\theta_t = \theta_0 \exp[k_1 t] \,, \,\,\,
m_t = m_0 \exp[-k_2 t]
\end{equation}
for $m_0, \theta_0, k_1$ and $k_2$ real and positive quantities.
Making $k_1=k_2$ and $m_0 \theta_0 = 1$,
we obtain the isotropic time-dependent harmonic oscillator mentioned above.

The Hessian in (\ref{ex:Ht}) for the chosen parameters in (\ref{ex:momega})
becomes
\begin{equation}\label{ex:Ht2}
{\bf H}_t =
\left(
\begin{array}{cc} h_{11} & h_{12} \\ h_{12} & h_{22} \end{array}
\right)
\end{equation}
with
\[
\begin{aligned}
& h_{11} :=
m_0\theta_0{\rm e }^{(k_1-k_2) t}
 \left[  {\rm e }^{k_1 t} k_1 \theta_0  +
       \tfrac{(k_1-k_2)}{2} \sin(2\theta_0{\rm e }^{k_1 t} )\right] \, ,\\
& h_{12} :=
(k_2-k_1) \sin^2(\theta_0{\rm e}^{k_1 t} ) \, , \\
& h_{22} :=\frac{{\rm e }^{(k_2-k_1) t}}{m_0\theta_0}
 \left[  {\rm e }^{k_1 t} k_1 \theta_0  -
       \tfrac{(k_1-k_2)}{2} \sin(2\theta_0{\rm e }^{k_1 t})\right] \, ,
\end{aligned}
\]
from where one can determine
\[
\det {\bf H}_t = {\rm e}^{2 k_1 t} k_1^2 \theta_0^2
-(k_1-k_2)^2 \sin^2(\theta_0 {\rm e}^{k_1 t} ) \, .
\]
The parameters must be chosen so that the Hessian matrix ${\bf H}_t$ remains
positive definite throughout the protocol, see Sec.\ref{sec:WR},
and further satisfies the initial condition ${\sf S}_0 = {\sf I}$, see Sec.\ref{sec:TDQH}.

\begin{figure}[htb]
\includegraphics[width=\columnwidth, trim=0 0 0 0]{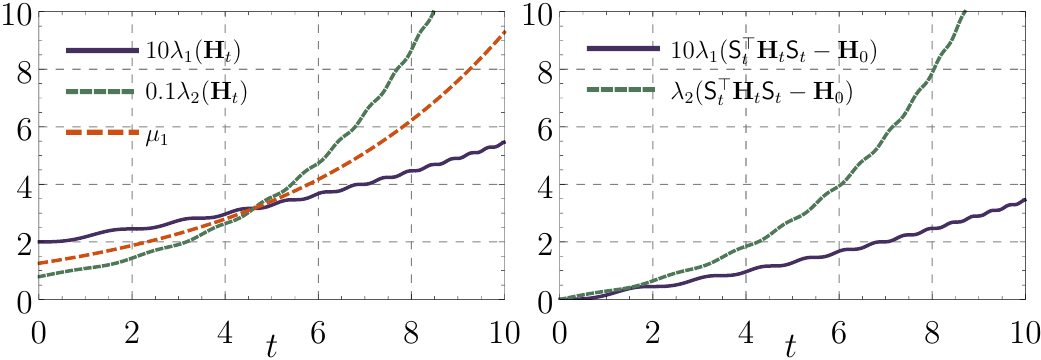}
\caption{{\bf Left:} Eigenvalues $\lambda_1$ and $\lambda_2$
and the symplectic eigenvalue $\mu_1$ of the Hessian matrix
${\bf H}_t$ in Eq.(\ref{ex:Ht2}).
The numerical value of $\lambda_1$ is multiplied by a factor $10$,
whereas that of $\lambda_2$ is divided by a factor $10$,
for visual clarity.
The crossing of the three curves around $t=4.5$ is a coincidence due
to the rescaling of the values.
{\bf Right:} Eigenvalues $\lambda_1$ and $\lambda_2$ of the Matrix
$({\sf S}_{\tau}^\top {\bf H}_\tau {\sf S}_{\tau}  - {\bf H}_0 )$
for the matrices ${\bf H}_t$ in Eq.(\ref{ex:Ht2}) and the symplectic matrix
${\sf S}_{t}$ in (\ref{ex:sympmat}).
The numerical value of $\lambda_1$ is multiplied by a factor $10$.
The parameters used in this figure are
$m_0 = 1, \, \theta_0 = 2\pi, \, k_1 = 0.2$, and $k_2 = 0.1$,
all written in arbitrary units.}                       \label{fig:eig}
\end{figure}

Considering a positive-definite Hessian in (\ref{ex:Ht2}),
it will be useful to perform its symplectic diagonalization, see Sec.\ref{sec:WRTO}.
Since this Hessian is a $2\times 2$ matrix, it has only one symplectic eigenvalue.
This considerably simplifies the Williamson decomposition,
since the symplectic spectrum is completely characterized by a single positive quantity
$\mu_1$, see Eq.(\ref{eq:sympdiag}).
Taking the determinant of Eq.(\ref{eq:sympdiag}),
one obtains $\det {\bf H}_t = \det {\bf \Lambda }_{{\bf H}_t} = \mu_1^2$ ,
where ${\bf \Lambda }_{{\bf H}_t} = \mu_1 {\sf I}_2$ with
$\mu_1(t) = \sqrt{\det {\bf H}_t}$.
Thus, the symplectic diagonalization of that Hessian is
\begin{equation} \label{ex:SympDiag}
{\sf S}_{{\bf H}_t}^\top {\bf H}_t {\sf S}_{{\bf H}_t}
= \mu_1(t) {\sf I}_2 \, , \,\,\,  \mu_1(t) = \sqrt{\det {\bf H}_t} \, .
\end{equation}
A particular set of parameters ensuring the required conditions
(${\bf H}_t >0$ and ${\mathsf S}_0 = {\sf I}_{2}$)
is used for the plot in Figure \ref{fig:eig}.

In this figure, the numerical values of the eigenvalues of
the matrix ${\bf H}_t$ in Eq.(\ref{ex:Ht2}) are plotted.
In the same plot one finds also the symplectic eigenvalue.
Since the ordinary eigenvalues are positive,
the matrix ${\bf H}_t$ remains positive definite during the entire protocol.
Consequently, the Williamson theorem ensures the existence of
a positive symplectic eigenvalue $\mu_1$.
All this guarantees the existence of a well-defined thermal equilibrium state
according to the discussion in Sec.\ref{sec:WRTO}.

For the Hessian in (\ref{ex:Ht2}),
using the definition in (\ref{eq:Sz_uindep}),
one can find, by expanding the exponential, that
\[
{\sf S}_u^{(t)} = {\rm e}^{u {\sf J}{\bf H}_t} =
 \cos(\mu_1 u) {\sf I}_{2} +
 {\mu_1}^{-1} \sin(u \mu_1) {\sf J}{\bf H}_t \, .
\]
The matrix above, together with $\xi_t$ in (\ref{ex:xit}), determines
the vector $z_u^{(t)}$ in (\ref{eq:Sz_uindep}).
These two quantities,
the symplectic evolution matrix in (\ref{ex:sympmat}),
and the vector in Eq.(\ref{ex:vecz})
completely determine the three parameters in (\ref{eq:chiparam})
and thus the CF in (\ref{eq:finalQuadChi}).
Note that the complex symplectic matrix in (\ref{eq:th_oper}) is given by
${\rm S}_\beta^{(0)} = {\sf S}_{u = i\hbar\beta}^{(0)}$.

\subsection{Pure Symplectic Dynamics}

\begin{figure}[bht]
\includegraphics[width=\columnwidth, trim=0 0 0 0]
{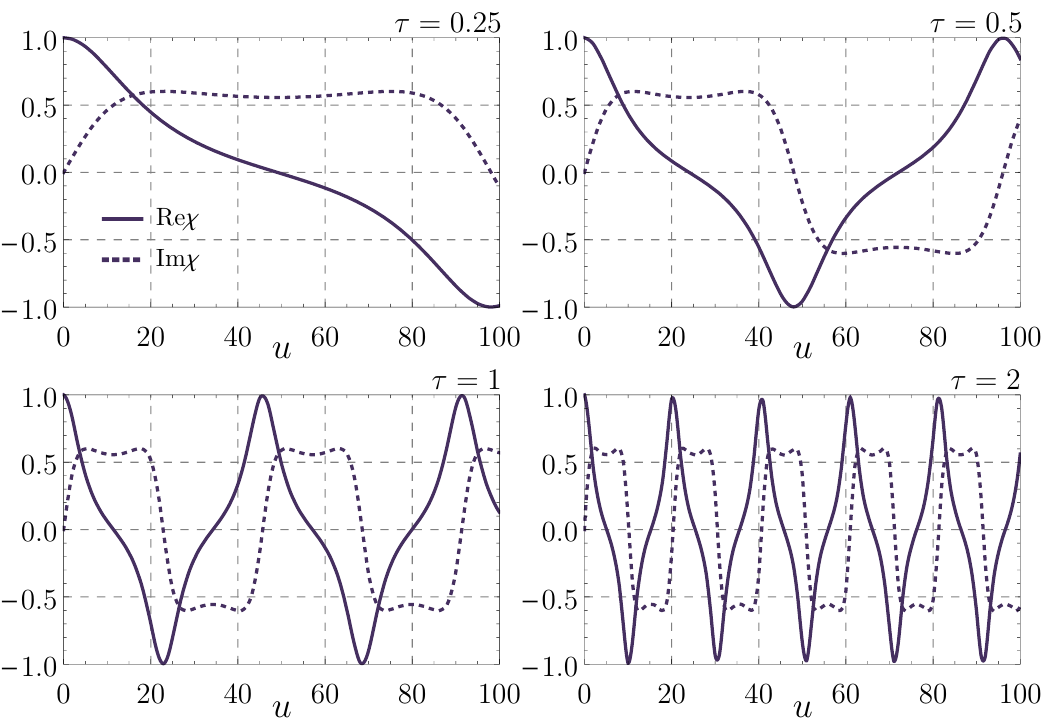}
\caption{CF as a function of $u$ under an {\it undisplaced Hamiltonian}
for four values of the final time of the protocol $\tau$.
The inverse temperature of the initial thermal
equilibrium state (\ref{eq:thermH0}) is $\beta = 1$.
The continuous (dotted) curves correspond to the
real (imaginary) part of function in (\ref{eq:chinullxi})
for the QH in (\ref{eq:QuadHam}) with
vector $\xi_t=0, \forall t$,
that is $a_0=0$ in (\ref{ex:xit}), and $h_t = 0, \forall t$.
The Hessian of Hamiltonian is defined in (\ref{ex:Ht}),
with mass and frequency both in (\ref{ex:momega}).
The parameters in (\ref{ex:momega})
used in this plot are
$m_0 = 1, \, \theta_0 = 2\pi, \, k_1 = 0.2$, and $k_2 = 0.1$
(The same as in Fig.\ref{fig:eig}).
All quantities are written in arbitrary units.}
\label{fig:ChiTau}
\end{figure}

We begin with the simplest situation,
namely an undisplaced Hamiltonian $(a_0=0)$.
In this regime, the work statistics are entirely determined by the symplectic evolution
generated by the Hessian matrix.
As the protocol duration grows,
the characteristic function develops increasingly rapid
oscillations\footnote{Let us emphasize that these oscillations
occur in the Fourier-conjugate variable $u$ of the work distribution
and should not be interpreted as physical oscillations in time.},
see Figure \ref{fig:ChiTau},
indicating larger departures from equilibrium,
as will be shown when analyzing the nonequilibrium lag. 

The CFs in this figure are computed for the QH in (\ref{eq:QuadHam})
for the Hessian in (\ref{ex:Ht2}) but setting $a_0 = 0$ in (\ref{ex:xit}),
that is, $\xi_t = z_t = 0, \forall t$,
see Eqs.(\ref{ex:xit}) and (\ref{ex:vecz}).
Under these conditions,
the plotted CF has the form in (\ref{eq:chinullxi})
and each graph in this figure corresponds to a different value of $\tau$.
The oscillations of the CF are not strictly periodic in $u$ and,
in particular, the local oscillation wavelength grows
and the spacing between successive maxima increases.

\begin{figure}[ht]
\includegraphics[width=\columnwidth, trim=0 0 0 0]
{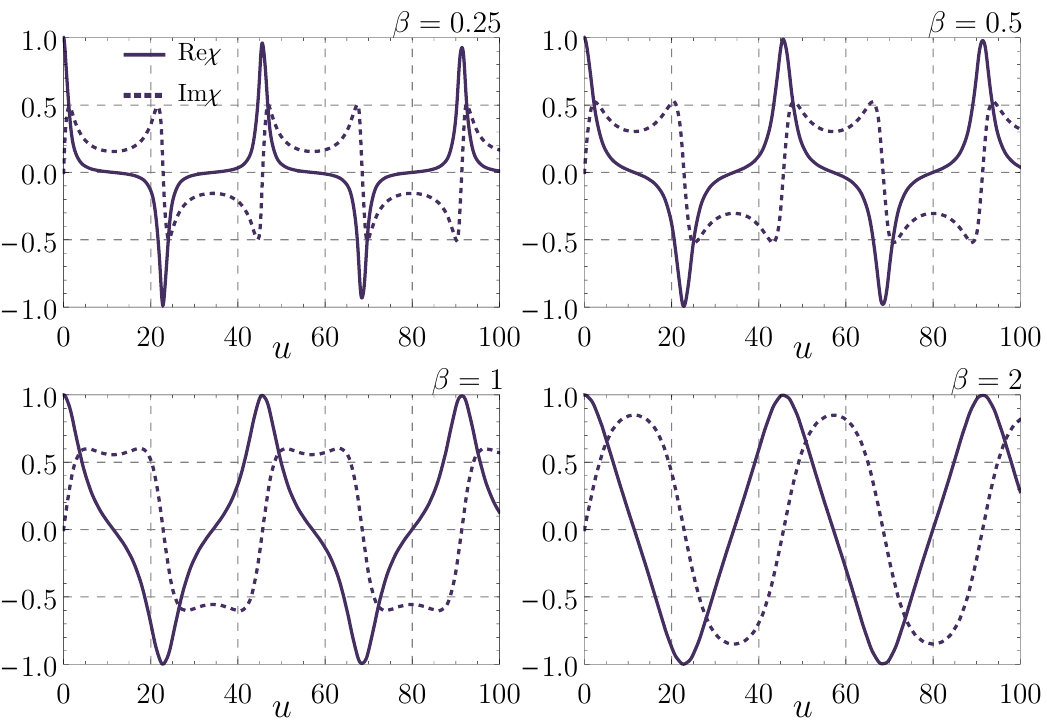}
\caption{CF as a function of $u$ under an
{\it undisplaced Hamiltonian}
for four values of the temperature of the initial state $\beta$.
The final time of the protocol is $\tau = 1$.
See Fig. \ref{fig:ChiTau} for the remaining parameters description.}
\label{fig:Chibeta}
\end{figure}

Having analyzed the dependence on the protocol duration,
we now examine the influence of the initial thermal state.
In Figure \ref{fig:Chibeta},
we consider the same CF but for four different values of the temperature
of the initial thermal equilibrium state, the state in Eq.(\ref{eq:thermH0}).
As $\beta$ increases (the temperature decreases),
the interval in $u$ between two complete successive oscillations does not change
(compare the position of the peaks in the four plots),
however the CF becomes progressively closer to a single-frequency sinusoidal
oscillation.
Although it is nonperiodic in $u$,
the dominant oscillation scale appears to be weakly dependent on temperature.

In the limit $\beta \to \infty$,
the equilibrium state $\hat \rho_\beta^\text{th}$
approaches the vacuum state of the Hamiltonian $\hat H_0$,
say $|\tilde{\mathfrak{0}} (0)\rangle$;
this is the vector in (\ref{eq:specQuad}) for $n = 1$, $t=0$, $\mathfrak{m} = 0$,
and $\xi_0 = 0$ (undisplaced Hamiltonian case).
Taking the trace in (\ref{eq:chi}) with
$\hat \rho^\text{th}_0 = \lim_{\beta\to\infty}
\hat M_{{\rm S}_{\beta}^{(0)}} =
| \tilde{\mathfrak{0}} \rangle\!\langle\tilde{\mathfrak{0}}|$,
one obtains
\[
\chi_\tau(u) = {\rm e}^{\frac{i}{2} u \mu_1(0)}
\langle \tilde{\mathfrak{0}}|\hat U_\tau^\dagger
\hat M_{{\sf S}_{-u}^{(\tau)}}
\hat U_\tau |\tilde{\mathfrak{0}} \rangle \, ,
\]
where we replaced
${\rm e}^{i u \hat {\mathcal H}_t /\hbar} = \hat M_{{\sf S}_{-u}^{(t)}} $
for a QH and used that
$\hat M_{{\sf S}_u^{(0)}} = {\rm e}^{-i u \hat {H}_0 /\hbar}$,
thus $\hat M_{{\sf S}_u^{(0)}}|\tilde{\mathfrak{0}}\rangle =
 {\rm e}^{- \frac{i}{2} u \mu_1(0)}|\tilde{\mathfrak{0}}\rangle$.
Writing $\hat U_\tau$ as in (\ref{eq:UtdepQuad})
but for $\xi_\tau = 0$ (undisplaced Hamiltonians),
performing the spectral decomposition
\[
\hat M_{{\sf S}_{-u}^{(\tau)}} = {\rm e}^{\frac{i}{\hbar} u \hat H_\tau} =
\sum_{\tilde{\mathfrak{m}} = 0}^\infty
{\rm e}^{i (\tilde{\mathfrak{m}} + \frac{1}{2}) \mu_1(\tau) u }
| \tilde{\mathfrak{m}}(\tau) \rangle\!\langle\tilde{\mathfrak{m}}(\tau)|,
\]
and using the Metaplectic composition in (\ref{eq:met_composition}),
the CF above becomes
\[
\chi_\tau(u) =
\sum_{\tilde{\mathfrak{m}} = 0}^\infty
{\rm e}^{i \nu(\tilde{\mathfrak{m}}) u }
|\langle \tilde{\mathfrak{m}}(\tau)|\hat U_\tau| \tilde{\mathfrak{0}} \rangle|^2 \,
\]
where we define
\[
\nu(\tilde{\mathfrak{m}}) :=
\tilde{\mathfrak{m}}\mu_1(\tau) + \tfrac{1}{2}[ \mu_1(\tau) - \mu_1(0)].
\]
The zero-temperature characteristic function can thus be interpreted
as a superposition of infinitely many oscillatory contributions,
each one with different amplitude\footnote{%
This amplitude may be determined using standard symplectic techniques,
since $\hat U_t$ is essentially a Metaplectic operator,
however, this is outside the scope of this paper.}
$|\langle \tilde{\mathfrak{m}}(\tau)|\hat U_\tau| \tilde{\mathfrak{0}} \rangle|$.
The period associated with each frequency
is given by
$\kappa(\tilde{\mathfrak{m}}) = 2\pi/\nu(\tilde{\mathfrak{m}})$ and,
using (\ref{ex:SympDiag}) and the values of the parameters in Fig.\ref{fig:Chibeta},
we find $\nu(0) \approx 0.14$ and $\nu(1) \approx 1.7$ respectively associated with
$\kappa(0) \approx 45.7$ and
$\kappa(1) \approx 3.8$.
There is one order of magnitude separating
these periods (and the frequencies)
of the first and the second term.
Note that the value of $\lambda(0)$ fits the separation between
two successive peaks in the plots in Fig.\ref{fig:Chibeta}.

\begin{figure}[htb]
\includegraphics[width=\columnwidth, trim=0 0 0 0]
{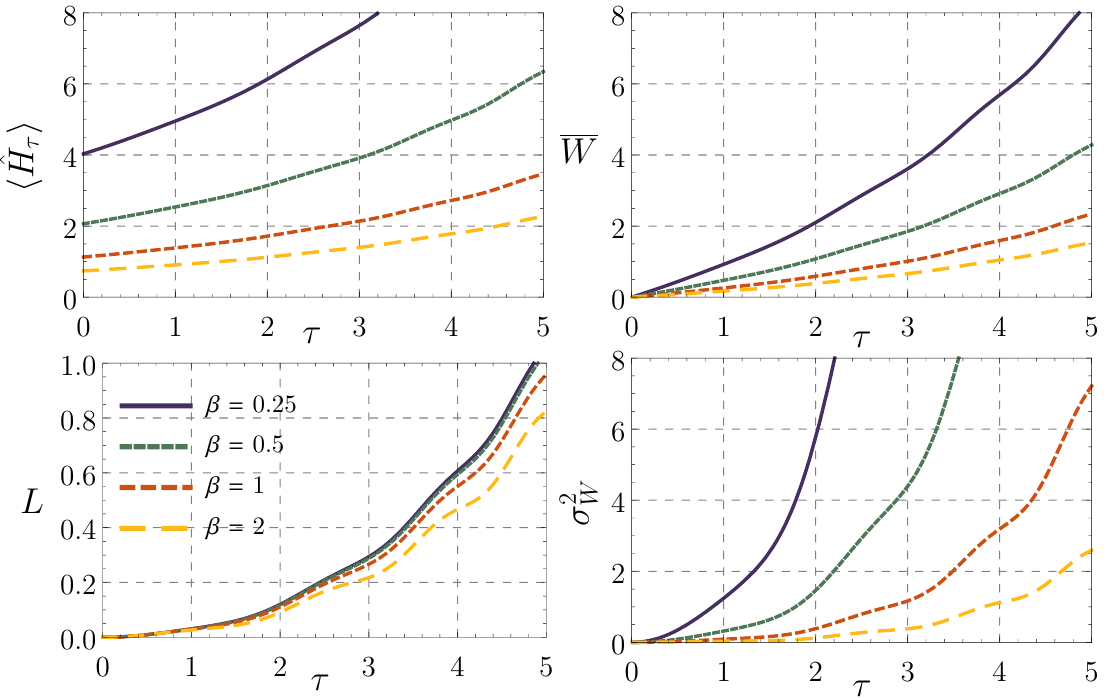}
\caption{
Mean Energy
$\langle {\hat H}_\tau \rangle$ in Eq.(\ref{eq:meanenerg}),
Mean work $\overline{W}$ in Eq.(\ref{eq:Wmv}),
Nonequilibrium Lag $L$ in Eq.(\ref{eq:noneqlag}),
and variance of the work distribution $\sigma_{W}^2$
in (\ref{eq:varW_nullxi})
as functions of the final time of the protocol $\tau$
under an {\it undisplaced Hamiltonian}
for four values of the inverse temperature
of the initial state $\beta$ in (\ref{eq:thermH0}).
See Fig. \ref{fig:ChiTau} for the remaining parameters description.}\label{fig:ThermBeta}
\end{figure}

In Figure \ref{fig:ThermBeta}, we show the behavior of the thermostatistical
properties of the system as a function of the final time of the work protocol.
The figure shows
the energy (mean-value of the Hamiltonian) ---
given by the expression in (\ref{eq:meanenerg}) with ${\bf V} = {\bf V}_\tau$
in (\ref{eq:mvcmEvol0}) and $\langle \hat x\rangle^\tau = 0$,
the mean-value of the work distribution [see Eq.(\ref{eq:Wmv})],
the nonequilibrium Lag [see Eq.(\ref{eq:noneqlag})],
as a measure of the irreversibility of the protocol, and
the variance of the work distribution [see (\ref{eq:varW_nullxi})].
Within the parameter regime considered here,
all quantities increase with $\tau$ and decrease with $\beta$ (increase with temperature).
Both the mean energy and the average work remain positive throughout the protocol.
For the latter, as discussed below equation (\ref{eq:Wmv}),
the positivity of the values is a consequence of the positive definiteness of the matrix
$({\sf S}_{\tau}^\top {\bf H}_\tau {\sf S}_{\tau}  - {\bf H}_0 )$,
see Fig.~\ref{fig:eig}.
The positivity of the mean energy follows directly from the positive definiteness of
${\bf H}_t$, as shown in Fig.\ref{fig:eig},
which makes the trace and the scalar product in Eq.(\ref{eq:meanenerg}) positive quantities.
As a comment, the positivity of $L$ for all values of $\tau$,
see Eq.(\ref{eq:noneqlag}),
imples the nonequilibrium version of the second law of thermodynamics
for the system in question: 
$\overline{W} \geq \Delta F$.

\subsection{Role of Displacements}
\begin{figure}[ht]
\includegraphics[width=\columnwidth, trim=0 0 0 0]
{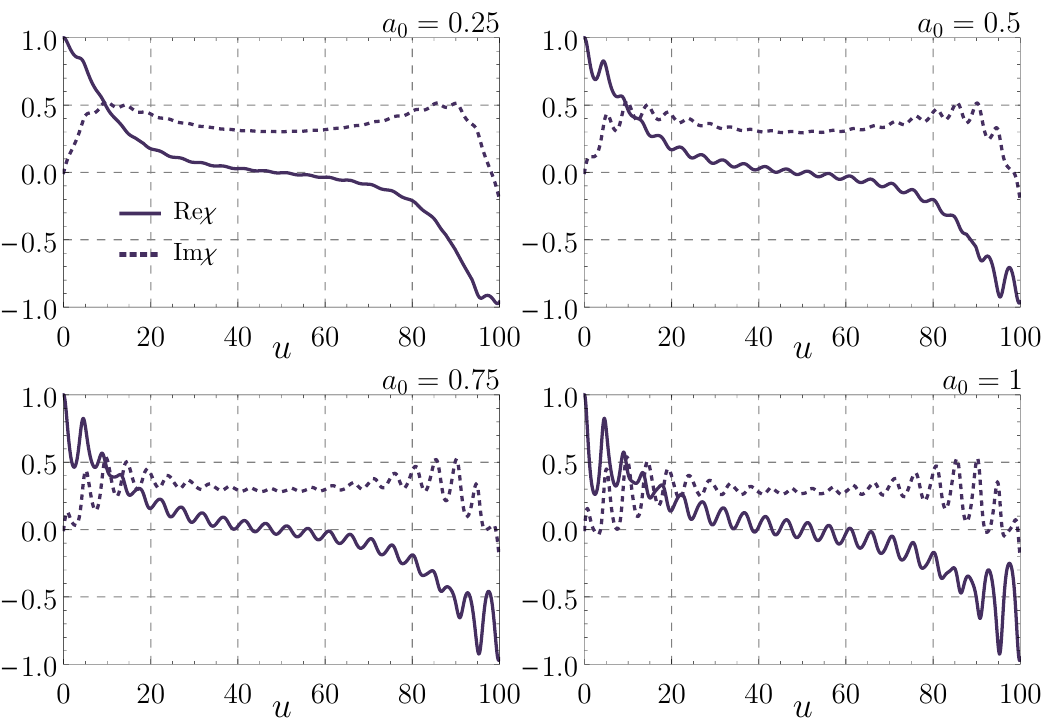}
\caption{CF as a function of $u$ under a {\it displaced Hamiltonian}
and for four values of the displacement parameter $a_0$.
The continuous (dotted) curves correspond to the
real (imaginary) part of the function in (\ref{eq:finalQuadChi})
for the QH in (\ref{eq:QuadHam}) with
vector $\xi_t$ in (\ref{ex:xit}), and $h_t = 0, \forall t$.
The Hessian of Hamiltonian is defined in (\ref{ex:Ht}),
with mass and frequency both in (\ref{ex:momega}).
The final time of the protocol is $\tau = 0.25$ and
the inverse temperature of the initial thermal
equilibrium state (\ref{eq:thermH0}) is $\beta = 0.5$.
See Fig.\ref{fig:ChiTau} for the remaining parameters description.} \label{fig:ChiXi}
\end{figure}

\begin{figure}[ht]
\includegraphics[width=\columnwidth, trim=0 0 0 0]
{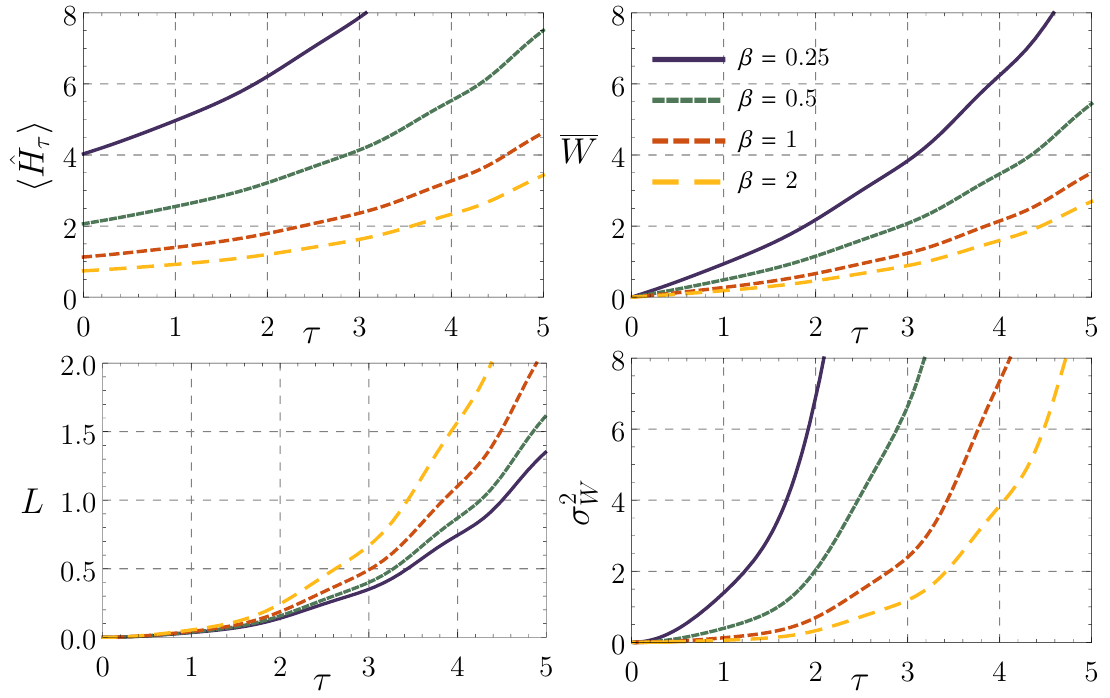}
\caption{Mean Energy  $\langle {\hat H}_\tau \rangle$,
Mean work $\overline{W}$,
Nonequilibrium Lag $L$, and
variance of the work distribution $\sigma_{W}^2$
as functions of the final time of the protocol $\tau$
under the {\it displaced Hamiltonian} and for
four values of the temperature of the initial state $\beta$.
The displacement is given by the vector $\xi_t$ in (\ref{ex:xit}) with
$a_0 = 0.05$.
The variance $\sigma_W^2 := \overline{ W^2 } - \overline{W}^2$
is determined using the definition (\ref{eq:Wmoments}).
The Hamiltonian of the system is given by (\ref{eq:QuadHam}) with
vector $\xi_t$ in (\ref{ex:xit}), $h_t = 0, \forall t$,
and Hessian defined in (\ref{ex:Ht}),
with mass and frequency both in (\ref{ex:momega}).
See Fig. \ref{fig:ChiTau} for the remaining parameters description.}
\label{fig:ThermBetaXi}
\end{figure}

\begin{figure}[htb]
\includegraphics[width=\columnwidth, trim=0 0 0 0]
{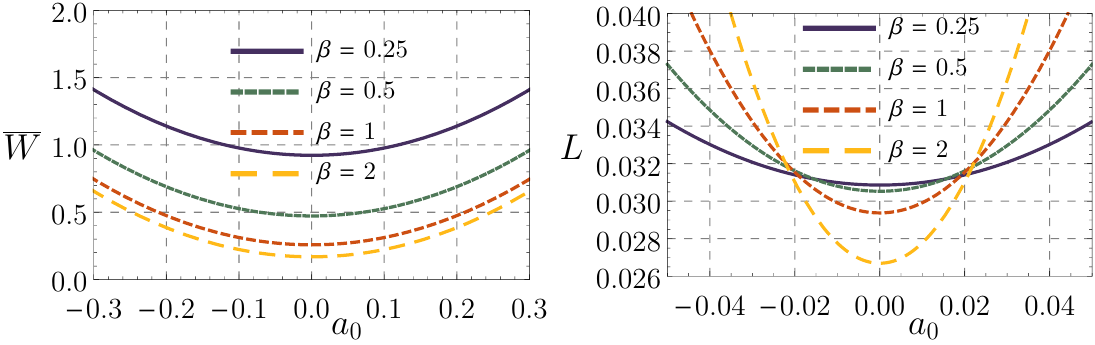}
\caption{Mean work $\overline{W}$ in Eq.(\ref{eq:Wmv})
and Non-equilibrium lag $L$ in Eq.(\ref{eq:noneqlag})
under the {\it displaced Hamiltonian},
as functions of the displacement parameter $a_0$,
and for four values of the inverse temperature of the initial state $\beta$.
The value of the final time protocol is $\tau = 1$.
See Fig. \ref{fig:ChiTau} for the remaining parameters description.}
\label{fig:ThermBetaXi2}
\end{figure}

Having established the behavior of the purely symplectic dynamics,
we now turn to the role of phase-space displacements.
We will consider $a_0 \ne 0$ in (\ref{ex:xit}),
that is, now the Hamiltonian has the form in (\ref{eq:QuadHam})
with $h_t = 0, \forall t$,
with Hessian in (\ref{ex:Ht2}) and $\xi_t$ in (\ref{ex:xit}).
Using the parameters in (\ref{eq:chiparam}),
the CF in (\ref{eq:finalQuadChi}) may be constructed.
Figure \ref{fig:ChiXi} reveals that displacements generate additional
oscillatory structures in the CF and this modification
is not merely cosmetic.
As shown in Fig.\ref{fig:ThermBetaXi},
the same displacements substantially increase the mean work
and the nonequilibrium lag.
The origin of this enhancement is explained in Fig.\ref{fig:ThermBetaXi2}.


In Figure \ref{fig:ChiXi},
we find the CF for four values of the parameter $a_0$.
Comparing with the plots in Fig.\ref{fig:ChiTau} (the case $a_0 = 0$),
the displacement generates additional oscillatory components in the
CF, leading to a richer interference pattern,
whose origin lies in the exponential term in (\ref{eq:chiQuadFunc}).
This term is absent in (\ref{eq:chinullxi}),
which is the case of Fig.\ref{fig:ChiTau}.
In Figure \ref{fig:ThermBetaXi}, we plot the thermostatistical functions for $a_0 = 0.05$.
Direct comparison with Fig.\ref{fig:ThermBeta}
shows that the displacement increases all the plotted quantities,
even for this small value of $a_0$.
Comparing (\ref{eq:Wmv_nullxi}) with (\ref{eq:Wmv}),
the corresponding displacement-added term for the mean work,
\[
\overline{W}_1(a_0) := \tfrac{1}{2} (\langle \hat x \rangle^\tau - \xi_\tau) \cdot
{\bf H}_\tau (\langle \hat x \rangle^\tau - \xi_\tau) \, ,
\]
is positive, since it is a quadratic form  with a positive-definite
${\bf H}_\tau$ (see Fig.\ref{fig:eig}),
and it does not depend on temperature, as discussed below Eq.(\ref{eq:Wmv}).
For the same reason,
this is also true for the respective term in the mean-value of energy
in (\ref{eq:meanenerg}), which now has $\langle \hat x\rangle^\tau \ne 0$.
Although the values of the nonequilibrium lag increase
with the presence of the displacement,
it behaves differently when compared with all other quantities:
there is a reversal in the ordering of the curves with respect to temperature
in comparison with the respective curves in Fig.\ref{fig:ThermBetaXi}.
This is associated with the term $\overline{W}_1(a_0)$ above, as we will explain.

Figure \ref{fig:ThermBetaXi2} reveals a crossover between two regimes.
For small displacements,
the temperature-dependent contribution dominates the nonequilibrium lag.
Beyond a critical displacement amplitude, the displacement-induced work
contribution becomes the leading mechanism governing irreversibility.
According to the definition in Eq.(\ref{eq:noneqlag}),
the nonequilibrium lag contains the free energy contribution,
which does not depend on the displacement, or on $a_0$,
but only on the temperature, see (\ref{eq:def_fe}).
Consequently, the only displacement contribution to the lag
in formula (\ref{eq:noneqlag}) is $\overline{W}_1(a_0)$ above, which
does not depend on temperature.
However, in the lag expression all terms are multiplied by $\beta$,
such that it contains the linear term $\beta\overline{W}_1(a_0)$
and two others (the free energy $\Delta F$ and $\overline{W}-\overline{W}_1$)
with more intricate $\beta$-dependence.
The competition between these terms is the reason behind the reverse ordering
and is plotted in Fig.\ref{fig:ThermBetaXi2}.
In this figure, one can note that both the work and the lag are
increasing functions of the modulus of the displacement parameter $|a_0|$.
However, for small values of $a_0$, the lag is smaller
for higher values of $\beta$ (smaller values of the temperature),
a region dominated by the $\beta$-dependent term in the mean-value of work.
After $|a_0|\approx 0.2$, the lag increases
for decreasing values of $\beta$,
where the displacement ($\beta$-independent)
contribution to the mean work dominates.
The irreversibility of the work protocol, measured by $L$,
is largely due to displacements in the Hamiltonian,
while quantum effects,
such as squeezing,
are due to the Hessian of the Hamiltonian ${\bf H}_t$,
the generator of the symplectic evolution\footnote{%
The Hessian generates the linear symplectic dynamics,
including squeezing transformations,
whose quantum consequences are encoded in the Metaplectic
representation.}, see Eq.(\ref{eq:dif_eq}).

\subsection{Sudden-Quench Limit}
This regime provides an interesting contrast with the continuous evolution
discussed above. In this case the displacement affects the work statistics
much less efficiently.
\begin{figure}[ht]
\includegraphics[width=\columnwidth, trim=0 0 0 0]
{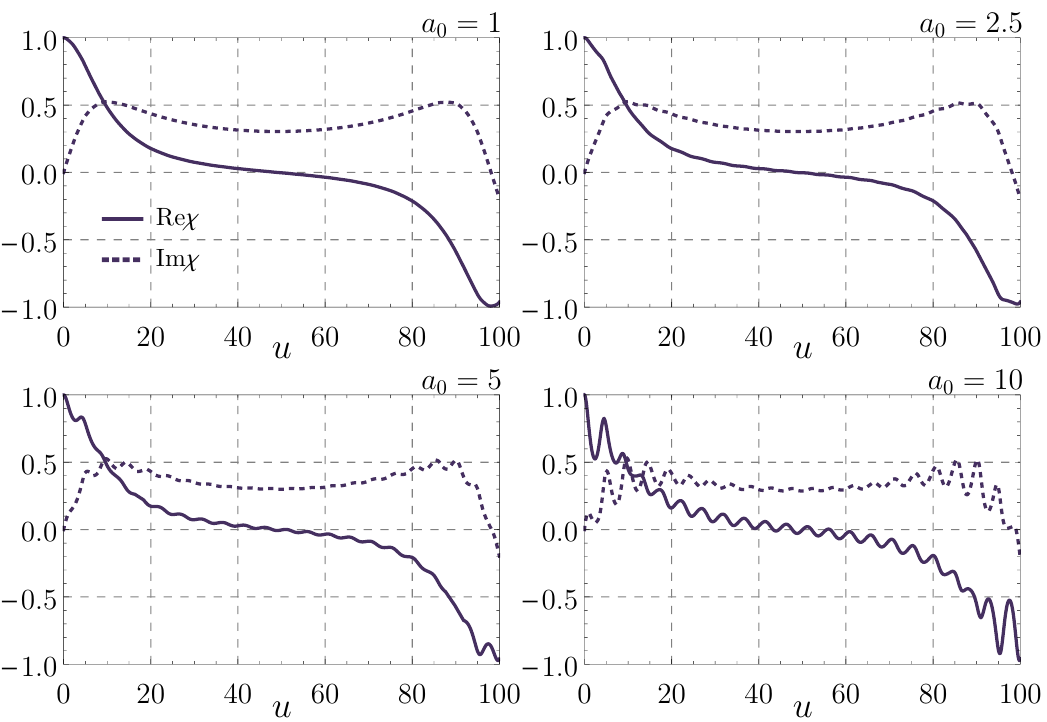}
\caption{CF as a function of $u$ under a {\it sudden quench}
of the system Hamiltonian for four values of the displacement
parameter $a_0$.
The final evolution time of the protocol is $\tau = 0.25$ and
the inverse temperature of the initial state is $\beta = 0.5$.
See Fig.\ref{fig:ChiXi} for details and
Fig. \ref{fig:ChiTau} for the remaining parameters description.}
\label{fig:Chia0Sd}
\end{figure}

\begin{figure}[ht]
\includegraphics[width=\columnwidth, trim=0 0 0 0]
{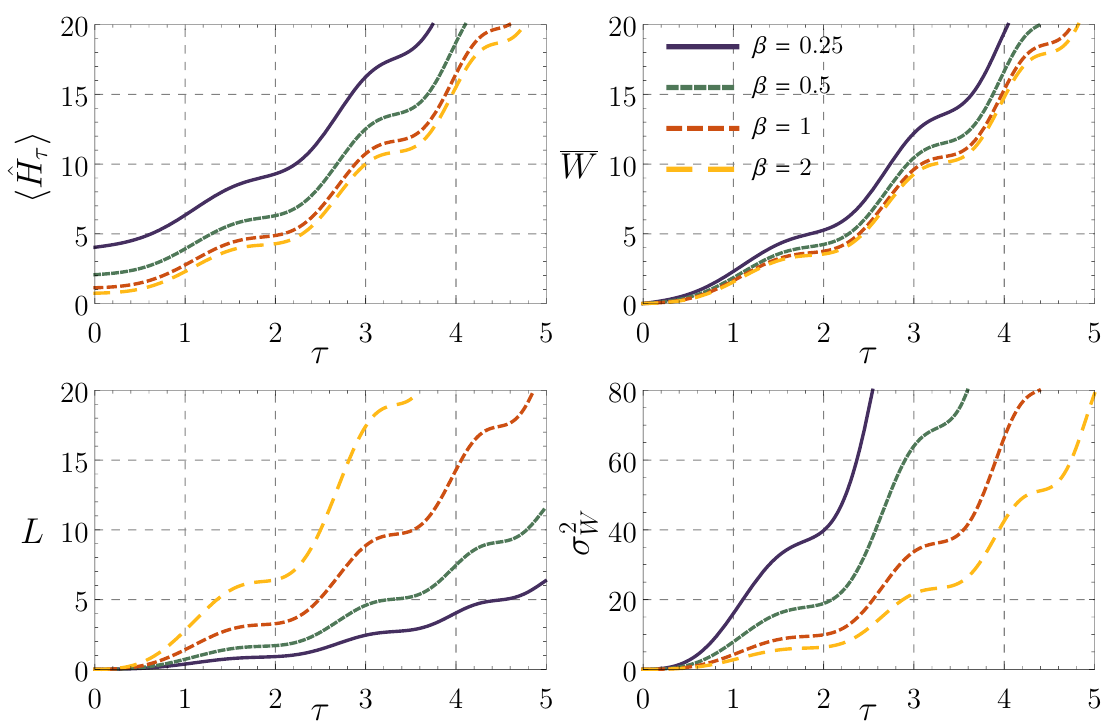}
\caption{Mean Energy  $\langle {\hat H}_\tau \rangle$,
Mean work $\overline{W}$,
Nonequilibrium Lag $L$, and
variance of the work distribution $\sigma_{W}^2$
as functions of the final time of the protocol $\tau$
under a {\it sudden quench} of the displaced Hamiltonian
for four values of the temperature of the initial state $\beta$.
The displacement is given by the vector $\xi_t$ in (\ref{ex:xit}) with
$a_0 = 5$.
Other details can be found in the legend of Fig.\ref{fig:ThermBetaXi}.}
\label{fig:ChiBetaSd}
\end{figure}

In Figure \ref{fig:Chia0Sd}, we plot the CF for a sudden evolution under a
displaced Hamiltonian.
In this case, we consider the sudden change of the protocol parameters
$\lambda_t = (\xi_t,{\bf H}_t, h_t)$ as
$\lambda_0 = (\xi_0,{\bf H}_0, 0) \to \lambda_\tau =
(\xi_\tau,{\bf H}_\tau,0)$, which determines the initial
and final Hamiltonian through (\ref{eq:QuadHam})
for ${\bf H}_0$ and ${\bf H}_\tau$ given by (\ref{ex:Ht2})
and for $\xi_t$ in (\ref{ex:xit}).
The four graphs in this figure correspond to different values of
the parameter $a_0$ entering in $\xi_t$, see Eq.(\ref{ex:xit}).
According to the recipe in Sec.\ref{sec:SE},
the sudden evolution replaces the matrix in (\ref{ex:sympmat})
by ${\sf S}_t = {\sf I_{2}}$ and the vector in (\ref{ex:vecz})
by $z_t = 0$.
The thermostatistical quantities are plotted in Figure \ref{fig:ChiBetaSd}
and exhibit behavior qualitatively similar to that shown in Fig.\ref{fig:ThermBetaXi},
however,
note that the value of $a_0$ of these figures differs by two orders of
magnitude.

In comparison to the plots in Fig.\ref{fig:ChiXi} (continuous time evolution),
much larger displacement amplitudes $a_0$ are required in the sudden-quench regime before
displacement-induced oscillations become clearly visible
in Fig.\ref{fig:Chia0Sd} (sudden quench).
The difference in behavior between these curves arises from the sudden change
prescription $z_t = 0$:
For the continuous time evolution, this vector depends linearly on $a_0$,
see Eq.(\ref{ex:vecz}), which contributes to the oscillatory phase
in (\ref{eq:finalQuadChi}) through the parameters in (\ref{eq:chiparam}),
while in the sudden regime, the only contribution depending on $a_0$ in
Eq.(\ref{eq:chiparam2}) is the vector $\xi_\tau$, see Eq.(\ref{ex:xit}).

\subsection{Adiabatic Regime}
We now turn to the adiabatic regime, which constitutes the opposite limit of the sudden
quench discussed above. In this regime, the system continuously adapts to
the instantaneous eigenstructure of the Hamiltonian.
As shown in Sec.\ref{sec:AA}, the resulting work statistics depend only
on the evolution of the symplectic spectrum and become insensitive
to phase-space displacements.

In the present model, the adiabatic Hamiltonian is
completely characterized by the effective Hessian ${\bf H}_\tau^0$ defined in
Eq.(\ref{eq:Hess_adiab}).
Therefore, the explicit construction of the adiabatic CF reduces to determining
the Williamson decomposition of the initial Hessian ${\bf H}_0$.

In order to determine the elements in Eq.(\ref{eq:Hess_adiab}),
we first proceed as done in Eq.(\ref{ex:SympDiag}),
where we found the symplectic spectrum of ${\bf H}_t$.
It remains to find the symplectic matrix diagonalizing the Hessian ${\bf H}_0$;
for this example, we can take advantage of the fact that this Hessian, given by
(\ref{ex:Ht2}) for $t=0$, is diagonal for the parameters in Fig.\ref{fig:ChiTau},
that is, ${\bf H}_0 = {\rm Diag}(h_{11}, h_{22})$ with
$h_{11} = 4\pi^2/5$ and $h_{22} = 1/5$.
In this case, resorting to (\ref{ex:SympDiag}) with $t=0$,
the determination of the symplectic diagonalizing matrix is simple:
${\sf S}_{{\bf H}_0} = {\rm Diag}(\sqrt[4]{h_{22}/h_{11}},\sqrt[4]{h_{11}/h_{22}})$,
as follows directly by evaluating the product in (\ref{ex:Ht2}) for $t=0$
using diagonal $2 \times 2$ matrices.
Ultimately, from Eq.(\ref{eq:Hess_adiab}),
${\bf H}_\tau^0 = \mu_1(t) {\rm Diag}(\sqrt{h_{11}/h_{22}},\sqrt{h_{22}/h_{11}})$,
where $\mu_1$ is plotted in Fig.\ref{fig:eig}.
With this matrix, the CF in Eq.(\ref{eq:adiabchi}) is determined.

\begin{figure}[hbt]
\includegraphics[width=\columnwidth, trim=0 0 0 0]
{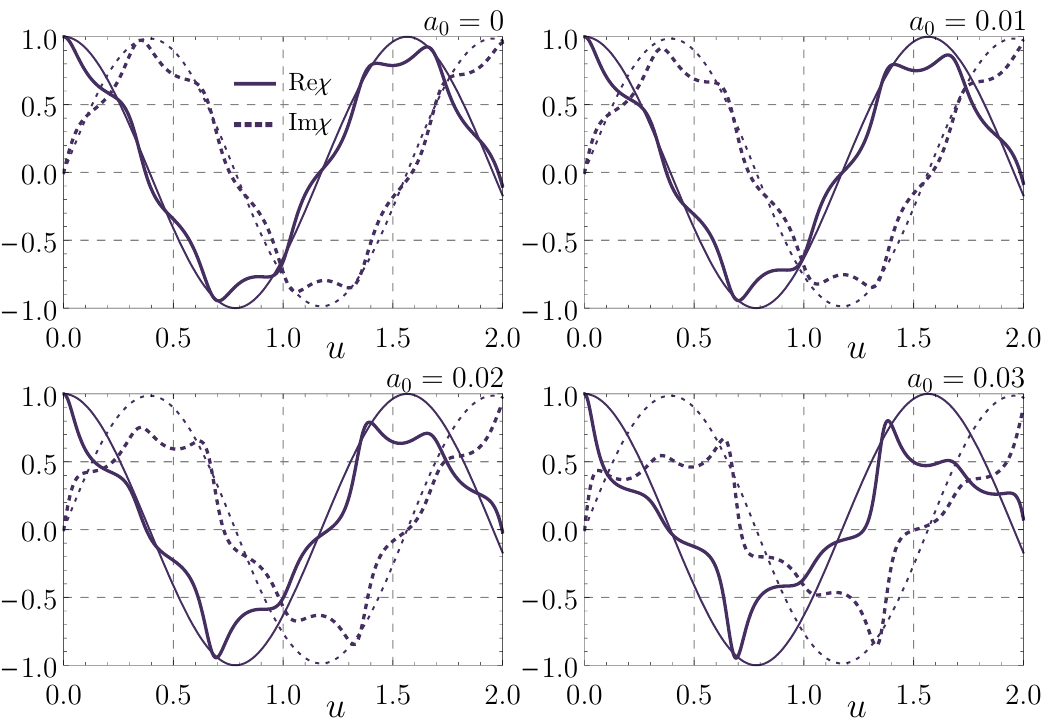}
\caption{CF as a function of $u$ for
four values of the displacement parameter $a_0$ (thicker lines)
and its {\it adiabatic version} (thin lines).
The adiabatic curves are the same in all four graphs.
The final evolution time of the protocol is $\tau = 10$ and
the inverse temperature of the initial state is $\beta = 4$.
See Fig.\ref{fig:ChiXi} for details and
Fig.\ref{fig:ChiTau} for the remaining parameters description.} \label{fig:Chia0Ad}
\end{figure}

The most remarkable feature of the adiabatic regime is the complete suppression of
displacement effects. This property is illustrated in Fig.\ref{fig:Chia0Ad}, where the
adiabatic characteristic function of Eq.(\ref{eq:adiabchi}) is compared with exact results
obtained for several displacement amplitudes $a_0$.
As explained in Sec.\ref{sec:AA},
the displacement dependence of any QH disappears in the adiabatic limit,
thus the adiabatic CF does not depend on it, that is,
it is the same for the four values of the parameter in the plot.
Consequently, the exact CF progressively departs from the adiabatic prediction
as the displacement amplitude $a_0$ increases,
owing to the appearance of displacement-induced oscillatory contributions.
Interestingly, small deviations from the adiabatic prediction remain visible
even in the absence of displacements ($a_0=0$).
These residual oscillatory structures arise from nonadiabatic corrections
that are not captured by the adiabatic approximation.
Note that, as discussed before (see Fig.\ref{fig:Chibeta}),
as the temperature decreases, the additional thermally populated oscillatory
components are suppressed,
and the characteristic function approaches its zero-temperature form.
In this limit, the initial thermal state becomes increasingly
concentrated around the ground state,
suppressing the contributions of excited transitions and
bringing the exact dynamics closer to the adiabatic prediction.

\begin{figure}[h]
\includegraphics[width=\columnwidth, trim=0 0 0 0]
{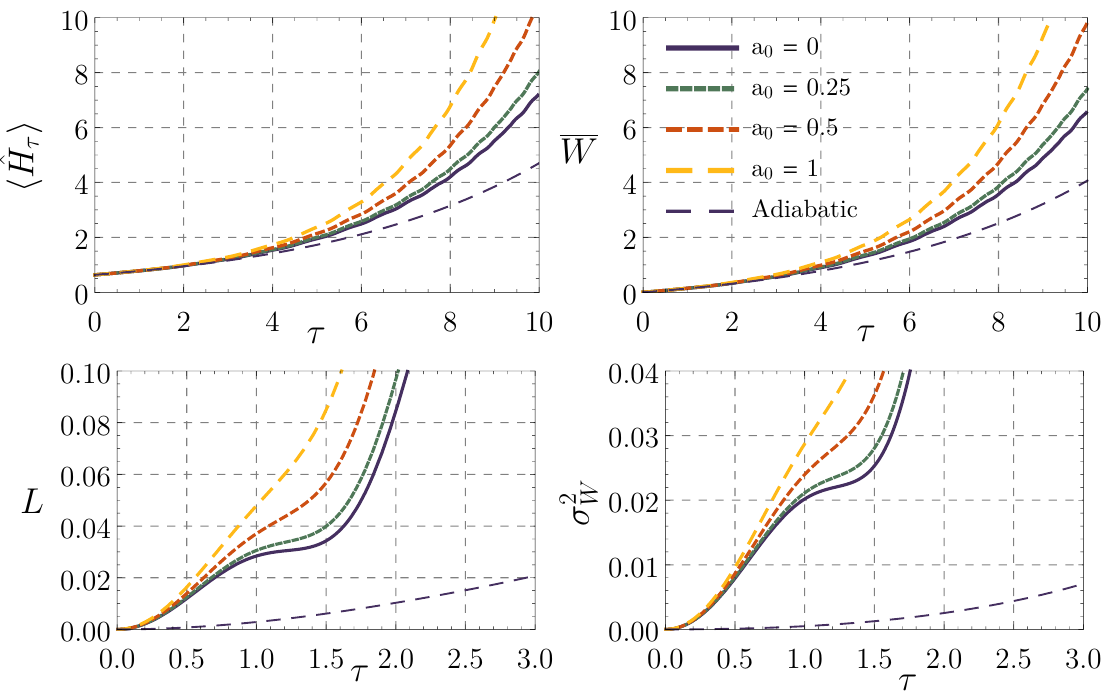}
\caption{Mean Energy $\langle {\hat H}_\tau \rangle$,
mean work $\overline{W}$ ,
Nonequilibrium Lag $L$, and variance of the work distribution $\sigma_{W}^2$
as functions of the final time of the protocol $\tau$
considering a displaced Hamiltonian
and for four values of the displacement parameter $a_0$ (see Fig.\ref{fig:ThermBetaXi})
and also for the adiabatic approximation (the thin dashed line).
In this case,
$\langle {\hat H}_\tau \rangle$ in Eq.(\ref{eq:admeanenerg}),
$\overline{W}$ and $\sigma_{W}^2$ both in Eq.(\ref{eq:adiabW}),
for the initial-state inverse temperature $\beta = 4$.
%
%
Other details can be found in the legend of Fig.\ref{fig:ThermBetaXi}.}
\label{fig:ChiAd}
\end{figure}

The disappearance of displacement dependence is not restricted to the CF.
It also affects the thermodynamic quantities derived from it.
Figure~\ref{fig:ChiAd} compares the exact dynamics with the
corresponding adiabatic predictions.

This figure shows that the discrepancy between the exact
and adiabatic descriptions increases with both the protocol duration
and the displacement amplitude.
This behavior reflects the growing importance of nonadiabatic effects as the system is
driven further away from the ideal adiabatic regime.
The non-adiabatic curves
are similar to those in Fig.\ref{fig:ThermBetaXi}
but for only one value of the temperature $\beta$ and different values of $a_0$.

Since the adiabatic characteristic function in Eq.(\ref{eq:adiabchi})
is independent of the displacement vector $\xi_t$,
all quantities derived from it inherit the same property.
In particular, the average work [Eq.(\ref{eq:adiabW})],
the work variance [Eq.(\ref{eq:adiabW})],
and the nonequilibrium lag [Eq.(\ref{eq:noneqlag})]
become independent of phase-space displacements in the adiabatic regime.

In summary, the adiabatic approximation removes all dis\-pla\-ce\-ment-induced
contributions from the characteristic function and from the thermostatistical quantities.
In this regime the work statistics are governed exclusively by the evolution of the
symplectic spectrum.
Deviations from the exact dynamics therefore provide a direct measure of
nonadiabatic effects and become particularly pronounced
when phase-space displacements play a significant role.

\section{Concluding Remarks} \label{conc}
We have derived an exact analytical expression for the characteristic function of work
in closed bosonic systems governed by general time-dependent quadratic Hamiltonians.
The central step is to represent every operator entering the two-projective-measurement
CF within the inhomogeneous metaplectic group,
associated with affine and linear transformations in phase space.
The composition of these operator then reduces to another element
within the group and the evaluation of the trace in the Weyl representation
yields Eq.(\ref{eq:finalQuadChi}),
which applies to an arbitrary number of degrees of freedom,
arbitrary time dependence of the Hamiltonian parameters,
and any initial thermal equilibrium state for which
the relevant expressions are well defined.

The inverse construction we adopt here enables a unified formulation for all
quadratic Hamiltonians: instead of confronting the explicit time-ordering problem,
we specify a differentiable symplectic path ${\sf S}_t \in {\rm Sp}(2n,\mathbb R)$,
and this path yields the corresponding quadratic Hamiltonian directly.
For a prescribed time-dependent quadratic Hamiltonian,
the same formulas remain applicable once we obtain
the associated symplectic matrix analytically or numerically.
Therefore, the method does not remove the underlying dynamical
problem in the conventional setting.
Rather, it reformulates that problem in terms of phase-space
differential equations and separates it from the subsequent operator
composition required to evaluate the work characteristic function.

The construction also extends to arbitrary nonequilibrium initial states
within the two-projective-measurement scheme.
In this case, we replace the initial density operator by its energy-dephased counterpart,
whose dependence enters the CF through a single Weyl symbol.
Further, the same operator structure of the CF
also adapts to Loschmidt amplitudes and mixed-state
interferometric total phases, placing these quantities and the work CF
within a common symplectic and metaplectic framework.

The general CF provides direct access to the thermostatistical properties of
the driven system, including the mean work, work variance,
the Jarzynski equality, and the nonequilibrium lag --- a
measure of the irreversibility of the protocol.
The symplectic formulation clarifies which Hamiltonian
parameters determine thermodynamical behaviour of the system:
the partition function, free energy and mean energy are all described in terms
of symplectic eigenvalues of the Hamiltonian, which are thus invariant under
symplectic transformations or metaplectic conjugations.
This invariance should be distinguished from the
statistics of a driven protocol,
which also depend on the evolution path and on the relation between
the initial state and the final Hamiltonian.

We illustrated the general construction using a one-di\-men\-sio\-nal model defined by a
nontrivial symplectic path.
The model includes physically relevant limiting configurations and permits a direct
comparison among the exact dynamics, the sudden-quench limit,
and the adiabatic approximation.
In the undisplaced case,
the symplectic evolution alone determines the work statistics.
The characteristic function displays a nonperiodic oscillatory structure.
These oscillations persist in the zero-temperature limit,
where the characteristic function can be interpreted as
a superposition of contributions weighted by
transition probabilities from the initial ground state.

In the presented example,
we showed that phase-space displacements and the purely symplectic dynamics
affect the work statistics in different ways.
The mismatch between the evolved first moment and
the final equilibrium displacement produces a nonnegative,
temperature-independent contribution to the mean work.
This contribution can substantially enhance the nonequilibrium lag.
By contrast, the Hessian
determines the linear symplectic flow and generates transformations such as squeezing,
whose quantum consequences appear through the metaplectic representation and the explicitly
$\hbar$-dependent terms in the work statistics.

Connecting with known results in the literature,
we obtain four particular regimes from the general CF,
these are cases of an undisplaced Hamiltonian, a sudden evolution,
a constant Hessian, and the adiabatic evolution.
These limiting cases further clarify the physical content of the exact result.
In the sudden-quench approximation,
the dynamical symplectic and affine evolution are suppressed
(the evolution unitary operator is the identity),
while the initial and final Hamiltonian parameters
still determine the work statistics.
In the adiabatic approximation, the work CF depends only on the
evolution of the instantaneous energy spectrum.
For quadratic Hamiltonians, the initial and final symplectic
eigenvalues encode the dependence,
while phase-space displacements do not contribute.
The difference between the exact and adiabatic results
therefore measures contributions that are not captured by ideal adiabatic transport.
In the studied example, these deviations become more pronounced as the displacement
contribution grows. Their detailed dependence on the protocol duration, however,
is model-specific and should not be interpreted as a universal monotonic
measure of nonadiabaticity.

Two assumptions delimit the scope of the present results.
Although the Wick rotation can be performed formally for a general quadratic Hamiltonian,
a normalizable thermal equilibrium state requires a Hamiltonian bounded from below.
We therefore restrict the thermodynamic interpretation to elliptic quadratic Hamiltonians
with positive-definite Hessians.
We also assume that the effective symplectic matrices entering the Weyl-symbol
and trace formulas have no eigenvalue equal to unity.
When such an eigenvalue occurs, the formulas used here are
no longer directly applicable in their present form,
even though the underlying operator need not be singular.
Finally, we have not explicitly tracked the Maslov-type indices associated
with the double-valued metaplectic representation.
Instead, the physical branches are selected by continuity and normalization of the
CF, and positivity of the partition function. A systematic treatment of these indices
would provide a more complete mathematical formulation of the construction.

The general time-dependent one-dimensional oscillator,
whose dynamics may be described through the Ermakov equation \cite{{onah2023}}
and the Lewis-Riesenfeld invariant \cite{lewis1969}.
Expressing this solution in the present symplectic language will yield explicitly the
CF for arbitrary frequency protocols.
Another direction concerns squeezing protocols \cite{galve2009},
for the derived CF provides a natural framework for separating
state-dependent fluctuations from the explicitly quantum contribution
to the work variance.

Furthermore,
the inverse construction naturally suggests an op\-ti\-mal-control formulation
in which trajectories on the symplectic group $\operatorname{Sp}(2n,\mathbb{R})$,
together with their affine components,
are varied to minimize the nonequilibrium lag or other suitable measures of energetic cost
\cite{liu2024,shackerleybennett2017,schmiedl2007,gomezmarin2008}.
This connection may also be useful in the design and optimization of shortcuts to
adiabaticity \cite{gueryodelin2019}.

Extensions to many-body systems are particularly promising.
For coupled bosonic systems \cite{paternostro2019}, the symplectic spectrum directly
determines the instantaneous normal-mode frequencies, while the characteristic function
derived here remains formally valid for any finite number of modes. This opens the
possibility of studying work fluctuations, irreversibility, critical behavior, and protocol
optimization in harmonic networks within a unified phase-space framework. It would also be
valuable to investigate whether the formalism can be extended beyond closed unitary
dynamics. A related direction concerns its connection with the Caldirola--Kanai system
\cite{Caldirola-Kanai}, which provides an effective Hamiltonian description of the
classically damped harmonic oscillator and a well-known framework for incorporating
dissipation at the level of the Schr\"odinger equation \cite{dekker}.

In summary, the inhomogeneous
metaplectic structure provides a unified and computationally
useful description of work statistics for driven quadratic bosonic systems.
It separates the dynamical construction of the symplectic
and affine trajectory from the algebraic evaluation of
the characteristic function, accommodates
both thermal equilibrium and general nonequilibrium initial states,
and recovers the principal quantum
thermodynamic relations and limiting regimes within a single framework.

\bigskip

\noindent \textit{Use of AI tools.---}
AI-assisted tools were used solely for language polishing,
including improvements to grammar, clarity, and readability.
No AI tool was used to generate scientific results, derivations,
data, figures, or conclusions.
The authors reviewed and edited all suggested changes
and take full responsibility for the manuscript.

\bigskip

\noindent\emph{Acknowledgments.---}
F.N. is a member of the Brazilian National Institute of
Science and Technology in Quantum Devices (INCT-DQ)
[CNPq Grant No. 408783/2024-9].

\section*{Appendices}  \appendix

\section{Derivation of \texorpdfstring{$\chi_\tau(u)$}{}} \label{ap:QuadChiFunc}
In this appendix we present the calculations leading to the expression of
CF in (\ref{eq:finalQuadChi}).
This equation is immediately determined after conveniently rewriting
the product of operators in (\ref{eq:prod1}).
We start from the operator product in (\ref{eq:prod1})
and derive the operational form (\ref{eq:chiQuadFunc})
with the parameters in (\ref{eq:chiparam}),
and finally obtain the closed form (\ref{eq:finalQuadChi})
via the Weyl symbol in (\ref{eq:Hsmet}).
%

Using the product in (\ref{eq:prod1}),
repeatedly using Eq.(\ref{MSTMS}) in the form
$\hat T_{z}\hat M_\mathsf{S} = \hat M_\mathsf{S} \, \hat T_{{\mathsf S}^{-1}z}$,
and the composition (\ref{eq:met_composition}),
we move all Metaplectic operators to the left and obtain
\begin{equation*} 
\begin{aligned}
& \hat T_{ z_\tau }^\dagger \hat M_{ {\sf S}_\tau }^\dagger
  \hat T_{ z_u^{(\tau)} }^\dagger \hat M_{ {\sf S}_u^{(\tau)} }^\dagger
  \hat M_{ {\sf S}_\tau } \hat T_{ z_\tau }
  \hat M_{ {\sf S}_u^{(0)} } \hat T_{ z_u^{(0)} }
  \hat T_{ \xi_0 }   \hat M_{ {\rm S}_\beta^{(0)} }
  \hat T_{ \xi_0 }^\dagger   = \\
& \hat M_{{\bf S}^{(u,\tau)}_\beta}\;
\prod_{k=1}^6 \hat T_{\eta_k}
\end{aligned}
\end{equation*}
with ${\bf S}^{(u,\tau)}_\beta$ in (\ref{eq:chiparam})
and the residual product $\prod_{k=1}^6\hat T_{\eta_k}$ of Weyl operators
having arguments
\begin{align*}
\eta_1 & = -[{\bf S}_\beta^{(u,\tau)}]^{-1} z_\tau \, , \, \,\,\,
\eta_2 =  -[{\bf S}_\beta^{(u,\tau)}]^{-1} {\sf S}_\tau^{-1} z_u^{(\tau)} ,  \\
\eta_3 &= ({\sf S}_u^{(0)}{\rm S}_\beta^{(0)})^{-1} z_\tau \, , \,\,\,
\eta_4 = ({\rm S}_\beta^{(0)})^{-1} z_u^{(0)}\, , \\
\eta_5 &= ({\rm S}_\beta^{(0)})^{-1} \xi_0 \, , \qquad \;
\eta_6 = - \xi_0 ,
\end{align*}
where we used that $\hat T_\xi = \hat T_{-\xi}^\dagger$.

Using the Heisenberg‑group law in (\ref{eq:WeylComp}),
we will perform five successive binary compositions,
accumulating elementary phases $\{\phi_i\}_{i=1}^5$
and intermediate arguments of the Weyl operators,
until a single $\hat T_{\zeta^{(u,\tau)}_\beta}$ is obtained:
\begin{align*}
\prod_{k=1}^6 \hat T_{\eta_k} =
{\rm e}^{\tfrac{i}{2\hbar}\sum_{i=1}^5 \phi_i}
\hat T_{\zeta^{(u,\tau)}_\beta} \, , \,\,\,
\zeta^{(u,\tau)}_\beta = \sum_{i=1}^6 \eta_i \, .
\end{align*}
Note that
$\zeta^{(u,\tau)}_\beta$
is simply the sum of the arguments of all
Weyl operators $\hat T_{\eta_i}$,
resulting in the expression in (\ref{eq:chiparam}),
after substituting $z_u^{(\tau)}$ and $z_u^{(0)}$
by their expressions, both defined by $z_u^{(t)}$
in (\ref{eq:Sz_uindep}).
It is not difficult to show that the composition
of Weyl operators is associative, in such a way that
the determination of the sum of the phases can be performed
in any binary order of compositions.
We will choose a particular ordering that will be useful for future analysis.

We first consider the composition
$\hat T_{\eta_1}\hat T_{\eta_2} =
{\rm e}^{\tfrac{i}{2\hbar}\phi_1} \hat T_{\eta_1 + \eta_2}$, then
according to (\ref{eq:WeylComp}), the phase is given by
\[
\begin{aligned}
\phi_1 & = {\sf J}\eta_1 \cdot \eta_2 =
{\sf J}[{\bf S}_\beta^{(u,\tau)}]^{-1} z_\tau \cdot
[{\bf S}_\beta^{(u,\tau)}]^{-1} {\sf S}_\tau^{-1} z_u^{(\tau)} \\
& = {\sf J} z_\tau \cdot {\sf S}_\tau^{-1} z_u^{(\tau)}
= {\sf J} z_\tau \cdot {\sf S}_\tau^{-1}
  ({\sf S}_{-u}^{(\tau)} - {\sf I})\xi_\tau  \\
& = {\sf J} {\sf S}_{u}^{(\tau)} {\sf S}_\tau z_\tau \cdot
  ({\sf I} - {\sf S}_{u}^{(\tau)})\xi_\tau  \, ,
 \end{aligned}
\]
where in the second equality we used the invariance of
the symplectic product under the symplectic matrix
${\bf S}_\beta^{(u,\tau)}$ and in the last one,
we replaced $z_u^{(\tau)}$ by its explicit expression
given by Eq.(\ref{eq:Sz_uindep});
now, consider
$\hat T_{\eta_1+\eta_2} \hat T_{\eta_3} =
{\rm e}^{\tfrac{i}{2\hbar}\phi_2}
\hat T_{\eta_1 + \eta_2 + \eta_3}$, where
\[\begin{aligned}
\phi_2 & = {\sf J}(\eta_1 + \eta_2) \cdot \eta_3 \\
& = - {\sf J} [{\bf S}_\beta^{(u,\tau)}]^{-1}
          [z_\tau +{\sf S}_\tau^{-1} z_u^{(\tau)}] \cdot
          ({\sf S}_u^{(0)}{\rm S}_\beta^{(0)})^{-1} z_\tau \\
& = - {\sf J} [z_\tau +{\sf S}_\tau^{-1} z_u^{(\tau)}] \cdot
           {\sf S}_\tau^{-1} {\sf S}_{-u}^{(\tau)} {\sf S}_\tau  z_\tau \\
& = - {\sf J} [z_\tau +{\sf S}_\tau^{-1} ({\sf S}_{-u}^{(\tau)} - {\sf I})\xi_\tau]
\cdot {\sf S}_\tau^{-1} {\sf S}_{-u}^{(\tau)} {\sf S}_\tau  z_\tau \\
& = - {\sf J} [{\sf S}_{u}^{(\tau)} {\sf S}_\tau z_\tau + ({\sf I} - {\sf
S}_{u}^{(\tau)})\xi_\tau]
\cdot {\sf S}_\tau  z_\tau\, ,
\end{aligned}\]
where we used the explicit formula for ${\bf S}_\beta^{(u,\tau)}$
and for $z_u^{(\tau)}$.
For the final result, let us simplify the sum of these two first phases:
\begin{widetext}
\[\begin{aligned}
\phi_1+\phi_2 & = {\sf J} {\sf S}_{u}^{(\tau)} {\sf S}_\tau z_\tau \cdot
({\sf I} - {\sf S}_{u}^{(\tau)})\xi_\tau
- {\sf J} [{\sf S}_{u}^{(\tau)} {\sf S}_\tau z_\tau
+ ({\sf I} - {\sf S}_{u}^{(\tau)})\xi_\tau] \cdot {\sf S}_\tau  z_\tau \\
& =
{\sf J} z_\tau \cdot [{\sf S}_{-\tau} {\sf S}_{u}^{(\tau)} {\sf S}_\tau z_\tau
- {\sf S}_{-\tau}({\sf S}_{-u}^{(\tau)} + {\sf I})
                 ({\sf S}_{u}^{(\tau)} - {\sf I}  ) \xi_\tau]
= {\sf J} {\sf S}_{\tau} z_\tau \cdot
[ {\sf S}_{u}^{(\tau)} {\sf S}_\tau z_\tau
- ({\sf S}_{u}^{(\tau)}  - {\sf S}_{-u}^{(\tau)}  )\xi_\tau] \, .
\end{aligned}\]
\end{widetext}

Let us now consider $\hat T_{\eta_5} \hat T_{\eta_6} =
\hat T_{\eta_5+\eta_6}{\rm e}^{\tfrac{i}{2\hbar}\phi_5}$ with
\[
\phi_5 = {\sf J} \eta_5 \cdot \eta_6 =
- {\sf J }({\rm S}_\beta^{(0)})^{-1} \xi_0 \cdot \xi_0 =
-{\sf J } \xi_0 \cdot {\rm S}_\beta^{(0)} \xi_0 \, ;
\]
and $\hat T_{\eta_4} \hat T_{\eta_5 + \eta_6} =
\hat T_{\eta_4 +\eta_5 +\eta_6}{\rm e}^{\tfrac{i}{2\hbar}\phi_4}$
with
\[
\begin{aligned}
\phi_4 &= {\sf J} \eta_4 \cdot  (\eta_5 + \eta_6)=
{\sf J} ({\rm S}_\beta^{(0)})^{-1} z_u^{(0)}
\cdot[ ({\rm S}_\beta^{(0)})^{-1} - {\sf I}]\xi_0 \\
&= {\sf J} z_u^{(0)}
\cdot[ {\sf I} - {\rm S}_\beta^{(0)}]\xi_0 \\
&= {\sf J} ({\sf S}_{-u}^{(0)} - {\sf I}) \xi_0
\cdot( {\sf I} - {\rm S}_\beta^{(0)})\xi_0 \, ,
\end{aligned}
\]
where we used the symplecticity of ${\rm S}_\beta^{(0)}$
and replaced $z_u^{(\tau)}$ by its explicit expression
given by Eq.(\ref{eq:Sz_uindep}).
It is convenient to present the sum
\[
\begin{aligned}
\phi_4 + \phi_5 & = {\sf J} {\sf S}_{-u}^{(0)} \xi_0 \cdot
( {\sf I} - {\rm S}_\beta^{(0)})\xi_0 \\
& = - {\sf J} \xi_0 \cdot
{\sf S}_{u}^{(0)} {\rm S}_\beta^{(0)}\xi_0 + {\sf J} \xi_0 \cdot
{\sf S}_{u}^{(0)} \xi_0 \, ,
\end{aligned}
\]
where we used that ${\sf J}\xi_0 \cdot \xi_0 = 0$.

Finally,
$\hat T_{\eta_1 +\eta_2 +\eta_3} \hat T_{\eta_4 +\eta_5 +\eta_6} =
\hat T_{\zeta^{(u,\tau)}_\beta} {\rm e}^{\tfrac{i}{2\hbar}\phi_3}$ with
\[
\begin{aligned}
\phi_3 &= {\sf J} (\eta_1 +\eta_2 +\eta_3) \cdot (\eta_4 + \eta_5 + \eta_6) \\
& = {\sf J} [\zeta^{(u,\tau)}_\beta -(\eta_4 + \eta_5 + \eta_6) ]
\cdot (\eta_4 + \eta_5 + \eta_6) \\
& = {\sf J} \zeta^{(u,\tau)}_\beta \cdot (\eta_4 + \eta_5 + \eta_6) \\
& = {\sf J} \zeta^{(u,\tau)}_\beta \cdot
[ ({\rm S}_u^{(0)}{\rm S}_\beta^{(0)})^{-1} - {\sf I} ] \xi_0 \\
& = {\sf J}\xi_0 \cdot
( {\sf I} - {\rm S}_u^{(0)}{\rm S}_\beta^{(0)})\zeta^{(u,\tau)}_\beta \, ,
\end{aligned}
\]
where we used the values of $\eta_i$ replacing $z_u^{(0)}$
by its explicit expression given by Eq.(\ref{eq:Sz_uindep}).
%
%

Taking the trace of the operator in (\ref{eq:prod1}),
we find Eq.(\ref{eq:chiQuadFunc}) with
\[
\Phi_\beta^{(u,\tau)} = \tfrac{1}{2} \sum_{i=1}^5 \phi_i +
\varphi_u^{(\tau)}-\varphi_u^{(0)}
\]
for
\[
\varphi_u^{(\tau)}-\varphi_u^{(0)}  = (h_\tau - h_0) u
+  \tfrac{1}{2}{\sf J} \xi_\tau \cdot {\sf S}_u^{(\tau)} \xi_\tau
-  \tfrac{1}{2}{\sf J} \xi_0 \cdot {\sf S}_u^{(0)} \xi_0 \, ,
\]
where we used the expression for $\varphi_u^{(t)}$ in (\ref{eq:Utindep3}).
Now, using the expressions for the phases calculated previously,
and noting that the last terms in $\varphi_u^{(\tau)}-\varphi_u^{(0)}$
cancels with the last term in $\phi_4 + \phi_5$,
the phase $\Phi_\beta^{(u,\tau)}$ becomes the one in
(\ref{eq:chiparam}).

\section{Jarzynski Equality} \label{ap:Jar}
In this Appendix, we will show that the final expression of the characteristic
function for a generic QH in (\ref{eq:finalQuadChi})
reproduces the Jarzynski equality in (\ref{eq:ebetaW}).

Applying the Wick rotation $u = i\hbar\beta$ to the symplectic matrix
${\sf S}_{u}^{(t)}$ defined in Eq.(\ref{eq:Sz_uindep}), we find
\begin{equation}\label{eq_ap:jar1}
{\sf S}_{u = i\hbar\beta}^{(t)} = {\rm S}_{-\beta}^{(t)} =
[{\rm S}_{\beta}^{(t)}]^{-1} \, , \forall t,
\end{equation}
where ${\rm S}_{\beta}^{(t)}$ is defined in (\ref{eq:th_oper}).
Consequently, the quantities in (\ref{eq:chiparam}) become
\begin{equation} \label{eq_ap:jar2}
\begin{aligned}
{\bf S} & :=
{\bf S}_\beta^{(u = i\hbar\beta,\tau)} =
{\sf S}_\tau^{-1} {\rm S}_{\beta}^{(\tau)}  {\sf S}_\tau \, ; \\
{\bm \zeta} & := \zeta_\beta^{(u= i\hbar\beta,\tau)}
%
= ( {\sf I} -{\bf S}^{-1} ) (z_\tau - {\sf S}_\tau^{-1}\xi_\tau) \, .  \\
\end{aligned}
\end{equation}

Using the expressions for the phases derived in Appendix \ref{ap:QuadChiFunc},
one can see that $\phi_3|_{u = i\hbar\beta} = \phi_4|_{u = i\hbar\beta} = 0$
solely due to (\ref{eq_ap:jar1}).
Thus, we can write $\Phi_\beta^{(u = i\hbar\beta,\tau)} =
(\phi_1 + \phi_2 + \phi_5)|_{u = i\hbar\beta}$,
which can help in the identification of null terms when substituting
$u = i\hbar\beta$ in the phase $\Phi_\beta^{(u,\tau)}$, defined in Eq.(\ref{eq:chiparam}),
which becomes
\begin{widetext}
\[\begin{aligned}
\Phi_\beta^{(u = i\hbar\beta,\tau)} & = i\hbar \beta (h_\tau - h_0)
-  \tfrac{1}{2}{\sf J} \xi_\tau \cdot {\sf S}_\beta^{(\tau)} \xi_\tau
+ \tfrac{1}{2}{\sf J} {\sf S}_{\tau}
z_\tau \cdot
[ {\rm S}_{-\beta}^{(\tau)} {\sf S}_\tau z_\tau
- ({\rm S}_{-\beta}^{(\tau)}  - {\rm S}_{\beta}^{(\tau)}  )\xi_\tau] \\
%
%
& = i\hbar \beta (h_\tau - h_0)
-  \tfrac{1}{2}{\sf J} \xi_\tau \cdot {\sf S}_\beta^{(\tau)} \xi_\tau
+ \tfrac{1}{2}{\sf J} z_\tau \cdot
[{\bf S}^{-1} {v} + {\bf S} \, {\sf S}_{-\tau}\xi_\tau] \\
& = i\hbar \beta (h_\tau - h_0)
- \tfrac{1}{2}{\sf J} \xi_\tau \cdot {\sf S}_\beta^{(\tau)} \xi_\tau
+ \tfrac{1}{2}{\sf J} {v} \cdot
[{\bf S}^{-1} {v} + {\bf S} \,  {\sf S}_{-\tau}\xi_\tau]
+ \tfrac{1}{2}{\sf J} {\sf S}_\tau^{-1}\xi_\tau \cdot
[{\bf S}^{-1} {v} + {\bf S} \,  {\sf S}_{-\tau}\xi_\tau]  \\
& = i\hbar \beta (h_\tau - h_0)
+ \tfrac{1}{2}{\sf J} {v} \cdot {\bf S}^{-1} {v}
+ \tfrac{1}{2}{\sf J} {v} \cdot {\bf S} {\sf S}_{-\tau}\xi_\tau
+ \tfrac{1}{2}{\sf J} {\sf S}_\tau^{-1}\xi_\tau \cdot {\bf S}^{-1} {v} \, ,
\end{aligned}\]
\end{widetext}
where
\begin{equation}\label{eq_ap:jar3}
v := (z_\tau - {\sf S}_\tau^{-1}\xi_\tau)
\end{equation}
is defined for convenience;
from the first to the second line, we used that
the expression for ${\bf S}$ in (\ref{eq_ap:jar2}),
after performing some algebraic manipulations with symplectic matrices;
from the second to the third,
we summed and subtracted the last term in the third line;
in the fourth line, we used the symplectic condition for the matrices
and recognized that
${\sf J} {\sf S}_\tau^{-1}\xi_\tau \cdot{\bf S}{\sf S}_\tau^{-1}\xi_\tau =
\tfrac{1}{2}{\sf J} \xi_\tau \cdot {\sf S}_\beta^{(\tau)} \xi_\tau $;
finally, noting that
${\sf J} {v} \cdot {\bf S} {\sf S}_{-\tau}\xi_\tau =
-{\sf J} {\sf S}_\tau^{-1}\xi_\tau \cdot {\bf S}^{-1} {v}$,
we attain the following simple form for the phase:
\begin{equation} \label{eq_ap:jar5}
\Phi_\beta^{(u = i\hbar\beta,\tau)} = i\hbar \beta (h_\tau - h_0)
+ \tfrac{1}{2}{\sf J} v \cdot {\bf S} v \, ,
\end{equation}
for $\bf S$ in (\ref{eq_ap:jar2}) and $v$ in (\ref{eq_ap:jar3}).

For $\bm \zeta$ and $\bf S$ both defined in (\ref{eq_ap:jar2}), $v$ defined in (\ref{eq_ap:jar3}),
and $\mathbf{C}_{\bf S}$ in (\ref{eq:Cayley}),
let us now deal with the following expression:
\begin{equation}\label{eq_ap:jar6}
\begin{aligned}
{\bm \zeta} \cdot {\sf J}\mathbf{C}_{\bf S}^{-1} {\sf J}{\bm \zeta} & =
( {\sf I} -{\bf S}^{-1} ) v \cdot {\sf J} \mathbf{C}_{\bf S}^{-1}{\sf J}
( {\sf I} -{\bf S}^{-1} ) v \\
& = v \cdot {\sf J} ( {\sf I} -{\bf S}) \mathbf{C}_{\bf S}^{-1}{\sf J}
({\sf I} - {\bf S}^{-1}) v \\
& = {\sf J} v \cdot ( {\bf S} + {\sf I})( {\sf I} - {\bf S}^{-1}) v \\
& = {\sf J} v \cdot ( {\bf S} - {\bf S}^{-1}) v = 2 {\sf J} v \cdot {\bf S} v \, ,
\end{aligned}
\end{equation}
where from the first to the second line we used
the symplecticity of ${\bf S}$
alongside the (bi)linearity of the dot product;
in the third line, we inserted the expression for $\mathbf{C}_{\bf S}$;
and in the last equality, again we employed the symplecticity of ${\bf S}$
alongside the (bi)linearity of the dot product.
As one can see, the above term is related to
the CF (\ref{eq:finalQuadChi}) due to
\[
[\zeta_\beta^{(u,\tau)} \cdot {\sf J}\mathbf{C}_{{\bf S}_\beta^{(u,\tau)}}^{-1}
{\sf J}\zeta_\beta^{(u,\tau)}]|_{u=i\hbar\beta}  =
{\bm \zeta} \cdot {\sf J}\mathbf{C}_{\bf S}^{-1} {\sf J}{\bm \zeta} \, .
\]

We can now evaluate the final form of
the CF (\ref{eq:finalQuadChi}) for $u = i\hbar\beta$:
\[
\begin{aligned}
\chi_\tau (i\hbar\beta)  &=
\frac{ {\rm e}^{ \frac{i}{\hbar} \Phi_\beta^{(u = i\hbar\beta,\tau)}
                 -\frac{i}{4\hbar} {\bm \zeta}_\cdot
                  {\sf J}\mathbf{C}_{\bf S}^{-1}
                  {\sf J}{\bm \zeta}}                }
     { \sqrt{\det[ {\bf S} - {\sf I}]
                 [ {\rm S}_\beta^{(0)} - {\sf I}]^{-1}}  } \\
&= {\rm e}^{-\beta(h_\tau-h_0)}
\sqrt{
\frac{
|\det[ {\rm S}_\beta^{(0)} - {\sf I}]|
}{
|\det[ {\bf S} - {\sf I}]|
}
}
= {\rm e}^{-\beta\Delta F} \, ,
\end{aligned}
\]
where we employed Eqs.(\ref{eq_ap:jar5}) and (\ref{eq_ap:jar6}) and
using the definition of $\bf S$ in the argument of the determinant
to find the expression in (\ref{expW:cl}),
which is the Jarzynski equality (\ref{eq:ebetaW}) for
QHs.

\section{Work Mean-Value} \label{ap:meanvalue}
In order to obtain the mean value of work from (\ref{eq:finalQuadChi}),
we need to perform its derivative with respect to $u$
and then evaluate at $u = 0$, see Eq.(\ref{eq:Wmoments}).

Basically, the only object in (\ref{eq:chiparam})
that depends on $u$ is the real symplectic matrix
${\sf S}_u^{(t)}$ ---
its expression is given in (\ref{eq:Sz_tindep}) and its derivative
with respect to $u$ is given in (\ref{eq:difSz_uindep}).
However, each object appearing in (\ref{eq:chiparam})
is a non-linear function of ${\sf S}_u^{(t)}$,
since there are instances for $t = 0$ and for $t = \tau$.
Apart from the derivatives,
any other manipulation to be performed in the calculations follows
the methods presented in Appendices \ref{ap:QuadChiFunc} and \ref{ap:Jar}.

First, we note for the parameters in (\ref{eq:chiparam}) that
\[
{\bf S}_\beta^{(0,\tau)} = {\rm S}_\beta^{(0)} \, , \,\,\,
\zeta_\beta^{(0,\tau)} = ( {\rm S}_{-\beta}^{(0)} - {\sf I} ) \xi_0.
\]
Taking now the derivative, we obtain
\[
\begin{aligned}
\tfrac{d}{du}{\bf S}_\beta^{(u,\tau)}\Big|_{u=0} & =
{\sf S}_{\tau}^{-1} {\sf S}_{0}^{(\tau)}
(-{\sf J}{\bf H}_\tau {\sf S}_\tau   +
{\sf S}_\tau {\sf J}{\bf H}_0)  {\sf S}_0^{(0)} {\rm S}_\beta^{(0)} \\
& = {\sf J} ({\bf H}_0 - {\sf S}_{\tau}^\top {\bf H}_\tau {\sf S}_{\tau})
{\rm S}_\beta^{(0)} \, ,
\end{aligned}
\]
where we used the expression of ${\sf S}_{u}^{(t)}$ in (\ref{eq:Sz_uindep})
and the symplecticity of ${\sf S}_\tau$.
Similar calculations are enough to determine
\[
\tfrac{d}{du} \zeta_\beta^{(u,\tau)} \Big|_{u=0} =
- {\rm S}_{-\beta}^{(0)} {\sf J}
[{\bf H}_0 \xi_0 + {\sf S}_{\tau}^\top {\bf  H}_{\tau} {\sf S}_{\tau}
( z_\tau - {\sf S}_{-\tau} \xi_\tau)]
\]

Differentiation of $\Phi_\beta^{(u,\tau)}$ using the same recipe furnishes
\[
\begin{aligned}
\tfrac{d}{du} \Phi_\beta^{(u,\tau)} \Big|_{u=0} &=
(h_\tau - h_0)
+  \tfrac{1}{2} \xi_\tau \cdot {\bf H}_{\tau} \xi_\tau
-  \tfrac{1}{2} \xi_0 \cdot {\bf H}_0 {\rm S}_\beta^{(0)}\xi_0 \nonumber\\
& - \tfrac{1}{2} ({\rm S}_\beta^{(0)} - {\sf I})\xi_0
\cdot {\sf S}_{\tau}^\top{\bf H}_{\tau}{\sf S}_{\tau}
( z_\tau - {\sf S}_{-\tau} \xi_\tau) \\
& + \tfrac{1}{2} z_\tau
\cdot {\sf S}_{\tau}^\top{\bf H}_{\tau}{\sf S}_{\tau}
( z_\tau - 2{\sf S}_{-\tau} \xi_\tau) \, .
\end{aligned}
\]

Using that
\begin{equation}\label{eq_ap:derinv}
\tfrac{d}{du} {\bf A}^{-1} = -{\bf A}^{-1}(\tfrac{d}{du}{\bf A}){\bf A}^{-1},
\end{equation}
we obtain the derivative of the inverse
of Cayley parametrization appearing in (\ref{eq:finalQuadChi}):
\[
\begin{aligned}
& \tfrac{d}{du}{\bf C}_{{\bf S}_\beta^{(u,\tau)}}^{-1}\Big|_{u = 0} = \\
&- 2 ({\rm S}_\beta^{(0)}-{\sf I})^{-1}
\tfrac{d}{du}{\bf S}_\beta^{(u,\tau)}\Big|_{u=0}
({\rm S}_\beta^{(0)}-{\sf I})^{-1} {\sf J} \, .
\end{aligned}
\]

The Jacobi identity for the derivative of a determinant \cite{horn2013},
\begin{equation}\label{eq_ap:jacobi}
\tfrac{d}{du}\text{det}(\mathbf{A}) =
\text{det}(\mathbf{A}) \text{Tr}(
\mathbf{A}^{-1} \frac{d}{du}\mathbf{A} ),
\end{equation}
enables us to determine
\[
\begin{aligned}
& \tfrac{d}{du} \det (\mathbf{S}_\beta^{(u,\tau)} - {\sf I})\Big|_{u=0} = \\
& \det({\rm S}_\beta^{(0)} - {\sf I})
{\rm Tr}\left[({\rm S}_\beta^{(0)} - {\sf I})^{-1}
\tfrac{d}{du}{\bf S}_\beta^{(u,\tau)}\Big|_{u=0}\right] \, .
\end{aligned}
\]

The derivative with respect to $u$ of the expression in (\ref{eq:finalQuadChi})
is now obtained by combining all above calculations and,
after some simplifications, we obtain
\[
\begin{aligned}
\overline{W} =
& - \tfrac{i\hbar}{2} \, \text{Tr}[ {\sf J}
          ( {\sf S}_{\tau}^\top  {\bf H}_\tau {\sf S}_{\tau} - {\bf H}_0 )
          ( {\sf I}_{2n} - {\rm S}_{-\beta}^{(0)})^{-1}]\\
&  + \tfrac{1}{2} (z_\tau + \xi_0 - {\sf S}_\tau^{-1} \xi_\tau) \cdot
{\sf S}_\tau^{\top}{\bf H}_\tau {\sf S}_\tau
(z_\tau + \xi_0 - {\sf S}_\tau^{-1} \xi_\tau) \\
& + (h_\tau-h_0) \, .
\end{aligned}
\]

Considering the identity in (\ref{eq:ident_S-V}),
using the expression for ${\mathbf V}_0^\text{th}$ given by Eq.(\ref{eq:mvcmDefs}),
and noting that ${\rm Tr}({\sf J}{\bf M}) = 0$, for a symmetric $\bf M$,
we find the expression in (\ref{eq:Wmv}).

\clearpage


\end{document}